\documentclass[prd,aps,twocolumn,a4paper,showkeys,nofootinbib,floatfix]{revtex4-1}

\usepackage{graphicx,psfrag}
\usepackage{mathrsfs}
\usepackage{amsmath,amsfonts,amssymb}
\usepackage{multirow}
\usepackage{comment}
\usepackage{tipa}
\usepackage{ulem}
\usepackage{hyperref}
\usepackage{enumitem}
\usepackage{tipa}
\usepackage{bm}
\usepackage{algorithm}
\usepackage{algpseudocode}
\usepackage{lipsum}

\newcommand{\be}{\begin{equation}}
\newcommand{\ee}{\end{equation}}
\newcommand{\bea}{\begin{eqnarray}}
\newcommand{\eea}{\end{eqnarray}}
\newcommand{\bel}{\begin{align}}
\newcommand{\eel}{\end{align}}

\def\i{{\rm i}}
\def\srate{F_{\rm s}}
\def\fNyq{f_{\rm Nyq}}
\def\fmin{f_{\rm min}}
\def\fmax{f_{\rm max}}

\def\GMc2{{\rm G M_{\odot} c^{-2}}}

\def\M{\mathcal{M}}
\def\B{\mathcal{B}}

\def\O{\mathcal{O}}

\def\Mo{{\rm M_{\odot}}}
\def\kt2{\kappa^\text{T}_2}

\def\params{\boldsymbol{\theta}}
\def\spin{\boldsymbol{\chi}}

\def\paramspace{\boldsymbol{\Theta}}

\def\data{\textbf{d}}

\def\d{\text{d}}
\def\tLam{{\tilde\Lambda}}
\def\dLam{{\delta\tilde\Lambda}}

\usepackage{pifont}

\newcommand{\bajes}{{\tt bajes}}
\newcommand{\py}{\textsc{Python}}

\usepackage{color}
\definecolor{cyan}{rgb}{0,0.9,0.9}
\definecolor{orange}{rgb}{0.9,0.5,0}
\definecolor{magenta}{rgb}{1,0,1}
\definecolor{purple}{rgb}{0.8,0.4,0.8}
\definecolor{gray}{rgb}{0.8242,0.8242,0.8242}

\begin{document}

\title{{\bajes}: Bayesian inference of multimessenger astrophysical data \\
  Methods and application to gravitational-waves}

\author{Matteo \surname{Breschi}}
\affiliation{Theoretisch-Physikalisches Institut, Friedrich-Schiller-Universit{\"a}t Jena, 07743, Jena, Germany}
\author{Rossella \surname{Gamba}}
\affiliation{Theoretisch-Physikalisches Institut, Friedrich-Schiller-Universit{\"a}t Jena, 07743, Jena, Germany}
\author{Sebastiano \surname{Bernuzzi}}
\affiliation{Theoretisch-Physikalisches Institut, Friedrich-Schiller-Universit{\"a}t Jena, 07743, Jena, Germany}

\date{\today}

\begin{abstract}
  We present {\bajes}, a parallel and lightweight
  framework for Bayesian inference
  of multimessenger transients.
  {\bajes} is a {\py} modular package with minimal dependencies on
  external libraries adaptable to the majority of the Bayesian models
  and to various sampling methods. 
  We describe the general workflow 
  and the parameter estimation pipeline for compact-binary-coalescence
  gravitational-wave transients. The latter is validated against
  injections  
  of binary black hole and binary neutron star waveforms, including
  confidence interval tests that demonstrates the inference is well-calibrated.     
  Binary neutron star postmerger injections are also studied 
  using a network of five detectors made of LIGO, Virgo, KAGRA
  and Einstein Telescope. Postmerger signals will be detectable
  for sources at ${\lesssim}80\,$Mpc,
  with Einstein Telescope contributing over 90\% of the total signal-to-noise ratio.
  As a full scale application, we re-analyze the GWTC-1 black hole transients using the 
  effective-one-body {\tt TEOBResumS} approximant, and reproduce
  selected results with other approximants.
  {\tt bajes} inferences are 
  consistent with previous results;
  the direct comparison of {\bajes} and {\tt bilby} analyses of GW150914 
  shows a maximum Jensen-Shannon divergence of $5.2{\times}10^{-4}$.
  GW170817 is re-analyzed using
  {\tt TaylorF2} with 5.5PN point-mass and 7.5PN tides,
  {\tt TEOBResumSPA},
  and {\tt IMRPhenomPv2\_NRTidal} with different cutoff-frequencies of $1024\,$Hz and $2048\,$Hz.
  We find that the former choice minimizes systematics on the reduced tidal
  parameter, while a larger amount of tidal information is gained with the latter choice. 
  {\bajes} can perform these analyses in about 1~day using 128 CPUs.
\end{abstract}

\pacs{
  04.25.D-,  
  04.30.Db,  
  95.30.Sf,    
  95.30.Lz,   
  97.60.Jd     
}

\maketitle

\section{Introduction} 
\label{sec:intro}

Bayesian inference has become a standard method for the analysis of
astrophysical and cosmological
events, e.g.~\cite{LIGOScientific:2018mvr,DelPozzo:2018dpu,Wang:2018lbp,Abbott:2020niy,Hort_a_2020},  
since it offers a generic statistical framework
to rigorously test hypothesis 
against observational information.
Given a set of parametric models (hypothesis) and assumptions on the
parameters (prior), Bayesian inference allows one to infer the
parameters in terms of probability distributions, 
and also to select the best-fitting model 
among competing hypotheses.
In particular, Bayesian methods are central tools used in 
gravitational-wave (GW) astronomy 
to determine the source properties of an observed signal~\cite{Veitch:2009hd,Veitch:2014wba,LIGOScientific:2019hgc} 
and the related applications.
Some examples are 
tests of General Relativity~\cite{LIGOScientific:2019fpa,Breschi:2019wki}, 
astrophysical population studies~\cite{Abbott:2020gyp}, 
inferences of the neutron star matter properties~\cite{Abbott:2018exr}
and cosmology~\cite{Abbott:2017xzu,Abbott:2019yzh}.
Furthermore,
Bayesian inference offers the optimal framework to combine different
observational datasets from multimessenger astronomical observations,
like GW170817 and the electromagnetic (EM) counterparts~\cite{TheLIGOScientific:2017qsa,Monitor:2017mdv,Savchenko:2017ffs,Pian:2017gtc,Smartt:2017fuw,Tanvir:2017pws,Tanaka:2017qxj,Villar:2017wcc}.
Multimessenger inference of astrophysical phenomena 
such as binary neutron star mergers (BNS) is a fundamental resource 
to clarify the mechanism at the origin of the radiation, 
to obtain accurate inferences on the properties of the source, 
and to improve theoretical models
gaining information from observational data \citep[e.g.][]{Radice:2018ozg,Dietrich:2020lps,Breschi:2021tbm}.

In the last years, many efforts have been made by the scientific community 
in the development of sophisticated parameter inference 
tools for astronomical observations.
In the context of GW data analysis,
{\tt LALSuite}~\cite{lalsuite} is the official software provided by the LIGO-Virgo collaboration~\cite{TheLIGOScientific:2014jea,TheVirgo:2014hva} 
and it offers a variegated framework for generic analysis of GW data.
Other mature software for parameter estimation of GW transients are also available; 
some example are 
the semi-analytical integrator {\tt RIFT}~\cite{Pankow:2015cra,Lange:2018pyp},
the user-friendly library {\tt bilby}~\cite{Ashton:2018jfp}
and the inference module of the {\tt pycbc} package~\cite{Biwer:2018osg}.
Bayesian software for parameter inference of 
other astrophysical transients have also been developed, such as tools for 
 high-energy photons from compact objects and galaxy clusters ~\cite{2014ascl.soft08004N,heasarc_software,Olamaie:2013vfa};
 neutrino radiation~\cite{Aartsen:2019mbc};
 supernovae transients~\cite{Shariff:2015yoa,Hinton_2019,M_ller_2019};
 pulsar arrival timings~\cite{Edwards:2006zg,Lentati_2013,luo2021pint};
 and for cosmological inferences~\cite{Lewis:2002ah,Das:2014mda,Ishida:2015wla,Handley_2015}.
Current pipelines 
for the analysis of astrophysical and cosmological observations
are targeted to specific applications.
However, within a multi-messenger framework,
it is essential to develop a flexible pipeline
capable of combining different datasets and physical models.
This issue can be tackled allowing the infrastructure to
merge different Bayesian models, extending the considered parameter space 
and generalizing the definition of the likelihood function.
This implies the use of large amounts of data and 
computationally expensive models. It follows that efficient parallelization
techniques and well-calibrated proposal methods are necessary to 
optimize the performances of such a flexible pipeline.

In this work, we present 
{\bajes} [\textipa{baIEs}], a {\py} package for Bayesian
inference developed at Friedrich-Schiller-Universt\"at Jena.
Our goal is to provide a simple, complete and reliable
implementation capable to robustly perform Bayesian inference on
arbitrary sets of data, with specific functionalities for multimessenger astrophysics.
The software is designed to be state-of-art, simple-to-use and light-weighted
with minimal dependencies on external libraries.
The paper is structured as follows.
In Sec.~\ref{sec:inference}, we recall the basic concepts of Bayesian theory of 
probability.
In Sec.~\ref{sec:design}, we describe the design and the workflow of the {\bajes}
software.
Sec.~\ref{sec:gw} describes the tools and the methods implemented in {\bajes}
for the analysis of GW transients,
including the available templates.
In Sec.~\ref{sec:pipe}, we describe the GW pipeline and 
the Bayesian framework of the GW model. 
In Sec.~\ref{sec:inj}, we present a survey of injection studies and 
validation tests performed with artificial binary merger signals.
Sec.~\ref{sec:lvc} shows the results of the {\bajes} pipeline on the 
GW events observed by the LIGO-Virgo interferometers~\cite{LIGOScientific:2018mvr,Abbott:2019ebz}.
Finally, we conclude in Sec.~\ref{sec:conclusion}.
The paper concludes with Appendices on the implemented sampling methods, the proposal methods and a simple use example.

\section{Bayesian inference}
\label{sec:inference}

The task of a Bayesian inference is 
the formulation and the computation of conditional probabilities.
It is possible to classify this topic in two main problems:
parameter estimation (PE) and model selection (MS).
With PE we mean the evaluation 
of the characteristic distribution for the parameters that define the model of interest. 
On the other hand, with MS we refer to
the discrimination between competing models in light of the data,
comparing the suitability of different assumptions directly on the observation.
In order to discuss how these tasks are achieved, in the following sections
we recall the basic concepts of Bayesian theory of probability.
By convention, we label the natural logarithm as $\log$ throughout all the paper.

\subsection{Bayes' theorem}
\label{sec:bayes}

Given a set of observed data $\data$
and a set of parameters $\params$, 
that characterizes our model within some background hypothesis $H$,
it is possible to estimate the posterior distribution for $\params$ using 
the Bayes' theorem~\cite{MacKay,gelmanbda04,Sivia2006},
\be
\label{eq:bayestheo}
p(\params|\data,H) = \frac{p(\data|\params,H) \, p(\params|H)}{p(\data|H)}\,,
\ee
where $p(\data|\params,H)$ is the likelihood function,
$p(\params|H)$ is the prior distribution 
and $p(\data|H)$ is the evidence.
The likelihood function describes 
the probability of observing the data $\data$ given $\params$ and assuming that the hypothesis $H$ is true.
Therefore, it encodes the observational information
and it predicts the agreement between the observed data $\data$ 
and the expected outcome for every given sample $\params$
of the parameter space.
The prior distribution $p(\params|H)$ depicts the knowledge on the 
parameters before performing the estimation. 
Usually, the functional form of this term is chosen in accordance with
geometrical and/or physically-motivated argumentation.
The term $p(\data|H)$ is 
labeled as evidence and it represents the probability 
of observing the data $\data$ given the hypothesis $H$.
The evidence is also called marginalized likelihood
since, according to the marginalization rule,
it can be expressed as
\be
\label{eq:evidence}
p(\data|H) = \int_{\paramspace} p(\data|\params,H) \, p(\params|H)\, \d\params \,,
\ee
where the integral is extended over the entire parameter space $\paramspace$.
Subsequently, the posterior distribution $p(\params|\data,H)$
represents the probability of the parameters $\params$ in light of the data
overhauled by our {\it a priori} information. 
The knowledge of $p(\params|\data,H)$ allows us 
to compute the expectation of the statistical quantities of interest.
For example, the mean values $E[\params]$ can be estimated as
\be
\label{eq:mean}
E[\params]= \int_{\paramspace} \params \, p(\params|\data,H) \, \d\params\,,
\ee
and, analogously, it is possible to infer the expectation of a generic function of the parameters $\params$,
\be
\label{eq:expect}
E\left[ f(\params) \right] = \int_{\paramspace} f(\params) \, p(\params|\data,H) \, \d\params\,.
\ee
From this argumentation it follows that,
in order to perform PE, we have to introduce a 
prior distribution $p(\params|H)$ 
and a likelihood function $p(\data|\params,H)$;
then, the properties of the model are encoded in the posterior 
distribution $p(\params|\data,H)$ that can be computed imposing Eq.~\eqref{eq:bayestheo}. 

\subsection{Model selection}
\label{sec:ms}

In Eq.~\eqref{eq:bayestheo}, for a fixed set of assumptions $H$, 
the evidence acts like a normalization constant;
however, this quantity plays a crucial role in the context of MS.
If we are interested in comparing two competing hypotheses, $H_A$ and $H_B$, 
quantifying which one is better explaining the data,
in the Bayesian framework it is natural to introduce the odds ratio,
\be
\label{eq:oddratio}
\O_A^B = \frac{p(H_B|\data)}{p(H_A|\data)} = \frac{p(H_B)}{p(H_A)} \, \frac{p(\data|H_B)}{p(\data|H_A)}\,.
\ee  
The term $p(H_i|\data)$ represents the posterior probability for the $i$-th hypothesis given the observed data
and the ratio $\O_A^B$ encodes the will of the data in 
favoring one model over another.
Assuming that the two hypotheses are equiprobable $p(H_B)=p(H_A)$, 
it is possible to reduce the computation to the ratio of the evidences, also known as Bayes' factor,
\be
\label{eq:bayesfact}
\B_A^B = \frac{p(\data|H_B)}{p(\data|H_A)}\,.
\ee  
If $\B_A^B < 1$ then the hypothesis $A$ is preferred by the data,
otherwise $B$ is favored if $\B_A^B > 1$.
However, this rule is not always straightforward since 
the estimation of the Bayes' factor might suffer of uncertainties~\cite{mattei2020parsimonious,Yao:2019}.
Then, in a realistic scenario,
more stringent bounds are required in order to prefer a hypothesis~\cite{Kass:1995}.

\subsection{Joint distributions}

Let us assume that we performed two independent 
observations, $\data_1$ and $\data_2$, 
and each of them can be modeled  using two sets of parameters, 
respectively $\params_1$ and $\params_2$.
In general, it is possible to apply Bayes' theorem, Eq.~\eqref{eq:bayestheo},
separately to every set of measurements.
However, if the two events are not independent
(e.g. the same physical process observed by two different observatories),
 the joint posterior distribution can be written as 
 \be
 \label{eq:bayesjoint}
 p(\params_1,\params_2|\data_1,\data_2,H) = \frac{p(\data_1,\data_2|\params_1,\params_2,H) \, p(\params_1,\params_2|H)}{p(\data_1,\data_2|H)}\,,
 \ee
where $p(\data_1,\data_2|\params_1,\params_2,H)$ is the joint likelihood function and
$p(\params_1,\params_2|H)$ is the joint prior distribution.
If the employed set of parameters $\params_1$ and 
$\params_2$ are independent, the joint probabilities simply correspond to the product of the single probability terms.
However, if $\params_1$ and $\params_2$ correlate,
the problem could easily become non-trivial; e.g.
 the intersection between the two parameter spaces $\paramspace_1 \cap \paramspace_2$ could be not empty,
 or the value of the two sets of parameters could be related with each other $\params_i\equiv \params_i(\params_j)$,
 or unexpected correlations might appear introducing a larger parameter space.
 
% A common situation is represented by the case in which 
% the intersection between the two parameter spaces is not empty,
% $\paramspace_1 \cap \paramspace_2\ne \varnothing$; i.e.
%the input parameter spaces contain a common parameter, 
% \be
% \params_1 = \{\theta^*_1,\theta^{a}_1,\theta^{b}_1,\dots\}\,, \quad  
% \params_2 = \{\theta^*_2,\theta^{\alpha}_2,\theta^{\beta}_2,\dots\}\,,
% \ee
% such that $p(\theta^*_1|H)=p(\theta^*_2|H)$. In this case,
%  the overall prior 
% can be computed as
% \be
% p(\params_1,\params_2|H)= p(\params_1|H)\,p(\params_2|H)\,\delta(\theta^*_1-\theta^*_2)\,,
% \ee
% where $\delta$ is the Dirac delta function.
% Analogously, a {\tt JointLikelihood} is  a collection of {\tt Likelihood} objects,
% t$assumes the single likelihood functions to be disjointed, i.e.
% \be
% \label{eq:disjointlike}
% p(\data_1,\data_2|\params_1,\params_2,H)  = p(\data_1|\params_1,H) \, p(\data_2|\params_2,H) \,.
% \ee
% The {\tt JointPrior} and {\tt JointLikelihood} 
% objects can be extended to an arbitrary number of models.

\subsection{Samplers} 
\label{sec:sampler}
 
 In a realistic scenario, the form of the likelihood function is not always analytically determinable and
 the parameter space has usually a large number of dimensions. For these reasons,
 the evaluation of the posterior distribution and the estimation of its integral 
 are performed with stochastic techniques. In particular,
 sampling methods have proven to be reliable and generic tools
for the analysis of non-analytical forms and multi-dimensional parameter
spaces~\cite{Allison_2013,Veitch:2014wba,Tichy_2015,Handley_2015}, capable to give robust and stable results.
 
Markov-chain Monte Carlo (MCMC) methods are suitable tools to perform PE,
exploring the parameter space through random-walks 
and collecting samples along the evolution.
Subsequently, the posterior distribution is estimated 
using the set of independent samples of the parameter space.
Nevertheless,
the nested sampling~\cite{Skilling:2006} is an advanced algorithm capable to 
extract posterior samples and perform accurate estimation of the evidence,
which is the key quantity for MS. In order to solve this task,
each sample is assumed to be the representative of a isoprobability contour.
Then, the evidence is computed as the sum of the likelihood values 
weighted on the respective prior volume, 
estimated resorting to Bayesian calculations.
The details of the sampling methods are discussed
in App.~\ref{app:mcmc} and App.~\ref{app:nest}.

\section{Design of the code} 
\label{sec:design}

{\bajes} is a pure {\py} software that aims to provide a versatile and robust framework 
for generic Bayesian inference 
within a simple and clear approach.  
In order to achieve this task, 
the software relies on a modular and composed 
architecture and it implements logically specialized objects.
Furthermore, we decide to keep a light-weight setup
with minimal
  dependencies on external libraries.
These properties make {\bajes} a simple and general tool with a wide range of applicability.
The body of the {\bajes} software is constituted by two components:
the {\tt inf} module, that represents an implementation of the Bayesian logic,
and the {\tt obs} module, that contains the tools to manage and process 
physical data. 

The {\tt inf} module is the Bayesian skeleton of the software
and contains the methods required to instantiate the model to be inferred.
Sec.~\ref{sec:workflow} describes the tools implemented in {\tt bajes.inf} 
and the general workflow of the module.
The Bayesian approach is 
constituted by three fundamental stages~\cite{gelman2008}: 
formulating a model, 
comparing the model with the data 
and inferring the properties of the model.
The goal of the {\tt inf} module is to provide a 
flexible and general interface capable to adapt itself to a broad variety of problems.
This structure promotes usability and applicability, 
supplying a comprehensive and unique architecture
that allows the user to tackle specific problems.
In order to enforce these concepts, 
the software is developed promoting an intuitive manageability 
and the simplicity of use:
providing the necessary basic information, the user can easily set up a full 
Bayesian analysis. 
App.~\ref{app:example} shows an example of a practical
PE analysis with {\tt bajes} tools.

In the context of data analysis,
the statistical infrastructure has to be flanked by the physical 
characterization of the experimental data with the purpose of defining a full Bayesian model. 
This is necessary in order to connect the statistical properties with the actual
observable quantities.
Obviously, the specific physical model to be used depends 
on the nature of the analyzed data, 
on the assumptions made to build the model,
and, in general, different physical events will require 
tailored treatments and specialized tools.
To address this,
{\bajes} provides the {\tt obs} module,
which is a container of methods needed to characterize and 
handle specific physical observations.
This module is designed aiming to the analysis of GWs
and EM counterparts.
Currently, {\tt bajes.obs} includes two sub-modules: {\tt gw} and {\tt kn}.
The {\tt gw} module contains the tools necessary to deal 
with GW analysis and it is 
described in Sec.~\ref{sec:gw}.
The {\tt kn} module supplies a framework
for the analysis of kilonovae light curves 
generated by BNS collisions, following our early work~\cite{Breschi:2021tbm}.
The implementation of the {\tt kn} module is inspired by
the approach presented in Ref.~\cite{Perego:2017wtu,Breschi:2021tbm} 
and it will be detailed in a followup work.

\subsection{Workflow} 
\label{sec:workflow}

\begin{figure*}[t]
	\centering 
	\includegraphics[width=0.90\textwidth]{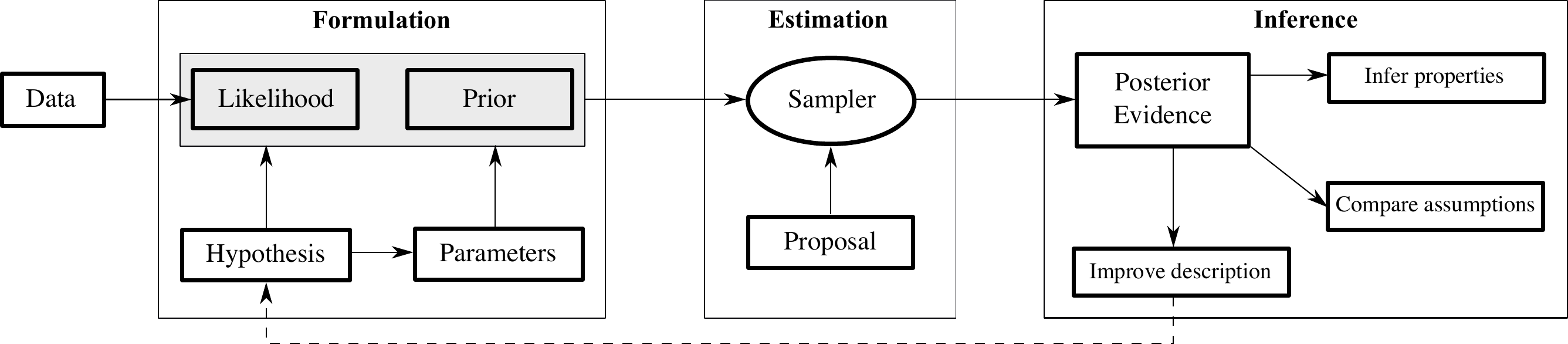}
	\caption{Schematic representation of the workflow 
					described in Sec.~\ref{sec:workflow}.
					The scheme highlights the three stages of the Bayesian formalism:
					formulation of the model,
					fitting the model to the data 
					and inference of the model properties.
					The gray box constituted by likelihood function and prior
					assumptions represents the Bayesian model to be inferred.
					The dashed back-propagating line refer to the case in which the 
					analysis is iterated with an improved description.
	}
	\label{fig:work}
\end{figure*}

 In order to fulfill the generic tasks of Bayesian inference, 
{\bajes} provides the tools necessary to instantiate a Bayesian model~\cite{gelman2020bayesian}
and to extract the statistical quantities of interest, such as posterior distribution or evidence. 

The {\tt bajes.inf} module supplies a {\tt Prior} and a {\tt Likelihood} objects,
and their combination defines the Bayesian model.
The {\tt Prior} provides the implementation of a generic prior distribution:
this object can be instantiated with list of {\tt Parameter},
that are going to define the parameter space of the model.
It is also possible to introduce constant and variable quantities 
to be carried along the sampling. 
Subsequently, it is possible to introduce a likelihood function 
customizing the specific {\tt log\_like} method of a {\tt Likelihood} object.
Furthermore, {\bajes} implements a {\tt JointPrior} and a {\tt JointLikelihood} objects
in order to handle joint posterior distributions.
Additionally, planned extensions include hierarchical models~\citep[e.g.][]{Hinton_2019,loredo2019multilevel}
and Bayesian methods to deal with error propagation.

Once the Bayesian model is defined,
it is possible to fit the model to the data estimating its properties.
The {\tt inf} module provides a {\tt Sampler} method
that wraps the provided model and initializes the requested sampling algorithm.
As mentioned in Sec.~\ref{sec:sampler},
the sampler explores the parameter space 
aiming to reconstruct the posterior distribution.
In order to conduct an accurate analysis, the {\tt Sampler} objects rely on
auxiliary functions such as the proposal methods 
described in App.~\ref{app:proposal}: 
these are statistical techniques that aim to extract independent samples 
and to conduct the sampler towards more likely regions of the parameter space.
(see App.~\ref{app:mcmc} and App.~\ref{app:nest}).

When the {\tt Sampler} completes the analysis, 
the algorithm returns the conditioned probability of interest
and the properties of the model can be inferred.
This information allows us to to compute the statistical quantities
that characterize the model and make it possible
to test competing hypotheses and verify different assumptions.
Moreover, from these results it is possible 
to understand the limits of the involved description;
then, in general, the model can be improved 
and the workflow can be iterated with the reviewed model,
improving the understanding and the modeling of the observed event.
Figure~\ref{fig:work} shows a schematic representation of the described workflow.

\subsection{Parallelisation} 
\label{sec:parallel}

\begin{figure}[t]
	\centering 
	\includegraphics[width=0.49\textwidth]{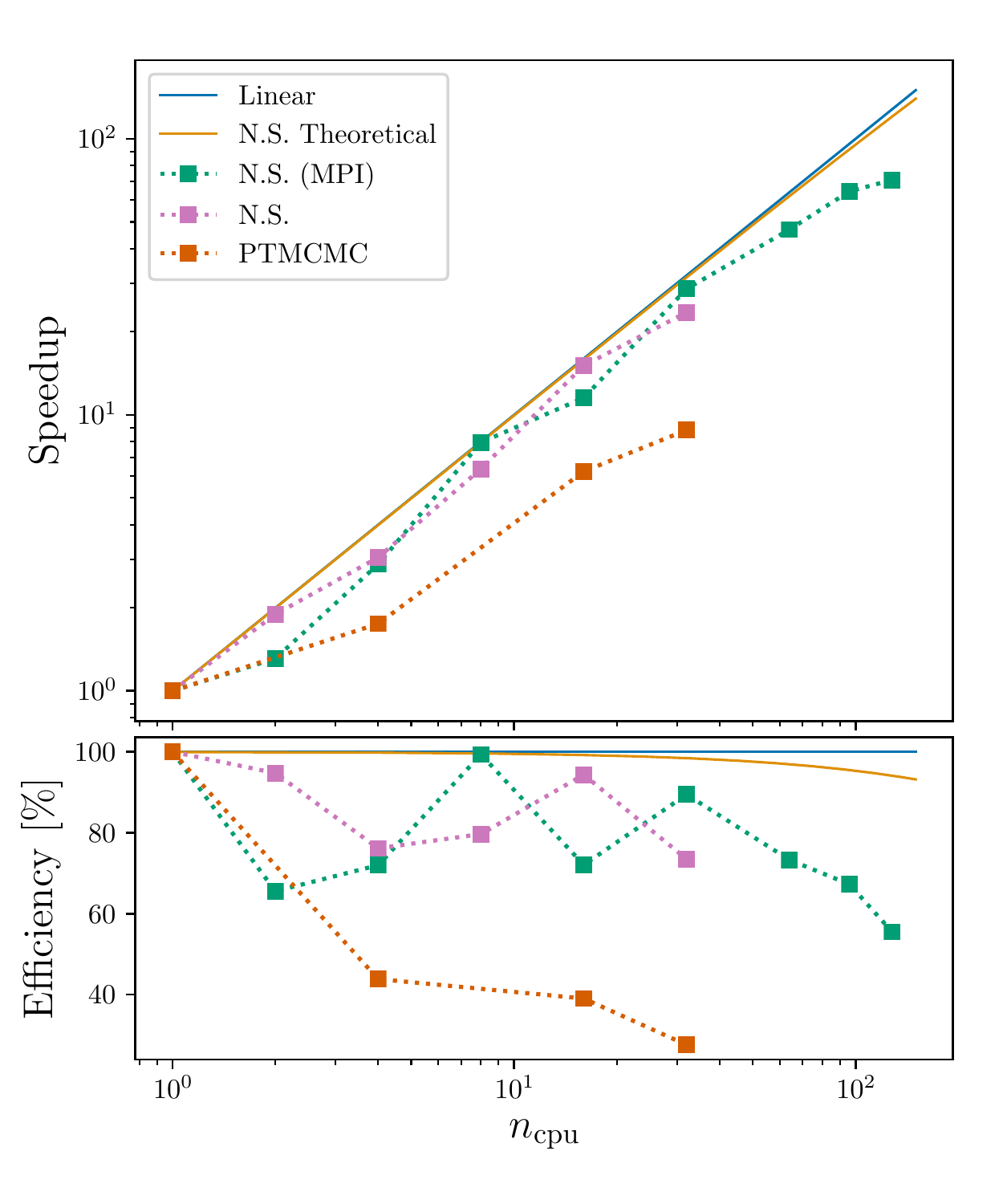}
	\caption{Scaling plot for the {\bajes} pipeline (see Sec.~\ref{sec:pipe}).
					Top panel: 
					The square markers are the measured
					speedup factors with respect to the serial execution-time.
					The execution-times for the PTMCMC (orange squares)
					algorithm are estimated performing $10^3$ iterations
					with $128\times 4$ tempered chains.
					The nested sampling execution-times  
					(pink squares for multi-threading 
					and green squares for MPI) are computed
					employing the {\tt dynesty} software 
					with 1024 live points and tolerance equal to 0.1.
						The blue solid line is the ideal linear scaling and
						the yellow solid line is the theoretical scaling of nested 
						sampling.
						Bottom panel:
						Same quantities discussed in the previous panel
						expressed in terms of efficiency.
						}
	\label{fig:scaling}
\end{figure}

By default, {\bajes} analyses can be performed
taking advantage of the parallel-threading {\tt multiprocess} {\py} library.
However, with this method the number of available processes is strictly 
limited by the size of the single machine 
and for non-trivial problems this could be a strong limitation.
For this reason, the {\bajes} software implements a customized method for multi-nodes
communication based on the message passing interface (MPI) protocol.

For ideal scaling, the execution-time of a machine computation is
inversely proportional to the number of central processing units (CPUs),
that leads to a linearly increase of the speedup.
However, in a realistic scenario, 
the scaling performances of sampling techniques 
are affected by unavoidable computational steps serially performed 
(e.g. temperature swapping in parallel chains and 
live points collection of nested sampling) and
by the continuous exchange of information 
between different processes,
required to adapt the evolution of the algorithm.

MCMC and nested sampling algorithms require separate treatments.
The performances of MCMC sampling are typically quantified 
in terms of proposal acceptance and 
correlation length of the chains~\cite{JSSv035i04,solonen2012,Foreman_Mackey_2013,Veitch:2014wba,Li_2014,Vousden_2015},
and generally the overall execution-time is determined 
by several contributions, 
such as the total number of chains,
the complexity of the parameter space and the employed proposal techniques.
Estimations of MCMC execution-times~\cite{sukys2017spux,robert2018accelerating,morzfeld2019localization} have shown that the efficiency
drastically decreases for an increasing number of parallel chains.
On the other hand, the parallelization performances of the nested sampling 
are well studied~\cite{feroz2008,Handley_2015,Brewer:2016scw, Higson_2018, Smith:2019ucc}
and the theoretical speedup factor $S_{\rm NS}$ of this algorithm
is expressed as a sub-linear scaling,
\be
\label{eq:nsscaling}
S_{\rm NS} (n_{\rm live},n_{\rm cpu}) = n_{\rm live}\cdot \log\left(1+\frac{n_{\rm cpu}}{n_{\rm live}}\right)\,.
\ee
For $n_{\rm live} \gg n_{\rm cpu}$, the values predicted by Eq.~\eqref{eq:nsscaling} are comparable with a linear trend.

Figure~\ref{fig:scaling}
shows the measured speedup factors in the execution-time
as a function of the number of CPUs
for different samplers and different parallelization methods.
The speedup factors are computed with respect to the execution-time
measured from the serial job.
The execution-times are estimated from
 the {\bajes} GW pipeline (see Sec.~\ref{sec:pipe})
 using GW150914~\cite{Abbott:2016blz} as target signal (see Sec.~\ref{sec:GW150914}).
 Moreover, Figure~\ref{fig:scaling} shows the respective efficiency rates, 
 that encode the deviation from the ideal linear scaling.
 The efficiency is computed as the percentage 
 ratio between the measure speedup 
 factor and the number of employed CPUs.
 Ideal linear scaling has an efficiency of 100\%.
 
 For the parallel-tempered MCMC (PTMCMC)
 algorithm implemented in {\tt bajes}, we estimate the speedup performing $10^3$ iterations
 with $128\times 4$ tempered chains, while, for nested sampling,
 we employ the {\tt dynesty} software
 with 1024 live points and tolerance equal to 0.1.
 The PTMCMC is not optimal in terms of execution-time scaling,
 mainly due to the serial swapping routine. 
 However, 
 it gives acceptable scaling performances
 with efficiency $\gtrsim 40\%$ up to $n_{\rm cpu} \simeq 16$ using multi-threading methods.
 The results with MPI are worst compared to
 multi-threading
 due to the data communication.
 
 Regarding the nested sampling,
 for a very small number of processes, roughly $n_{\rm cpu} \lesssim 2$, the multi-threading 
 method gives more efficient results, since the MPI protocol requires additional time for 
data communication.
 For an increasing number of CPUs, roughly $n_{\rm cpu} \gtrsim 6$,
 the 
 two parallelization methods give comparable results.
 However, the strength of MPI parallelization is the capability of accessing multiple 
 CPUs located in different physical machines:
 the MPI interface implemented in {\bajes} 
 gives an efficiency greater than 70\% up to $n_{\rm cpu} \simeq 100$, 
 that is the typical size of standard PE job.

\subsection{Dependencies} 
\label{sec:depend}

The {\bajes} software is compatible with {\py}~3.7 or higher versions,
and it can be easily installed 
using the standard {\py} setup tools.
It relies on essential modules, preferring 
those that can be easily installed via {\tt pip}.
The {\bajes} software mandatory requires the following libraries:
{\tt numpy}~\cite{NumPy:2020} for algebraic computations and array managing,
{\tt scipy}~\cite{SciPy:2020} for advanced mathematical tools,
and {\tt astropy}~\cite{Price-Whelan:2018hus} is invoked for astrometric computations.
However, in order to execute the pipeline supplied with {\bajes} (see Sec.~\ref{sec:pipe}), 
other libraries might be required:
{\tt gwpy}~\cite{GWpy,2019ascl.soft12016M} 
is used to access the GWOSC archive~\cite{Abbott:2019ebz,Trovato:2019liz,gwosc},
{\tt matplotlib}~\cite{MPL} and {\tt corner}~\cite{corner} are employed
for plotting.
Moreover, if the MPI parallelisation is requested, the software need the
installation of the {\tt mpi4py} library~\cite{mpi4py1,mpi4py2,mpi4py3}.

In order to perform the sampling, {\bajes} implements a PTMCMC
algorithm based on {\tt ptemcee}~\cite{Vousden_2015} or can use
additional external packages. In particular, we interface
the MCMC {\tt emcee}~\cite{Foreman_Mackey_2013} and the nested
sampling algorithms of {\tt cpnest}~\cite{cpnest} and {\tt dynesty}~\cite{Speagle_2020}.

\section{Gravitational-waves} 
\label{sec:gw}

The {\tt bajes.obs.gw} module contains the functions and the tools needed to deal
with gravitational-wave (GW) transients, that are mainly provided
 by signal processing and matched-filtered techniques~\cite{LIGOScientific:2019hgc}.
 
Ground-based GW detectors measure the local 
perturbations of the spacetime as time series. Then,
if we can believe that a physical GW transient is 
recorded in the data,
the detector output $d(t)$ is assumed to be 
the sum of the noise contribution $n(t)$ and 
the GW strain $s(t)$, such as
 \be
\label{eq:signal}
d(t) = n(t) + s(t)\,.
\ee
If $n(t)$ respects the conditions of Gaussianity and stationarity
and if we dispose of a template $h(t)$ able to reproduce 
the {\it real signal} $s(t)$,
then it is possible to filter out the noise contribution revealing the presence 
of a signal in the observed data.
 It is useful to introduce the inner product between 
 two time-domain series, $a(t)$ and $b(t)$, as 
 \be
 \label{eq:innerprod}
 (a|b) = 4 \Re \int_0^{\infty} \frac{a^*(f)\,b(f)}{S_n(f)}\,\d f\,,
 \ee
where $a(f)$ is the Fourier transform of the time series $a(t)$,
 \be
\label{eq:fourier}
a(f) = \int_{-\infty}^{+\infty} a(t)\, e^{-2\pi \i f t }\,\d t \,,
\ee
and analogously for $b(f)$, 
while $S_n(f)$ is the power spectral density (PSD) of the noise $n(t)$.

In order to perform inference on GW data,
it is necessary to provide an infrastructure capable
to process data segments, 
characterize the noise contamination,
localize sources and detectors and 
to generate waveform templates.
In the following sections, 
we discuss how these tasks are achieved 
by the {\tt obs.gw} module.

\subsection{Time and frequency series} 
\label{sec:series}

A realistic portion of data measured by an interferometer is 
a time series with constant sampling rate $\srate$ and finite duration $T$.
The restriction to a finite domain leads to approximate 
Eq.~\eqref{eq:innerprod} and Eq.~\eqref{eq:fourier} numerically, 
taking the advantage of the fast Fourier transform (FFT) algorithm~\cite{FFT}. 
Within this framework, the time step $\Delta t = 1/\srate$ and the duration $T$ of the
time-domain series are reflected in the spacing $\Delta f =1/T$ 
of the frequency bins and in the maximum frequency accessible from the data $\fNyq = \srate/2$,
known as Nyquist's frequency.
Then, we can approximate Eq.~\eqref{eq:innerprod} as
 \be
\label{eq:innerapprox}
(a|b) \approx \frac{4}{T} \,\Re \sum_{i} \frac{a^*(f_i)\,b(f_i)}{S_n(f_i)}\,,
\ee
where $f_i = i \cdot\Delta f $ and the index $i$ takes integer values from 0 to $\srate T/2$.
Generally, this summation is performed on a restricted frequency band,
identified by a lower and an upper cutoff-frequencies, respectively $\fmin$ and $\fmax$,
in order to neglect irrelevant portion of the spectrum. 
From Eq.~\eqref{eq:innerprod}, 
or its approximation Eq.~\eqref{eq:innerapprox},
it is possible to introduce the signal-to-noise ratio (SNR) as
\be
\rho = \frac{(d|h)}{\sqrt{(h|h)}}\,,
\ee
that estimates the power of the signal $h(t)$ enfolded in the data $d(t)$
weighted on the variance of the background noise.

{\bajes} implements a {\tt Series} object
 designed to manage time-domain and frequency-domain series.
 This instance stores the input series and it 
 computes the FFT (or the inverse-FFT)
 of the given data, in order to provide both 
 representations of the series.
 The {\tt Series} object supplies also an interface 
 capable to to perform tapering, filtering and padding of the input series:
we make use of the Tukey window for the tapering,
 while the filtering is performed using a Butterworth filter.
 Furthermore, the {\tt Series} object implements a summation 
 and a product between objects of the same type, defined in the frequency-domain,
 and contains methods to compute inner products and SNRs.

\subsection{Noise characterisation} 
\label{sec:noise}

As shown in Eq.~\eqref{eq:signal}, the measured data $d(t)$ are
intrinsically related with the noise fluctuations $n(t)$.
The noise of a GW detector is represented by
stochastic fluctuations~\cite{LIGOScientific:2019hgc} 
that propagate to the output.
The primary noise sources in a ground-based interferometer
are gravity gradients and seismic motions~\cite{Losurdo:2001bi,Daw:2004qd,Accadia:2011dg}, 
thermal agitation~\cite{Harry:2001iw,Somiya:2011np}, 
quantum perturbations~\cite{Lyons:00,Buonanno:2003ch},
and internal optomechanical resonances~\cite{Nachman,Bond:2017}.
Moreover, 
the time series recorded by a GW detector are also affected 
by external non-gravitational signals~\cite{Cirone:2018guh},
such as the AC power grid transmission 
responsible for the 60~Hz peak of LIGOs 
and the 50~Hz one of Virgo.
The noise fluctuations are assumed to be Gaussian distributed 
and stationary on a relatively large time-scale~\cite{LIGOScientific:2019hgc}.
Then, the PSD shows the distribution of the noise power for every frequency component 
and
it can be computed as
\be
\label{eq:psd}
 E\Big[|n(f)|^2  \Big]  = \frac{T}{2} \,S_n(f)\,,
\ee
where the expectation is computed with the temporal average over 
a period $T$.
In other words, the PSD characterizes the uncertainty of the observed data
in frequency-domain as the variance of the associated noise spectrum.
The amplitude spectral density (ASD) is usually defined as the square
root of the PSD, $\sqrt{S_n(f)}$.

The PSD is a key quantity in order to estimate the product Eq.~\eqref{eq:innerprod},
since it describes the weight of each frequency component.
It follows that a full characterization of the noise sources 
and a proper estimation of its contribution 
are essential in order to perform accurate measurements of GW transients.
A general tool for estimating PSD is the Welch's method~\cite{Welch:67}, consisting in the 
average of the spectral densities computed on chunks of the full noise segment.
However, this is not the only technique suitable for this task~\cite{Blais:1996, Cornish:2014kda}.

\begin{figure}[t]
	\centering 
	\includegraphics[width=0.49\textwidth]{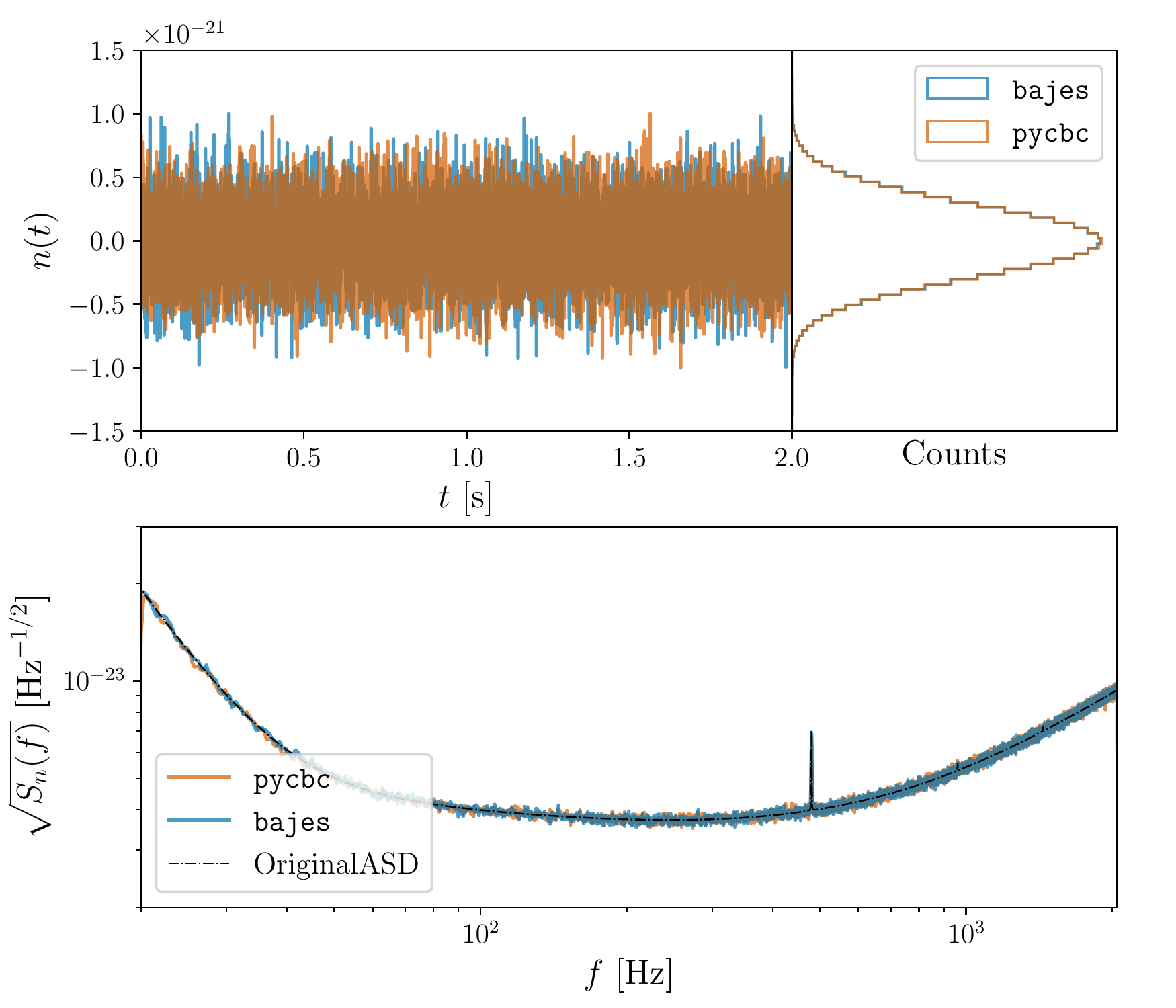}
	\caption{Top left panel: 
					Comparison of artificial noise segment produced using 
					{\bajes} (blue) 
					and {\tt pycbc} (orange).
					The segments are computed using 
					LIGO design sensitivity P1200087~\cite{Aasi:2013wya}
					with a lower cutoff-frequency at 20~Hz and a sampling rate of 4096~Hz.
					The panel shows a 2~s segment extracted from a
					segment with total duration of $T=1024~{\rm s}$.
					Top right panel: 
					Histograms of the time-domain samples
					computed using the whole artificial segment of length $T=1024~{\rm s}$.
				 	Bottom panel:
				 	ASDs reconstructed 
				 	from the artificial noise segments using the Welch's method.
				 	The spectra are computed using the whole segments
				 	with a chuck length of 4~s and an overlap fraction of 99\%.
				 	The black dashed line represents the original ASD. 
				}
	\label{fig:noise}
\end{figure}

Aiming to data analysis purposes, 
{\bajes} implements a {\tt Noise} object.
This component is
capable to estimate the PSD of a given noise time-domain 
series using the Welch's method,
generate an artificial segments of stationary and Gaussian noise from a given PSD,
and it disposes of methods for PSD interpolation.
Figure~\ref{fig:noise} shows a comparison of artificial noise segments 
generated with {\tt bajes} 
and with the {\tt pycbc}
routines~\cite{Usman:2015kfa,Biwer:2018osg,Nitz:2018rgo, alex_nitz_2021_4556907},
where the total length of the artificial segment is 1024~s. 
The histograms and the PSDs 
show that the generated noise fluctuations respect the conditions of 
Gaussianity and stationarity with a frequency spectrum described 
by the requested PSD.

\subsection{Ground-based interferometers} 
\label{sec:det}

The current ground-based observatories for GW detection 
are L-shaped Michelson interferometers
with Fabry-Perot cavities~\cite{Dooley:2014nga, Akutsu:2018axf}. 
Each arm has a length of $L \approx 3{-}4~{\rm km}$,
depending on the actual detector~\cite{TheLIGOScientific:2014jea,TheVirgo:2014hva},
and it is constituted of two mirrors acting like test-masses.
The detector performs measurements of the gravitational strain as a difference in length of the orthogonal arms,
\be
\label{eq:dL_L}
d(t) = \frac{\Delta L(t)}{L} \,,
\ee
where $\Delta L = \Delta L_x - \Delta L_y$ is the difference between the displacements 
along the two orthogonal arms.
The projection of the signal on the detector 
can be computed from the GW polarization components $h_{+,\times}$ as 
\be
\label{eq:pol}
h(t) = D_{ij} h_{ij}(t) = F_+ h_+(t) + F_\times h_\times(t) \,,
\ee
where $D_{ij}$ is labeled as detector tensor and it depends on the geometry of the interferometer,
while $F_{+,\times}$ are the {\it antenna pattern} functions for each polarization.
The antenna patterns $F_{+,\times}$
characterize the detector sensitivity in the different regions of the sky for every given time~\cite{Raymond:2014uha}. 

Generally, in standard observing conditions, 
the observations of GW transients are performed simultaneously by
a worldwide network of ground-based interferometers.
Thanks to this, 
it is possible to correlate strains observed independently 
in different locations,
improving the estimation of the 
the astrophysical origin of the transients~\cite{Nitz:2018imz,Pankow:2018phc}.

The necessity to localize a GW observatory in a fixed frame arises,
in order to project the expected signal on the detector and to 
estimate the light travel time from each detector in the network. 
For these tasks, {\bajes} disposes of a {\tt Detector} class 
able to characterize a ground-based interferometer.
This object is identified with the 
coordinates of the site of the interferometer 
(latitude, longitude and altitude) and the angles of the two arms (azimuth and tilt).
It is also possible to initialize the {\tt Detector} object
to precomputed detector configurations
using the two-digits string 
identifying a ground-based interferometer, e.g. 
{\tt H1} for LIGO-Hanford~\cite{TheLIGOScientific:2014jea},
{\tt L1} for LIGO-Livingston~\cite{TheLIGOScientific:2014jea},
{\tt V1} for Virgo~\cite{TheVirgo:2014hva},
{\tt G1} for GEO600~\cite{Luck:2010rt,Dooley:2014nga},
{\tt K1} for KAGRA~\cite{Aso:2013eba,Akutsu:2020his}
and {\tt ET} for Einstein Telescope (configuration D)~\cite{Punturo:2010zz,Hild:2010id}.

\subsection{Waveform templates} 
\label{sec:wave}

The last ingredient necessary to complete the framework is a waveform template,
i.e. a model able to compute the expected representation of the signal $h(t,\params)$ 
(or ${h}(f,\params)$)
for every given set of parameters $\params$.
The {\tt Waveform} object is a class that access the methods 
disposable in the {\tt bajes.obs.gw.approx}
sub-module,
and computes the expected
GW polarization components for every given set of parameters. 
The frequency-domain waveform in Eq.~\eqref{eq:pol}
can be written in terms of the 
amplitude $A(f)$ and the phase $\phi(f)$, 
\be
\label{eq:h_ampphi}
 h(f) = A(f) \, e^{-\i \big[\phi_0 + 2 \pi t_0 f + \phi(f)\big]}\,, 
\ee
where $\phi_0$ and $t_0$ are respectively phase and time references. 

{\bajes} directly implements and interfaces 
	with all the most advanced GW
  templates for quasi-circular compact binary mergers, and includes
  state-of-art templates for eccentric and hyperbolic binary black hole (BBH) mergers
  \cite{Chiaramello:2020ehz,Nagar:2020xsk} as
  well as for BNS postmerger~\cite{Breschi:2019srl}. In particular:
\begin{itemize}
	\item {\tt TaylorF2}: The post-Newtonian (PN) framework
	~\cite{Blanchet:1989fg,Faye:2012we,Levi:2015ixa,Levi:2015uxa,Mishra:2015bqa,Levi:2016ofk}
	represents a milestone for the description of the two-body problem.
	This approach solves the dynamical evolution 
	of a compact binary with a perturbative method 
	assuming low velocities and weak field approximations,
	which are reflected in the condition $v=(\pi\, G M f)^{1/3}\ll c$,
	where $v$ is the characteristic velocity in the binary,
	$M=m_1+m_2$ is the total mass and $f$ is the GW frequency.
	The exact analytic solution of the gravitational radiation emitted
	by a point-particle compact binary is known up to the 3.5PN
        order.
	{\bajes} also implements a pseudo-5.5PN accurate description of the point mass baseline, as 
	derived in Ref.~\cite{Messina:2019uby}.
	Pure tidal corrections are implemented up to
        7.5PN~\cite{Vines:2011ud,Damour:2012yf} and include the 
        recently computed tail terms (6.5, 7.5 PN) of \cite{Henry:2020ski},
	whereas spin-quadrupole terms are included up to 3.5PN~\cite{Nagar:2018plt}.    
	\item {\tt TEOBResumS}: The effective-one-body (EOB) formalism
	~\cite{Buonanno:1998gg, Buonanno:2000ef,Damour:2000we,Damour:2001tu,Damour:2008qf,Damour:2015isa,Bini:2019nra,Bini:2020wpo,Bini:2020nsb} 
	is an Hamiltonian approach that reduces the two-body problem to
	a single orbiting object in an effective potential.
	{\tt TEOBResumS}~\cite{Nagar:2018zoe} is an EOB
	approximant for spinning coalescing compact binaries~\cite{Damour:2014sva,Nagar:2017jdw} (black holes or neutron stars)
	inspiralling along generic orbits. 
	It includes 
	tidal effects~\cite{Bini:2014zxa, Bernuzzi:2014owa, Akcay:2018yyh} 
	and subdominant modes~\cite{Nagar:2019wds, Nagar:2020pcj} up to $\ell =5$, as well as a
	description of precession-induced modulations up to merger~\cite{Akcay:2020qrj}. 
	When considering systems evolving along quasi-circular orbits, the computational performance of 
	the model is enhanced by means of a post-adiabatic description 
	of the early inspiral~\cite{Nagar:2018gnk}. 
	Moreover, the model includes the implementations of 
	hyperbolic encounters~\cite{Nagar:2020xsk}, 
	eccentric mergers~\cite{Chiaramello:2020ehz} and 
	a frequency domain approximation valid for
	quasi-circular BNS coalescences, labeled as {\tt TEOBResumSPA}~\cite{Gamba:2020ljo}.
        The model is publicly available at~\cite{teobresums}
        and used in {\bajes} from the provided {\py} interface.	 
      \item {\tt NRPM}: BNS merger remnants are expected to be loud sources of GWs~\cite{Takami:2014tva,Clark:2015zxa,Radice:2018xqa},
	that convey unique information on the equation of state of hot matter at extreme
	densities~\cite{Radice:2016rys,Radice:2017lry,Abbott:2018exr}. 
	{\tt NRPM}~\cite{Breschi:2019srl} is a analytical model based on numerical relativity BNS simulations.
	The model is tuned on a set of simulations covering the range $q\le 1.5$ and $\Lambda_{1,2} \gtrsim 400$.
	For smaller values of the tidal parameters, 
	the model is identically zero since it is not expected to have a post-merger signal in these cases, due to prompt black-hole formation~\cite{Zappa:2017xba,Agathos:2019sah,Bernuzzi:2020txg}.
	{\tt NRPM} is directly implemented in {\bajes} and
	it can be attached to the {\tt TEOBResumS} pre-merger template, 
	obtaining a complete model for the gravitational radiation
        expected from BNS coalescences.
      \item {\tt MLGW}: 
	Machine learning tools can be employed to construct
	accurate representations of GW signals.
	The {\tt mlgw} package~\cite{mlgw,Schmidt:2020yuu} takes advantage of these methods 
	 to generate fast and reliable GW templates for BBH coalescences.
	The model is composed by contributions extracted with
	principal component analysis
	and a linear combination of regression models, 
	which maps the orbital parameters of the black holes 
	to the reduced representation of the wave. 
	A complete model includes two principal component models,
	for both phase and amplitude of the wave, 
	and a mixture of regression models
	for each of the principal components considered. 
	The algorithm is trained on time-domain models
	and tested only for aligned spin BBHs. 
	Currently, the released software provides 
	the representations of EOB templates, {\tt TEOBResumS}~\cite{Nagar:2018zoe}
	and {\tt SEOBNRv4}~\cite{Bohe:2016gbl}.
      \item {\tt GWSurrogate}: The templates provided by the {\tt
        gwsurrogate} package~\cite{gwsur}
	implements fast waveforms based on reduced-order models~\cite{Field:2013cfa} 
	trained on numerical relativity simulations.
	The {\tt NRSur7dq4} model~\cite{Varma:2019csw} is
	a precessing extension of the model presented in Ref.~\cite{Blackman:2017pcm}
	trained on a set of simulations with $q \le 4$ and $\chi_1,\chi_2\le 0.8$ 
	that contains all higher-order modes with $\ell\le 4$.
	On the other hand, {\tt NRHybSur3dq8}~\cite{Varma:2018mmi}
	and its tidal version~\cite{Barkett:2019tus} are calibrated using hybridized waveforms in order to increase 
	the number of orbits of the training templates, 
	improving the quality of the approximation.
	This model is tuned in a wider range in the mass ratio, $q \le 8$,
	but it does not include precession contributions.
	\item {\tt LALSimulation}: 
	The LIGO Algorithm Library {\tt LALSuite}~\cite{lalsuite}
	is the official LIGO-Virgo Collaboration software and it provides 
	the largest variety of waveform template models.
	{\bajes} implements the waveform generator 
	of {\tt LALSimulation}, a module of {\tt LALSuite}.
	For the results of this paper, we make use of this implementation
	in order to employ {\tt IMRPhenomPv2} approximant~\cite{Husa:2015iqa,Khan:2015jqa} 
	and its tidal extension, {\tt IMRPhenomPv2\_NRTidal}~\cite{Dietrich:2017aum}.
	A list of all the approximants available through {\tt LALSimulation} can be found at \cite{lal_approximants}
\end{itemize}

\section{Pipeline} 
\label{sec:pipe}

{\bajes} provides a customized 
and automatized pipeline for the 
analysis of GW transients and EM counterparts.
In this section, we discuss the model implemented to 
perform PE analysis on GW transients
with {\bajes}.

In the context of GW data analysis,
we introduce the working hypotheses 
that are going to define the employed Bayesian model.
We call the assumption that the data contains a 
non-vanishing GW transient {\it signal hypothesis} $H_{\rm S}$
i.e. $d(t)=n(t)+s(t)$ with $s(t)\ne 0$.
On the other hand, the {\it noise hypothesis} $H_{\rm N}$ is the condition
for which the recorded data corresponds only to pure noise realization, i.e. $d(t)=n(t)$.
Then, in the signal hypothesis condition,
a GW signal emitted by a quasi-circular compact binary coalescence (CBC)
can be fully characterized with a set of 17 parameters according to General 
Relativity. Precisely,
\be
\label{eq:gwparams}
\params_{\rm cbc} = \{ m_1, m_2, \spin_1 , \spin_2, \Lambda_1, \Lambda_2,
									D_L, \iota, \alpha, \delta, \psi, t_0, \phi_0\}
\ee
where: 
\begin{itemize}
\item $m_{1,2}$ refer to the detector-frame masses of the two objects, 
taken with the convention $m_1{\ge} \,m_2$;
\item $\spin_{1,2}$ are the dimensionless spin vectors,
\be
\label{eq:chii}
\spin_{i}=\frac{c\, \textbf{S}_{i}}{Gm_{i}^2}\,,\quad  \,i=1,2\,,
\ee
where $\textbf{S}_{1,2}$ are the spins of the two objects,
 $c$ is the speed of light and $G$ is the gravitational constant.
\item $\Lambda_{1,2}$ are the dimensionless tidal polarizability parameters
that encode the quadrupolar deformability of the $i$-th object under the effect of an external force,
\be
\label{eq:lambdai}
\Lambda_{i}=\frac{2}{3}\,k_{2,i}\,\left(\frac{c^2\, R_{i}}{Gm_{i}}\right)^5\,,\quad  \,i=1,2\,,
\ee
where $k_{2,i}$ and $R_i$ are respectively the second Love number and the radius of the $i$-th object
($k_2$ is identically zero for black holes).
\item $D_L$ is the luminosity distance of the source from the observer;
\item $\iota$ is the angle between the line of sight and the total angular momentum of the system and it takes value in the range $[0,\pi]$,
\item $\{\alpha,\delta\}$ are respectively 
right ascension and declination angles that identify the sky position of the source;
\item $\psi$ is the polarisation angle and it takes value in the range $[0,\pi]$;
\item $\{t_0, \phi_0\}$ are respectively reference time and reference phase.
\end{itemize}
The sampling is performed promoting the chirp mass $\M$ and the mass ratio $q$,
\be
\label{eq:mcq}
	\M = \frac{(m_1m_2)^{3/5}}{(m_1+m_2)^{1/5}}\,, \quad
	q=\frac{m_1}{m_2} \ge 1\,,
\ee
over the single mass components, since they appear to be less correlated for this type of signals~\cite{Veitch:2009hd,Veitch:2014wba}.
For spinning binary mergers,
we define the effective spin parameter $\chi_{\rm eff}$ as
\be
\label{eq:chieff}
\chi_{\rm eff} =\frac{m_1\chi_{1,z}+m_2\chi_{2,z}}{m_1+m_2}\,,
\ee
that encodes the aligned-spin contribution and it 
can lead to narrower uncertainties
than the single spin components~\cite{Ng:2018neg}.
Furthermore, in the context of BNS mergers, 
it is useful to introduce the reduced tidal parameter $\tilde \Lambda$,
\be
\label{eq:lambdat}
\tilde \Lambda =\frac{16}{13}\left[\frac{(m_1+12m_2)m_1^4\Lambda_1}{M^5} + (1\leftrightarrow 2)\right]\,
\ee
and 
the asymmetric tidal parameter $\dLam$,
\be
\label{eq:dlt}
\dLam = 
\left[1 -\frac{7996}{1319}  \frac{m_2}{m_1} -\frac{11005}{1319} \left(\frac{m_2}{m_1} \right)^2\right]\frac{m_1^6\Lambda_1}{M^6} - (1\leftrightarrow 2)\,,
\ee 
where $M= m_1+m_2$. 
The tidal parameters $\tLam$ and $\dLam$ are respectively
proportional to the leading order and the next-to-leading order tidal contributions
according with PN expansion.

Generally, concerning GW analysis, it is common to label 
$\params_{\rm int} = \{ m_1, m_2, \spin_1 , \spin_2, \Lambda_1, \Lambda_2\}$
as intrinsic parameters, since they affect the physical dynamics of the system;
while, the extrinsic parameters $\params_{\rm ext} =\{D_L, \iota, \alpha, \delta, \psi, t_0, \phi_0\}$
are related with the observed signal by trivial geometrical argumentation. 
Moreover, it is possible to include in the pipeline further parameters
in order to take into account the calibration of the input quantities, 
such as calibration envelopes and PSD uncertainties.
For a detailed discussion about these topics, see Sec.~\ref{sec:impl}.

In the following sections, we present the implementations available in the {\bajes} GW pipeline.

\subsection{Data segments}
\label{sec:data}

The default GW routine implemented in {\bajes} is designed
for binary mergers analyses.
In general, the pipeline is able to access to the 
open-source database of GWOSC~\cite{Abbott:2019ebz,Trovato:2019liz,gwosc},
to all the data released with GWTC-1~\cite{LIGOScientific:2018mvr}
and to the expected PSD curves for current and next-generation detectors~\cite{,Punturo:2010zz,Hild:2010id}.

The input data to be analyzed by the pipeline 
can be gathered in different ways.
The GW pipeline provides a customized {\it injection} generator
capable to produce artificial data given a prescribed set of 
parameters and the detector sensitivity curves.
Moreover, the {\bajes} pipeline allows to access 
the observational data recorded by the LIGO-Virgo 
interferometers~\cite{Aasi:2013wya,TheLIGOScientific:2014jea,TheVirgo:2014hva} from the GWOSC,
specifying the central value of the GPS time and the duration of the segment.

When the data information is gathered, 
the pipeline initializes the {\tt Likelihood} function 
and the {\tt Prior} with the requested parameters belonging to the set $\params_{\rm cbc}$,
and it passes these arguments to the requested sampler which performs the Bayesian inference.

\subsection{Prior distributions} 
\label{sec:prior}

The prior distribution for the masses is chosen flat in the components $\{m_1,m_2\}$,
that can be written in terms of $\{\M,q\}$ as 
\be
\label{eq:prior-mass}
p(\M,q|H_{\rm S}) = \frac{\M}{\Pi_{\M}\Pi_q} \, \left( \frac{1+q}{q^{3}} \right)^{2/5}\,,
\ee
where $\Pi_{\M}$ and $\Pi_q$ are the prior volumes 
limited by the bounds $[\M_{\rm min}, \M_{\rm max}]$ in chirp mass
and $[1,q_{\rm max}]$ in mass ratio,
\be
\label{eq:prior-mass-norm}
\begin{split}
\Pi_{\M} &= \frac{1}{2}\left(\M^2_{\rm max} - \M_{\rm min}^2\right)\,, \\
\Pi_{q} &=
5\left[\frac{2^{2/5}\sqrt{\pi}\,\Gamma\left(\frac{4}{5}\right)}{\Gamma\left(\frac{3}{10}\right)}
-\frac{{}_{2}F_1\left(-\frac{2}{5},-\frac{1}{5},\frac{4}{5},-q_{\rm max}\right)}{q_{\rm max}^{1/5}}\right]\,,\\
\end{split}
\ee
where ${}_{2}F_1(a,b,c;z)$
is the hypergeometric function 
and $\Gamma(x)$ is the Euler function. 
 
 The spin vectors can be written in the polar frame of the binary as 
 $\spin_i = \{\chi_i , \vartheta_i, \varphi_i\}$, where $\chi_i$ is the spin magnitude, $\vartheta_i$ 
 is the tilt angle and $\varphi_i$ is the complimentary azimuthal angle between $\spin_i$ and the 
 orbital angular momentum $\textbf{L}$ of the binary.
 The prior distribution for these quantities is specified 
 by the maximum value of spin magnitude $\chi_{\rm max}\ge 0$ and it 
 can be chosen between the following:
 \begin{itemize}
 	\item {\it Isotropic prior with precessing spins}: 
 			the prior on the angular components $\{\vartheta_i,\varphi_i\}$ 
 			is isotropic over the solid angle,
 			while the spin magnitude is uniformly distributed in the range $[0,\chi_{\rm max}]$,
 			\be
 			\label{eq:prior-precess-isotropic}
 			p(\chi_i , \vartheta_i, \varphi_i|H_{\rm S}) = \frac{\sin\vartheta_i}{4\pi \chi_{\rm max}}\,.
 			\ee
	\item {\it Isotropic prior with aligned spins}:
			this case is identical to the isotropic one 
			except for the assumption of aligned spins, $\vartheta_i = 0,\pi$. 
			The $xy$ components of the spin vectors are marginalized, 
			obtaining the form~\cite{Lange:2018pyp}
			\be
			\label{eq:prior-aligned-isotropic}
			p(\chi_{i,z}|H_{\rm S}) = \frac{1}{2 \chi_{\rm max}} \log \left|\frac{\chi_{\rm max}}{\chi_{i,z}} \right|\,.
			\ee
 	\item {\it Volumetric prior with precessing spins}:
 				the distribution is taken uniform in all Cartesian components,
 				i.e. flat over the sphere with radius $\chi_{\rm max}$. 
 				This prior can be written as 
 			\be
			\label{eq:prior-precess-volumetric}
			p(\chi_i , \vartheta_i, \varphi_i|H_{\rm S}) =\frac{3}{4\pi} \,
			\frac{ \chi_i^2 \sin\vartheta_i}{\chi^3_{\rm max}}\,.
			\ee
 	\item {\it Volumetric prior with aligned spins}:
 			the same of volumetric case with aligned components; 
 			the marginalization over the $xy$ components leads to the form
 						\be
 			\label{eq:prior-aligned-volumetric}
 			p(\chi_{i,z}|H_{\rm S}) = \frac{9}{16\pi} \frac{ \chi^2_{\rm max} - \chi_{i,z}^2}{ \chi^3_{\rm max}}\,.
 			\ee
 	\end{itemize} 
 
 The prior distribution for the sky position parameters $\{\alpha,\delta\}$
 is taken isotropic over the entire solid angle,
 such that $\alpha \in [0, 2\pi]$ and $\delta \in [-\pi/2,+\pi/2]$,
  						\be
 \label{eq:prior-skypos}
 p(\alpha,\delta|H_{\rm S}) = \frac{\cos\delta}{4\pi}\,,
 \ee
 and analogously for the inclination $\iota$ in the range $[0,\pi]$,
  	\be
 \label{eq:prior-iota}
 p(\iota|H_{\rm S}) =\frac{\sin\iota}{2}\,.
 \ee
 Regarding the luminosity distance the bounds are specified by the lower and the upper bounds 
 $[D_{\rm min}, D_{\rm max}]$, and the analytic form of the prior can be chosen between the followings:
  \begin{itemize}
 	\item {\it Volumetric prior}: 
 	general analysis assumes that the source is uniformly distributed over the 
 	sphere centred around the detectors, then 
 	\be
 	\label{eq:prior-dist-vol}
 	p(D_L|H_{\rm S}) = \frac{3 D_L^2}{D_{\rm max}^3-D_{\rm min}^3}\,.
 	\ee
 	\item {\it Comoving-volumetric prior}: 
 	in order to take into account the cosmological expansion of the Universe,
 	a prior uniformly distributed over the comoving volume $V_C$ is a more suitable
 	physically-motivated choice.
 	Within this assumption, the prior on the luminosity distance can be written as  
 	\be
 	\label{eq:prior-dist-com}
 	p(D_L|H_{\rm S}) \propto \frac{\d V_C}{\d D_L}\,.
 	\ee
 	The luminosity distance $D_L$ and the comoving volume $V_C$
 	 are related through the redshift $z$ for a fixed cosmological model;
 	by default, {\bajes} acquires the values of the cosmological constants 
 	from Ref.~\cite{Aghanim:2018eyx}.
	\item {\it Source-frame prior}:
	as shown in Ref.~\cite{Abbott:2019yzh,Abbott:2020niy},
	Eq.~\eqref{eq:prior-dist-com} does not take into account 
	contributions due to time dilatation. Then, we can introduce 
	a prior distribution uniformly distributed in the source-frame volume as
	\be
	\label{eq:prior-dist-source}
	p(D_L|H_{\rm S}) \propto \frac{1}{1+z}\cdot \frac{\d V_C}{\d D_L}\,,
	\ee
	where the factor $(1 + z)^{-1}$ converts the merger rate from the
	source frame to the detector frame.
 	\item {\it Log-uniform prior}:
 	it could be useful to introduce a uniform prior in $\log D_L$, then
 	\be
 	\label{eq:prior-dist-log}
 	p(D_L |H_{\rm S}) = \frac{D_L^{-1}}{\log\left(D_{\rm max}/D_{\rm min}\right)}\,.
 	\ee
 \end{itemize} 

For the remaining parameters, i.e. $\{\psi,t_0,\phi_0\}$ and $\{\Lambda_1,\Lambda_2\}$ (if required),
their prior distributions are taken uniform within the provided bounds.
Then, the overall prior is the product of the priors of the single parameters.

\subsection{Likelihood function} 
\label{sec:like}

The key ingredient of the inference is the likelihood function,
that encodes the capability of a given model $h(t,\params_{\rm cbc})$ to match the observed data $d(t)$. 
For Gaussian and stationary noise $n(t)$, 
we expect the mean of the noise fluctuations to be centered around 
zero with a variance described by the PSD in the frequency-domain,
i.e.
\be
\label{eq:noiseprob}
p(n|H_{\rm N}) \propto e^{-\frac{1}{2} (n|n)}\,.
\ee
It follows that, within the signal hypothesis $H_{\rm S}$
and supposing that we dispose of a template $h(t,\params_{\rm cbc})$ 
capable to reproduce the {\it real signal} $s(t)$ for a given set of $\params_{\rm cbc}$,
the log-likelihood function can be written 
as the frequency-domain residuals 
between the recorded data and the template
with the product defined in Eq.~\eqref{eq:innerprod}, 
\be
\label{eq:gwlike}
p(d|\params_{\rm cbc},H_{\rm S}) = \frac{1}{\mathcal{N}}\,  e^{-\frac{1}{2}(d-h|d-h)}\,.
\ee
where $\mathcal{N}$ is the normalisation constant,
that can be expressed in terms of the PSD
using the numerical approximation Eq.~\eqref{eq:innerapprox},
\be
\label{eq:gwlike_norm}
\mathcal{N} \approx  \prod_{i} \sqrt{ \frac{\pi \,T\, S_n(f_i)}{2}}\,.
\ee
Then, the Bayes's factor of the signal hypothesis against
the noise assumption can be computed as
\begin{widetext}
\be
\label{eq:B_signal}
\B^{\rm S}_{\rm N} = \frac{p(d|H_{\rm S})}{p(d|H_{\rm N})}=\int_{\paramspace} \exp\left[\big(d\big|h(\params_{\rm cbc})\big)-\frac{1}{2}\big(h(\params_{\rm cbc})\big|h(\params_{\rm cbc})\big)\right]\, p(\params_{\rm cbc}|H_{\rm S})\,\d\params_{\rm cbc}\,.
\ee
\end{widetext}
	
Combining Eq.~\eqref{eq:gwlike} with Eq.~\eqref{eq:h_ampphi},
it is possible to write the explicit 
dependency of the likelihood with respect to the reference parameters $\{\phi_0,t_0\}$,
since these
  values have no physical relevance, we marginalize the posterior
distribution with respect to $\{\phi_0,t_0\}$ in order to increase the
efficiency of the sampling.
The marginalization over $\phi_0$ can be computed analytically
and the solution can be written in terms of the 
modified Bessel function of the first kind~\cite{margphi}.
For the time-shift $t_0$, the computation is semi-analytical
since the values of the likelihood are evaluated on a equally-spaced grid
resorting on the FFT computation~\cite{margtime}.

\subsection{Additional implementations} 
\label{sec:impl}

In order to perform accurate and reliable 
inferences of GW transients, the pipeline requires refinements
and auxiliary control systems. 
In this section we discuss some of the 
additional tools implemented in
the GW pipeline supplied with {\bajes}.

\subsubsection{Calibration envelopes} 
\label{sec:calib}

The necessity of calibration envelopes~\cite{Vitale:2011wu} arises due to 
imperfect knowledge of the interferometer response to differential arm length changes~\cite{Abbott:2016jsd,PhysRevD.96.102001,Acernese_2018},
which affects the transfer functions of the detector components
introducing systematic errors that propagate to the recorded data.
These uncertainties are estimated by inspecting the detector control systems
and propagating the measurements into a frequency-dependent probability distribution.
Subsequently, the information on calibration errors must be taken into account when inferring
the astrophysical parameters of GW signals.
In order to achieve this task, it is useful to introduce two auxiliary functions 
$\delta A(f)$ and $\delta \phi (f)$ that characterise respectively the amplitude 
and the phase uncertainties of the measured data segments.
Then, the calibration envelopes $\{\delta A(f),\delta \phi (f)\}$ can be 
taken into account in the Bayesian model as
\be
\label{eq:calib}
h(f) \to \big[1+\delta A(f)\big]\,e^{\i \delta\phi(f)}\,h(f)\,.
\ee
This procedure is accomplished 
specifying the values of the calibration envelopes at predefined logarithmic-spaced 
frequency nodes $f_j$ and linearly interpolated over the interested frequency axis.
The calibration parameters $\{\delta A(f_j),\delta \phi(f_j)\}$ are introduced in the sampling
and estimated along with the signal parameters. 
The prior for calibration envelopes $\{\delta A(f_j),\delta \phi(f_j)\}$ is a multivariate 
normal distribution with variance specified by the measured calibration errors. 

\subsubsection{PSD uncertainties} 
\label{sec:psderr}

The usage of a fixed estimated PSD 
might generate biases due to non-stationary effects 
and unaccounted slow variations in the noise spectrum.
Then, it arises the necessity to take into account the 
uncertainty of the PSD estimate during the inference of the properties of a GW signal.
For this reason,
the pipeline allows the possibility to include PSD uncertainty weights $\eta_j$~\cite{Veitch:2014wba}:
the Fourier domain is dived in predefined logarithmic-spaced
bins $[f_j ,f_{j+1}]$ and the weights are included such that 
\be
\label{eq:psdweight}
S_n(f) \to \eta_j\,S_n(f)\,,\quad{\rm for }\, f_j \le f  <f_{j+1}\,,
\ee
where $\eta_j$ is taken constant in respective frequency bin.
The full set of $\{\eta_j\}$ parameters, one for every frequency bin,
is introduced in the sampling 
and they are estimated during the exploration with the signal parameters.
The prior distribution for the PSD uncertainty $\eta_j$ is take normal with mean zero and 
variance $1/N_j$, where $N_j$ is the number of data sample enclosed in the bin $[f_j,f_{j+1}]$.
This scheme has shown to improve the robustness of the 
GW inference~\cite{Littenberg:2013gja} and it 
offers a flexible model capable to quantify the differences between the 
estimated PSD and the spectrum of the analyzed data.

\section{Injection studies} 
\label{sec:inj}

In this section, we show the results coming from a set 
of {\it injection} studies 
performed with the {\bajes} pipeline,
in order to test the sampling routines with the GW infrastructure.
An injection is a simulated GW signal that has 
been added into a time-domain segment. 
For our studies, we generate artificial noise segments
according with a prescribed PSD assuming 
Gaussian and stationary fluctuations,
as discussed in Sec.~\ref{sec:noise}. 
Subsequently, 
the artificial signal $h(t)$ is simulated 
according with the input parameters $\params_{\rm cbc}$, 
projected on the requested detectors and included in the data segment.
Finally, the data are analyzed by the {\bajes} pipeline
resorting to the framework described above.
The sensitivity curves employed for these studies correspond to
the noise spectra expected at design sensitivity for current ground-based detector~\cite{TheLIGOScientific:2014jea,Aasi:2013wya,Harry:2010zz,TheVirgo:2014hva,Aso:2013eba,Akutsu:2018axf} 
and for next-generation interferometers~\cite{Punturo:2010zz,Hild:2010id}.
The properties of the injected signals are described in the following paragraphs 
depending on the particular kind of source.

\begin{table*}[t]
	\centering    
	\caption{Recovered parameters during the BBH injections studies.
				The signal has been injected in {\tt H1}{+}{\tt L1} using 
			design sensitivity curves, with an overall network SNR of 14.
			The data has been analyzed with the {\bajes} PTMCMC sampling.
				The reported values correspond to the medians with the 90\% credible regions.
			The last column reports the estimated logarithmic Bayes' factor 
			and the associated standard deviation.}
			\resizebox{\textwidth}{!}{
	\begin{tabular}{c|ccccccccc|cc}        
		\hline
		\hline
		Approximant & 
		$\M$ & 
		$q$ & 
		$\chi_{1,z}$ & 
		$\chi_{2,z}$ & 
		$\chi_{\rm eff}$ & 
		$D_L$ & 
		$\iota$ & 
		$\alpha$ & 
		$\delta$ & 
		$\log\B^{\rm S}_{\rm N}$\\
		& $[\Mo]$ &&& && $[{\rm Gpc}]$ &$[{\rm rad}]$ &  $[{\rm rad}]$ & $[{\rm rad}]$ & \\
		\hline
		\hline
		Injected&  30.0&2.0&0.3&0.0&0.2  & 3.0 &0.0 &0.372 &0.811 &--&\\
		\hline
		{\tt TEOBResumS}& ${30.63}^{+1.84}_{-1.64}$ &
		${1.56}^{+0.78}_{-0.49}$&  
		${0.24}^{+0.36}_{-0.41}$&
		${0.17}^{+0.59}_{-0.63}$&
		${0.23}^{+0.14}_{-0.15}$&
		${2.49}^{+1.29}_{-1.17}$&
		${0.82}^{+2.05}_{-0.63}$&
		${2.17}^{+4.05}_{-2.13}$&
		${0.21}^{+1.10}_{-1.09}$&
		${73.29}^{+2.64}_{-2.64}$  \\
		{\tt IMRPhenomPv2}& ${30.34}^{+1.77}_{-1.69}$ &
		${1.50}^{+0.83}_{-0.46}$&  
		${0.17}^{+0.43}_{-0.43}$&
		${0.17}^{+0.57}_{-0.60}$&
		${0.19}^{+0.14}_{-0.16}$&
		${2.47}^{+1.30}_{-1.22}$&
		${0.80}^{+2.06}_{-0.62}$&
		${2.34}^{+ 3.88}_{- 2.31}$&
		${0.17}^{+1.13}_{-1.05}$&
		${73.16}^{+2.66}_{-2.66}$ \\
		{\tt NRSur7dq4}& ${30.35}^{+1.75}_{-1.60}$ &
		${1.56}^{+0.82}_{-0.53}$&  
		${0.18}^{+0.37}_{-0.39}$&
		${0.17}^{+0.59}_{-0.63}$&
		${0.19}^{+0.56}_{-0.63}$&
		${2.49}^{+1.25}_{-1.23}$&
		${0.84}^{+2.02}_{-0.64}$&
		${2.11}^{+4.12}_{-2.07}$&
		${0.21}^{+1.10}_{-1.09}$&
		${73.11}^{+2.72}_{-2.72}$ \\
		\hline
		\hline
			\end{tabular}
	}
\label{tab:inj_bbh}
\end{table*}

\subsubsection{Binary black holes} 
\label{sec:bbh}

The first example corresponds to an aligned-spinning BBH coalescence
with intrinsic parameters 
\{$\M = 30~\Mo$, $q=2$, $\chi_{1,z} = 0.3$, $\chi_{2,z} = 0.$\} 
located at a luminosity distance $D_L=3~{\rm Gpc}$ with inclination angle $\iota=0$.
The signal is injected such that the merger occurs at GPS time 1126259462.0
with a sampling rate of 4096~Hz and a signal length of
$16~{\rm s}$.
The data are analyzed using two detectors, {\tt H1} and {\tt L1}, with LIGO design sensitivity P1200087~\cite{TheLIGOScientific:2014jea,Aasi:2013wya,Harry:2010zz}.
The sky location of the source corresponds to the position of 
maximum sensitivity for the detector {\tt H1}, 
$\{\alpha=0.372, \delta=0.811\}$.
The injected signal is generated with {\tt TEOBResumS} waveform model
(employing only the dominant mode)
with a network SNR of 14, 
corresponding to 11 in {\tt H1} and 9 in {\tt L1}. 

The recovery of the posterior distribution is performed with PTMCMC sampling
with 8 tempered ensembles and 128 chains per ensemble.
Moreover, we requested $8{\times} 10^3$ burn-in iterations.
The injected strain is analyzed 
in the frequency domain from 20~Hz to 1~kHz
employing three different templates:
{\tt TEOBResumS}, {\tt IMRPhenomPv2} and {\tt NRSur7dq4}.
The likelihood is marginalized over reference time and phase.
We set the chirp mass prior in $[23, 37]\,\Mo$ and the mass ratio in $[1,8]$.
The spins are kept aligned with an isotropic prior in the range $[-0.9,+0.9]$ for every component.
We employ volumetric prior for the luminosity distance
in the support $[100~{\rm Mpc}, 5~{\rm Gpc}]$
and the prior distributions for the remaining parameters are chosen 
according with Sec.~\ref{sec:prior}.

\begin{figure}[t]
	\centering 
	\includegraphics[width=0.49\textwidth]{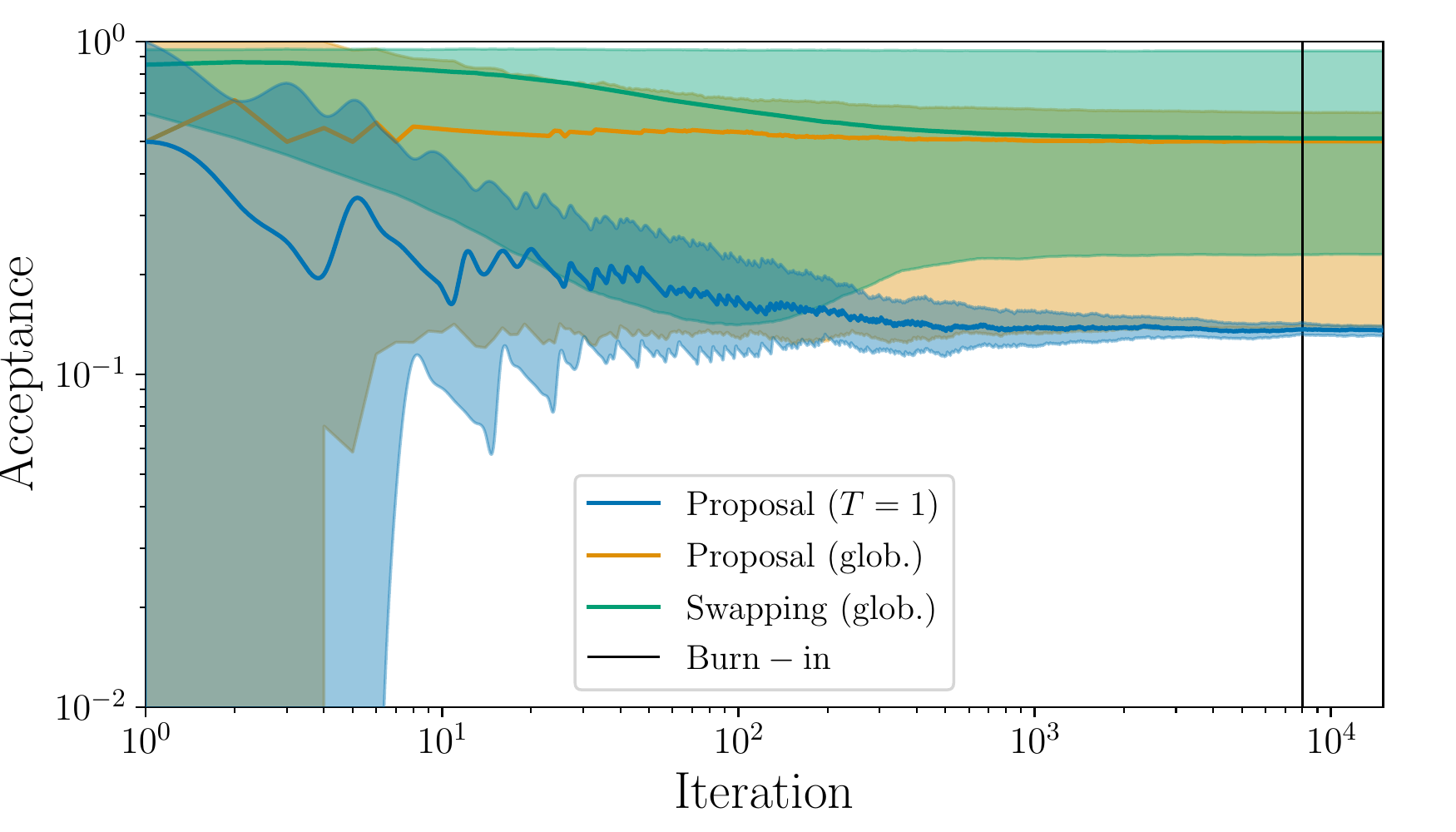}
	\caption{Sampler acceptances during the BBH injection study
		performed with {\tt TEOBResumS} described in Sec.~\ref{sec:bbh}.
		The blue line is the proposal acceptance of the  
		untempered ensemble averaged over the chains,
		the yellow and the green lines are 
		respectively
		the proposal and the swapping acceptances of whole sampler   
		averaged over all the tempered ensembles.
		The solid lines represent the median values 
		and the shadowed areas are the 90\% credible regions.
		The vertical black line is the requested last burn-in iteration.}
	\label{fig:inj_bbh_chain}
\end{figure}

Figure~\ref{fig:inj_bbh_chain} shows the acceptance fractions 
for the analysis performed with {\tt TEOBResumS} waveform model.
The inferences performed with other approximants shown similar behaviours. 
The untempered ensemble required less then $10^{4}$ iterations to
converged to the maximum-posterior value.
After the requested burn-in, 
the untempered ensemble shown an average acceptance of $15\%$
and, averaging over all the tempered ensembles, 
the sampler advanced with a global proposal acceptance of ${\sim}45\%$
and with a global swapping acceptance of ${\sim}50\%$.
The final autocorrelation length (ACL) 
of the untempered ensemble corresponds to a lag of 70 iterations
and the sampler collected a final amount of $1.5{\times} 10^4$ 
independent posterior samples.

\begin{figure*}[t]
	\centering 
	\includegraphics[width=0.49\textwidth]{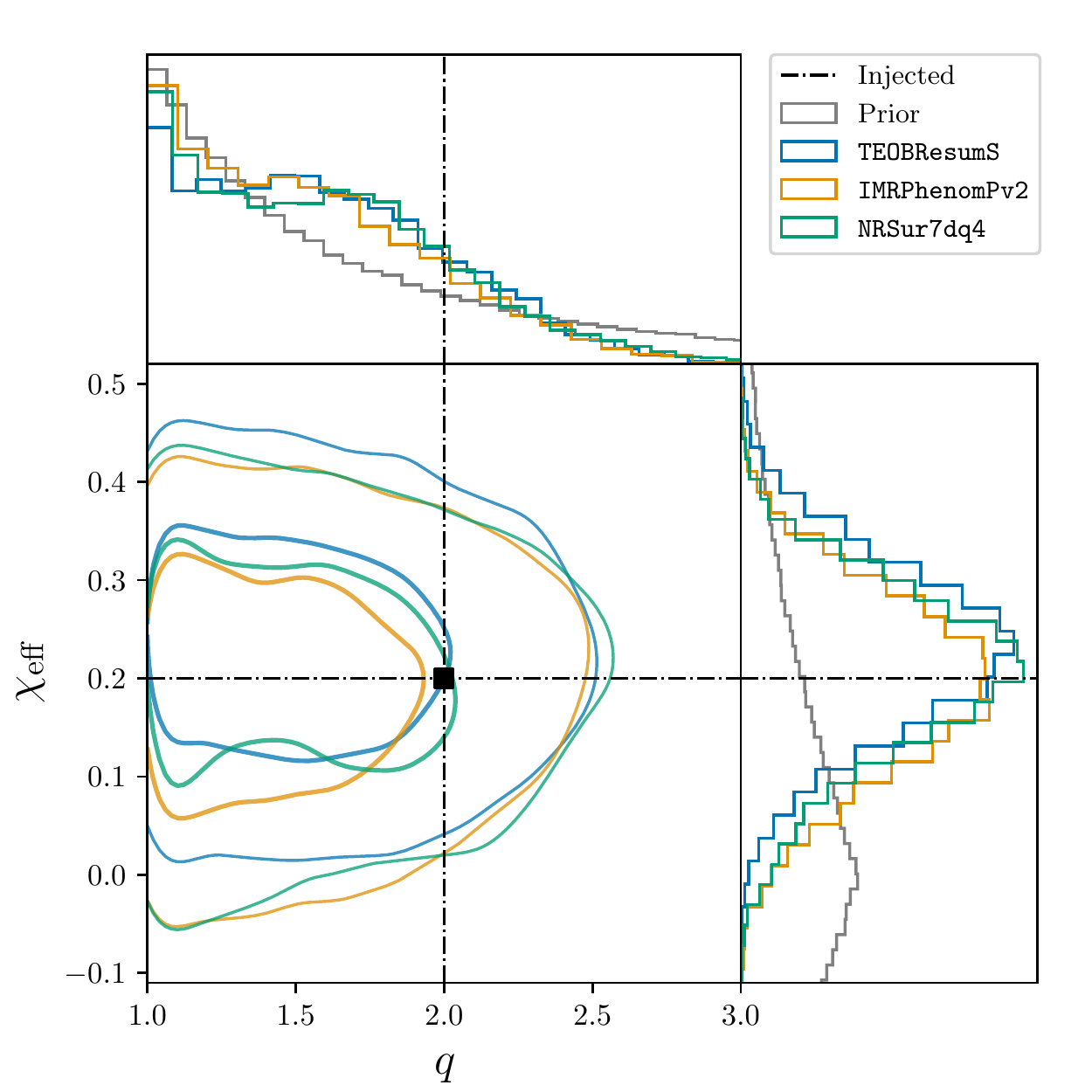}
	\includegraphics[width=0.49\textwidth]{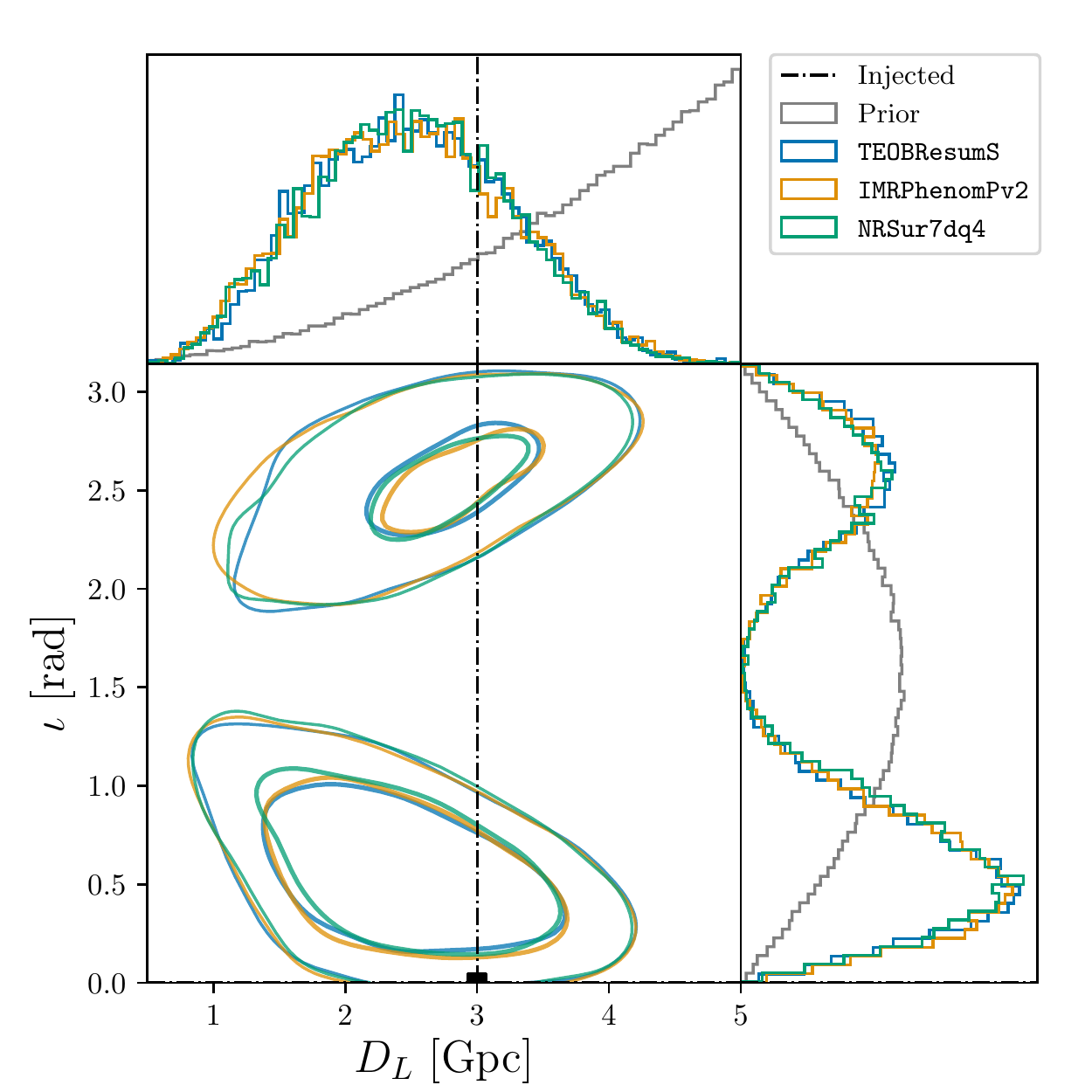}
	\caption{Posterior distributions for $\{q,\chi_{\rm eff}\}$ and $\{D_L, \iota\}$ recovered from the injection studies 
					performed on a BBH signal with two inteferometers ({\tt H1}+{\tt L1}) at design sensitivities
					with network SNR of 14.
					The artificial signal has been generated with {\tt TEOBResumS} model and the injected parameters 
					are marked with black lines and squares.
					The contours represent the 50\% (thick) and the 90\% (thin) credible regions.
					The recovery has been performed with four different approximants 
					analysing the frequency range from 20~Hz to 1~kHz.
					The estimation of the luminosity distance is affected by the 
					degeneracy with the inclination angle~\cite{TheLIGOScientific:2016wfe}, 
					due to the correlations in the strain amplitude for aligned-spin sources.}
	\label{fig:inj_bbh}
\end{figure*}

Table~\ref{tab:inj_bbh} shows the recovered mean values
and Figure~\ref{fig:inj_bbh} shows the recovered marginalized posterior distribution
for some exemplary parameters.
The marginalized posterior distributions 
enclose the injected values within 90\% credible intervals
for all the waveform approximants.
The estimated evidences slightly prefer {\tt TEOBResumS} waveform,
accordingly with the injected template.
However, these values lie in the same range for all the analyses,
leading to a not fully resolved model selection.
This is due to the large uncertainties associated to the evidence estimation
of the PTMCMC and with the relatively low SNR of the 
injected signal.
For the latter reason, it is also 
not possible to reveal systematic 
differences between the different approximants~\cite{Abbott:2016wiq},
and the results of the employed templates 
are largely consistent between each other.

\subsubsection{Binary neutron stars inspiral} 
\label{sec:bnsinsp}

\begin{table*}[t]
	\centering    
	\caption{Recovered parameters 
		during the inpiralling BNS injections studies.
		The signal has been injected in {\tt H1}{+}{\tt L1} using 
		design sensitivity curves, with an overall network SNR of 20.
		The data has been analyzed with the nested sampling 
		provided by {\tt dynesty}.
		The reported values correspond to the medians with the 90\% credible regions.
		The last column reports the estimated logarithmic Bayes' factor 
		and the associated standard deviation.}
	\resizebox{0.99\textwidth}{!}{
		\begin{tabular}{c|cccccccc|cc}        
			\hline
			\hline
			Approximant & 
			$\M$ & 
			$q$ & 
			$\chi_{\rm eff}$ & 
			$\tLam$& 
			$D_L$ & 
			$\iota$ & 
			$\alpha$ & 
			$\delta$ & 
			$\log\B^{\rm S}_{\rm N}$\\
			& $[\Mo]$ & &&& $[{\rm Mpc}]$ &$[{\rm rad}]$ &  $[{\rm rad}]$ & $[{\rm rad}]$ & \\
			\hline
			\hline
			Injected&  1.188&1.00&0.00 & 600  & 120 &0.00 &0.372 &0.811 &--&\\
			\hline
			{\tt TEOBResumSPA}& ${1.1880}^{+0.0002}_{-0.0002}$ &
			${1.20}^{+0.42}_{-0.17}$&  
			${0.01}^{+0.05}_{-0.01}$&
			${435}^{+305}_{-248}$ &
			${113}^{+19}_{-38}$&
			${0.60}^{+1.98}_{-0.42}$&
			${0.48}^{+0.26}_{-0.12}$&
			${0.86}^{+0.20}_{-0.21}$&
			${564.6}^{+0.3}_{-0.3}$ \\
			{\tt TaylorF2}& ${1.1880}^{+0.0005}_{-0.0001}$ &
			${1.28}^{+1.16}_{-0.25}$&  
			${0.01}^{+0.10}_{-0.01}$&
			${392}^{+415}_{-260}$ &
			${106}^{+26}_{-38}$&
			${0.82}^{+1.91}_{-0.56}$&
			${0.75}^{+3.51}_{-0.33}$&
			${0.81}^{+0.28}_{-1.56}$&
			${564.3}^{+0.3}_{-0.3}$ \\
			{\tt IMRPhenomPv2NRT}& ${1.1880}^{+0.0002}_{-0.0001}$ &
			${1.25}^{+0.36}_{-0.21}$&  
			${0.01}^{+0.02}_{-0.01}$&
			${316}^{+304}_{-215}$ &
			${111}^{+20}_{-50}$&
			${0.76}^{+1.98}_{-0.55}$&
			${0.63}^{+3.54}_{-0.21}$&
			${0.79}^{+0.24}_{-1.49}$&
			${563.9}^{+0.3}_{-0.3}$ \\
			\hline
			\hline
		\end{tabular}
	}
	\label{tab:inj_bns}
\end{table*}

In this section we analyse an inspiralling nonspinning BNS merger with intrinsic parameters are
\{$\M = 1.188~\Mo$, $q=1$, $\Lambda_1=\Lambda_2 = 600$\}.
The source is located at a luminosity distance 
$D_L=120~{\rm Mpc}$ with inclination angle $\iota=0$.
The signal is injected such that the merger occurs at GPS time 1126259462.0
with a sampling rate of 4096~Hz and a signal length of 128~s.
The data are analyzed using two detectors, {\tt H1} and {\tt L1}, with LIGO design sensitivity P1200087~\cite{TheLIGOScientific:2014jea,Aasi:2013wya,Harry:2010zz}.
The sky location of the source corresponds to the position of 
maximum sensitivity for the detector {\tt H1}, 
$\{\alpha=0.372, \delta=0.811\}$.
The injected signal is generated with {\tt TEOBResumS} waveform model
with a network SNR of 20, 
corresponding to 15 in {\tt H1} and 13 in {\tt L1}.

The recovery of the posterior distribution is performed employing nested sampling algorithm
with 1024 live point and tolerance equal to $0.1$.
Furthermore, we set respectively the minimum and the maximum number
of iterations for every MCMC sub-chain to 32 and $4{\times} 10^{3}$.
The injected strain is analyzed 
in the frequency domain from 20~Hz to 1~kHz
employing three different templates:
{\tt TEOBResumSPA}, {\tt IMRPhenomPv2\_NRTidal} and 
{\tt TaylorF2} including 5.5PN point-mass corrections 
and 6PN tidal contributions.
We set the chirp mass prior in $[0.8, 2]\,\Mo$ and the mass ratio in $[1,4]$.
The spins are kept aligned with an isotropic prior in the range $[-0.9,+0.9]$ for every component.
The tidal parameters are extracted in the range $[0,5000]$.
We employ volumetric prior for the luminosity distance 
in the support $[10,400]\,{\rm Mpc}$
and the prior distributions for the remaining parameters are chosen 
according with Sec.~\ref{sec:prior}.

\begin{figure}[t]
	\centering 
	\includegraphics[width=0.49\textwidth]{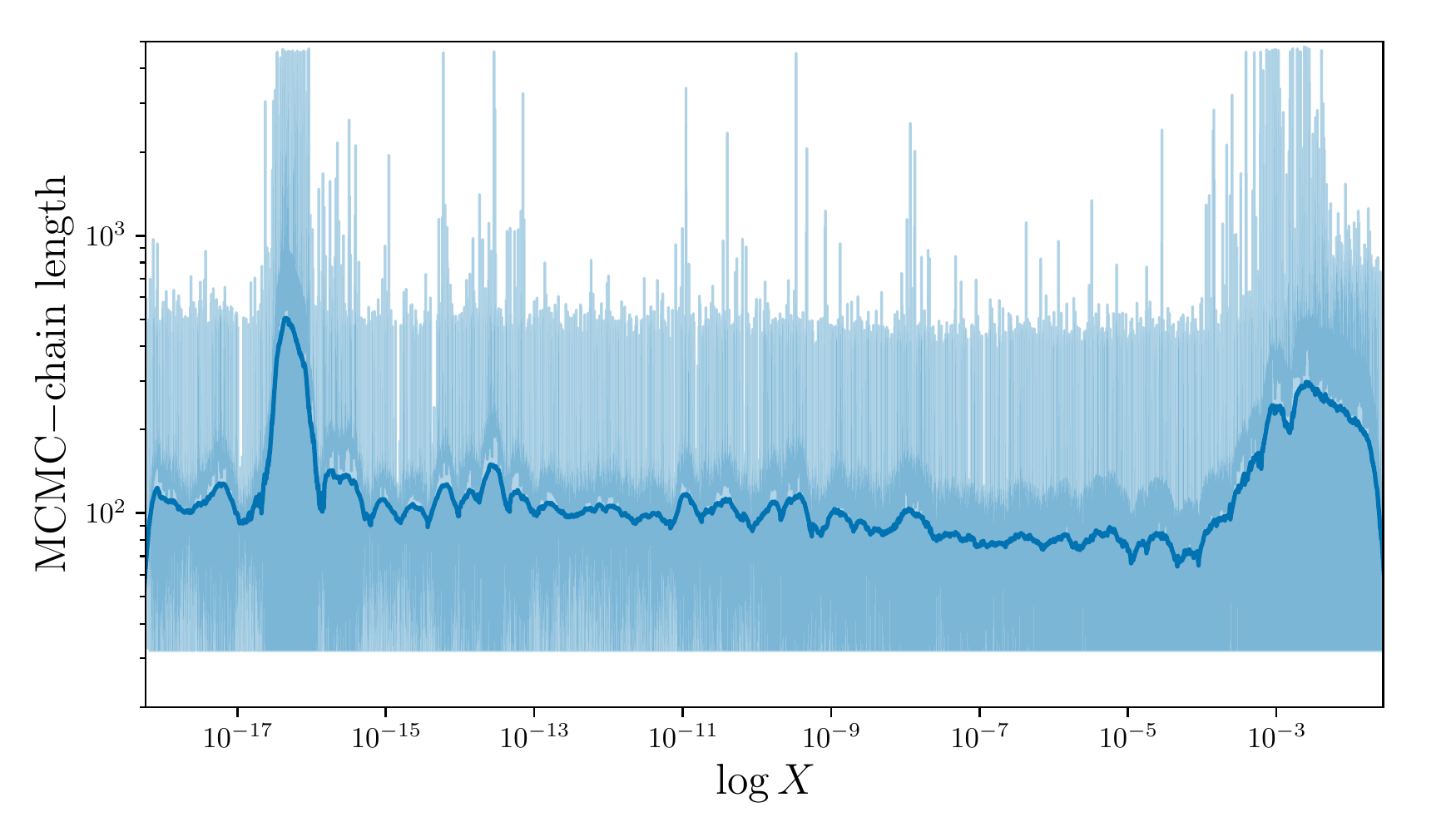}
	\caption{Length of the MCMC sub-chains during the nested sampling 
		performed with {\tt TEOBResumSPA} described in Sec.~\ref{sec:bnsinsp}.
		The actual values extracted from the sampler 
		have been smoothed with a Savitzky-Golay filter for visualisation purposes.
		The shadowed region shows the difference between the actual lengths
		and the filtered estimation.
		The values are lower-bounded by the requested minimum value, set equal to 32.}
	\label{fig:inj_bns_chain}
\end{figure}

Figure~\ref{fig:inj_bns_chain} shows the number of iterations of the 
MCMC sub-chains employed to propose a new point as a function of the 
prior mass $X$ (see App.~\ref{app:nest} for the definition).
The actual values extracted from the sampler 
have been smoothed with a Savitzky-Golay filter for visualisation purposes.
The shadowed region shows the difference between the actual MCMC-chain lengths
and the filtered estimation.
Large values of MCMC-chain length (say $\gtrsim 250$) characterize
the more expensive steps, where the proposal method requires
more iterations in order to propose a new samples and estimate the 
boundaries of the current contour.
During the initial stages ($X {\approx} 1$), the boundaries defined by the current live points 
were comparable with the prior and 
the sampler required many iterations in order to propose new samples.
Subsequently, for $\log X \lesssim 10^{4}$, 
the sampler identified the region that encloses the majority of the posterior volume
and the algorithm advanced spending ${\sim}100$ iterations to propose a new sample. 
The length of the MCMC-chain slightly increases again 
during the latest stages, since the sampler has to reach
the bulk of the posterior distribution
restricting the boundaries to a neighborhood 
of the maximum-likelihood values.

\begin{figure*}[t]
	\centering 
	\includegraphics[width=0.49\textwidth]{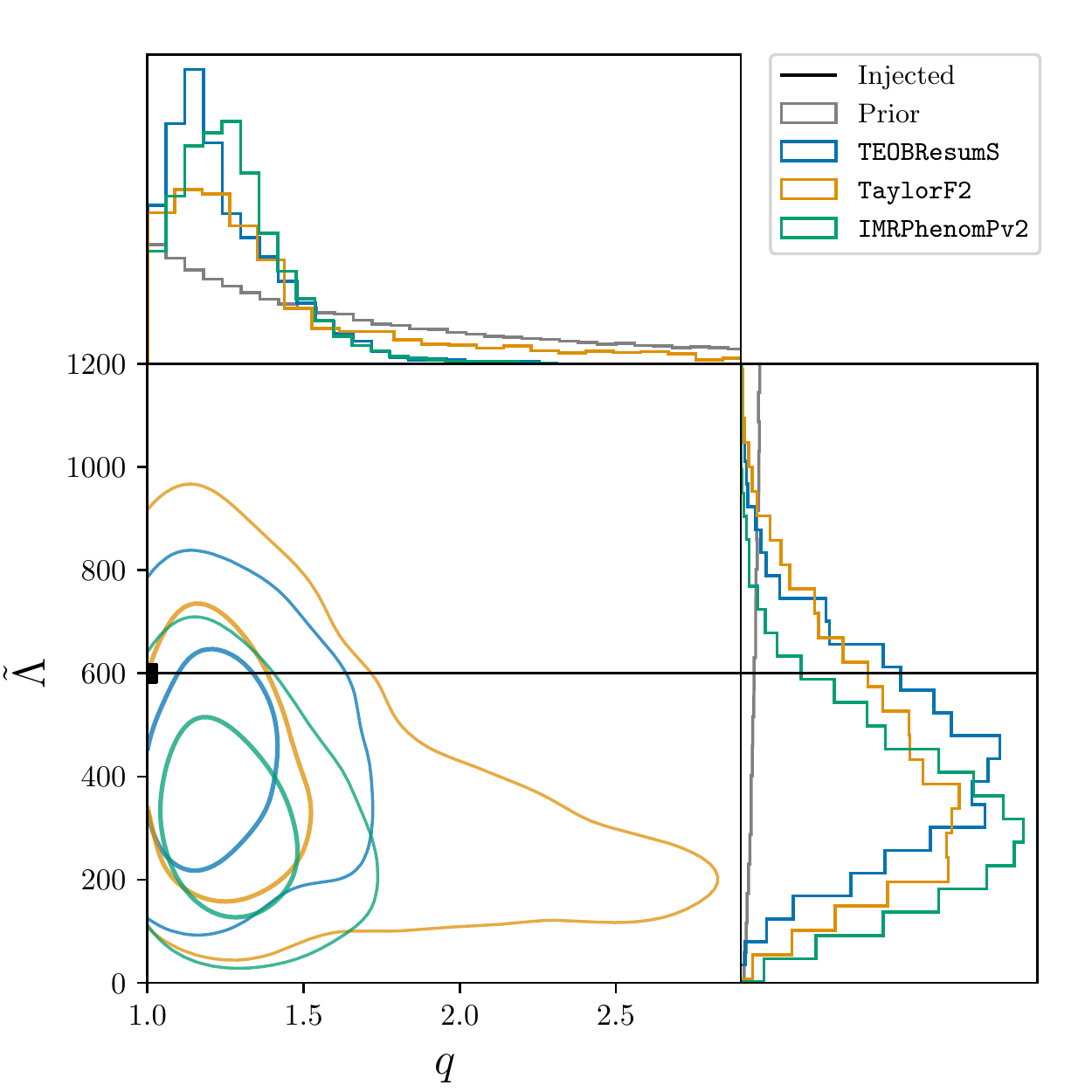}
	\includegraphics[width=0.49\textwidth]{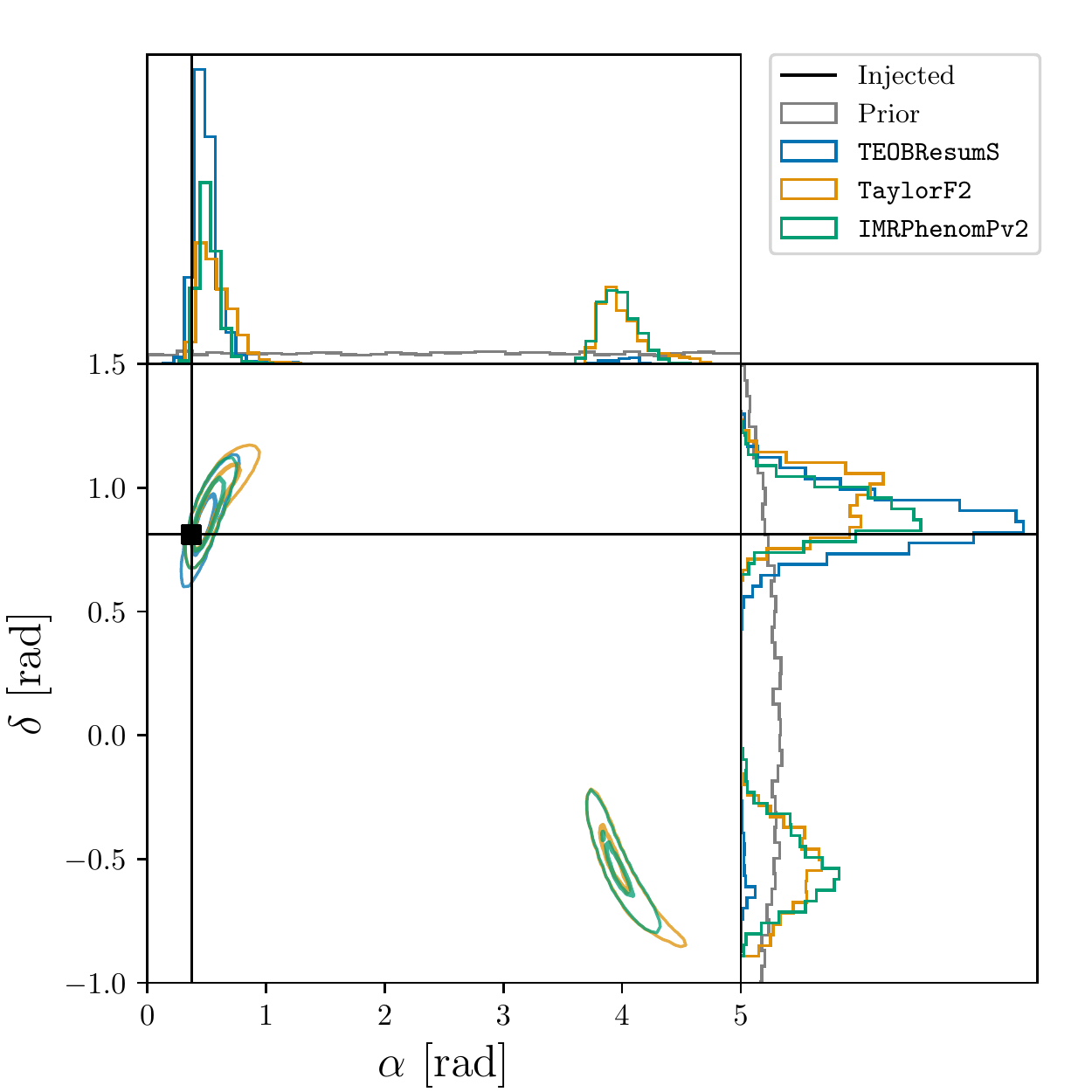}
	\caption{Posterior distributions for $\{q,\tLam\}$ and 
		$\{\alpha, \delta\}$ recovered from the injection studies 
		performed on an inspiralling BNS signal with two inteferometers ({\tt H1}+{\tt L1}) at design sensitivities with network SNR of 20.
		The artificial signal has been generated with {\tt TEOBResumS} model and the injected parameters 
		are marked with black lines and squares.
		The recovery has been performed with three different approximants analysing frequency range from 20~Hz to 1~kHz. 
		The degeneracy in the sky location can be removed introducing a third detector~\cite{Abbott:2017oio} 
		and it is due to the correlations 
		between longitudinal and latitudinal angles 
		that concur in the estimation of times of arrival in the different interferometers.}
	\label{fig:inj_bns}
\end{figure*}

Table~\ref{tab:inj_bns} shows the recovered median values 
and Figure~\ref{fig:inj_bns} shows the marginalized posterior distribution
for some exemplary parameters. 
The recovered values are in agreement with the properties of the
injected signal: the posterior distribution encloses the injected sample
for all the parameter in the 50\% credible region, with a
small bias in the in the maximum-posterior value  
for the reduced tidal parameter $\tLam$, corresponding to
roughly ${\sim}150$. However,
this behavior is expected~\cite{Wade:2014vqa,Abbott:2018wiz,Dudi:2018jzn,Samajdar:2019ulq,Gamba:2020wgg}
considering that we employed an upper cutoff-frequency of 1~kHz
\footnote{For typical BNS, the information on the tidal parameters 
	is gathered in frequency range above 800~Hz~\cite{Damour:2012yf,Gamba:2020wgg}.}
combined with the large aligned-spin prior
\footnote{Large spin effects can mitigate the tidal contributions,
	leading to an underestimate the tidal parameters~\cite{Samajdar:2019ulq}.}.
The estimated evidences slightly prefer {\tt TEOBResumSPA} approximant,
accordingly with the injected model; however, due to the low SNR,
they do not show any strong preference.

\subsubsection{Binary neutron stars postmerger} 
\label{sec:bnspm}

\begin{table}[t]
	\centering    
	\caption{Recovered parameters during the BNS postmerger injection survey.
		The reported values correspond to the medians with the 90\% credible regions.}
	\resizebox{0.49\textwidth}{!}{
	\begin{tabular}{c|cccc|c}        
		\hline
		\hline
		SNR & 
		$\M$ & 
		$\tLam$&
		$f_2$ & 
		$R_{\rm max}$ & 
		$\log\B^{\rm S}_{\rm N}$\\
		& $[\Mo]$ & &$[{\rm kHz}]$&$[{\rm km}]$& \\
		\hline
		Injected & 1.188 & 600 & 2.94 & 10.8 & -- \\
		\hline
		8&
		${2.2}^{+0.8}_{-1.5}$  &
		${2700}^{+1800}_{-1800}$ &  
		${0.8}^{+2.1}_{-0.3}$& 
		${27}^{+10}_{-19}$& 
		${0.2}^{+0.2}_{-0.2}$ \\
		9&
		${1.3}^{+1.6}_{-0.7}$  &
		${2350}^{+1900}_{-1700}$ &  
		${1.7}^{+1.2}_{-1.1}$& 
		${14}^{+22}_{-6}$& 
		${2.6}^{+1.3}_{-1.3}$ \\
		10&
		${0.8}^{+2.2}_{-0.2}$  &
		${2830}^{+1630}_{-1910}$ &  
		${2.7}^{+0.3}_{-2.0}$& 
		${8.9}^{+2.8}_{-0.9}$& 
		${11.0}^{+4.8}_{-4.8}$ \\
		12&
		${0.8}^{+2.1}_{-0.2}$  &
		${1860}^{+2570}_{-1410}$ &  
		${2.9}^{+0.7}_{-2.3}$&  
		${9.0}^{+2.7}_{-0.8}$& 
		${13.9}^{+6.3}_{-6.3}$  \\
		16&
		${0.78}^{+0.43}_{-0.18}$  &
		${1780}^{+2640}_{-1220}$ &  
		${2.93}^{+0.02}_{-0.39}$&  
		${8.9}^{+2.0}_{-0.8}$& 
		${45.2}^{+4.2}_{-4.2}$ \\
		32& 
		${0.80}^{+0.45}_{-0.12}$  &
		${1730}^{+730}_{-1220}$ &  
		${2.93}^{+0.02}_{-0.03}$&  
		${9.0}^{+2.1}_{-0.6}$& 
		${271}^{+29}_{-29}$ \\
		\hline
		\hline
	\end{tabular}
	}
	\label{tab:inj_pm}
\end{table}

\begin{figure*}[t]
	\centering 
	\includegraphics[width=0.99\textwidth]{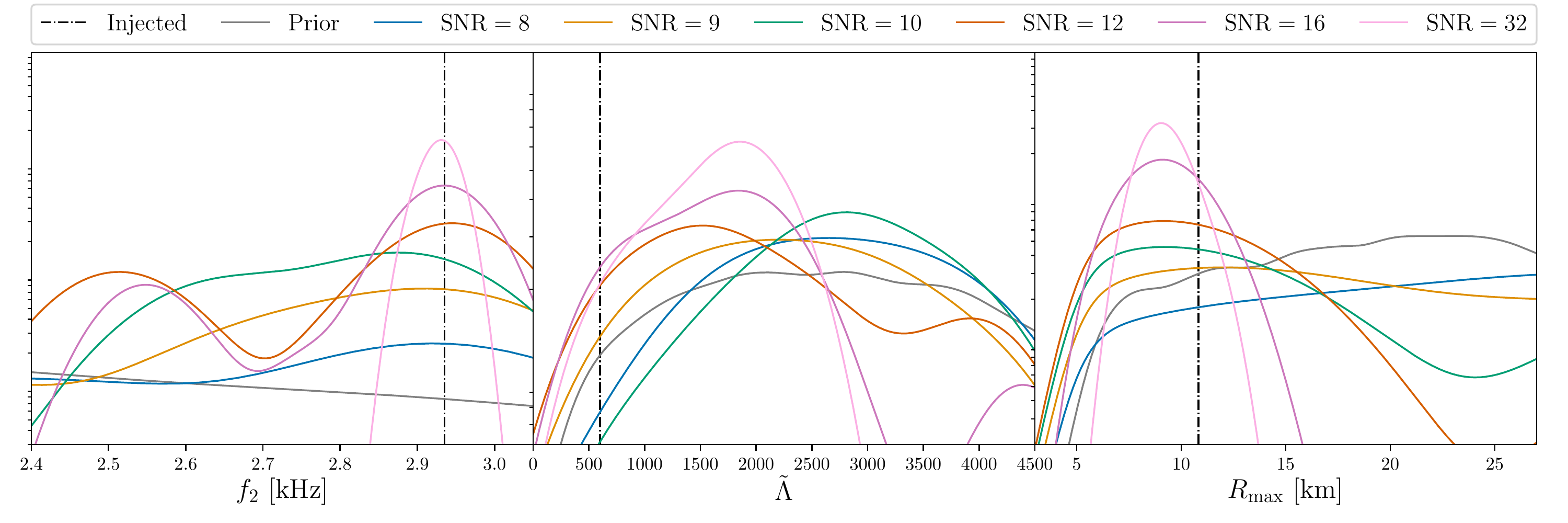}
	\caption{
		Marginalized posterior distributions for $\{f_2, \tLam, R_{\rm max}\}$ recovered
		in the BNS postmerger injection survey
		using a five-detector-network {\tt H1}+{\tt L1}+{\tt V1}+{\tt K1}+{\tt ET}
		at design sensitivities 
		varying the luminosity distance
		and locating the source in the position of maximum sensitivity for {\tt ET}.
		The injected signals have been generated with {\tt NRPM} template and recovered
		with the same model. 
	}
	\label{fig:inj_pm}
\end{figure*}

\begin{figure}[t]
	\centering 

	\includegraphics[width=0.49\textwidth]{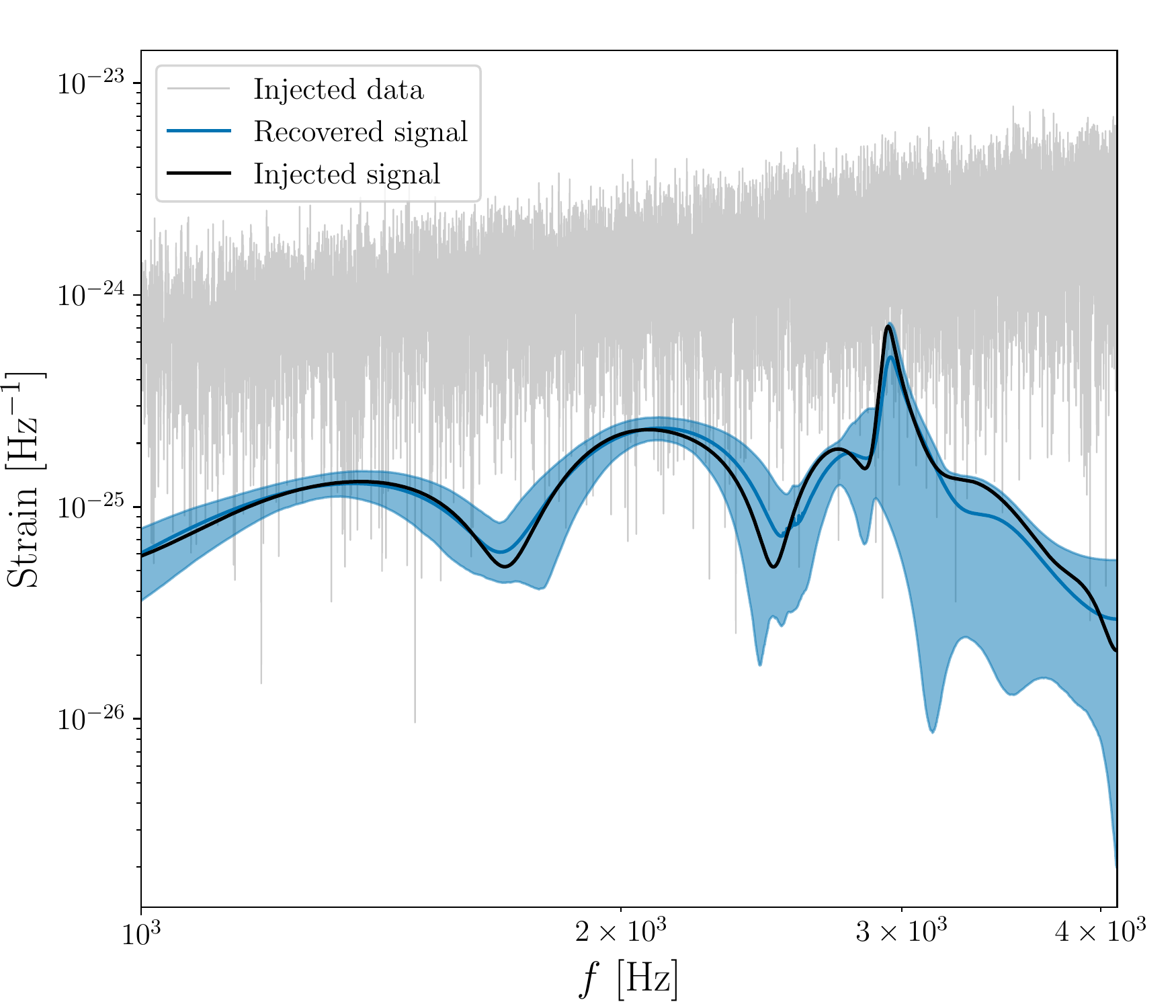}	
	\caption{Frequency-domain amplitude strains
		in the {\tt ET} detector
		for the analysis at SNR 16 corresponding to a 
		luminosity distance of 40~Mpc.				
		The plotted quantities correspond to 
		the injected signal (black), 
		the full artificial data strain (grey) and
		the recovered template (blue),
		where the solid line is the mean value 
		and the shadowed area is the 90\% credible region. 
	}
	\label{fig:injrec_pm}
\end{figure}

We perform a set of BNS postmerger injections 
using a five-detector network which includes: 
{\tt H1} and {\tt L1} with LIGO design sensitivity P1200087~\cite{TheLIGOScientific:2014jea,Aasi:2013wya,Harry:2010zz},
 {\tt V1} with Virgo design sensitivity P1200087~\cite{TheVirgo:2014hva},
 {\tt K1} with KAGRA design sensitivity T1600593~\cite{Aso:2013eba,Akutsu:2018axf}
 and the third-generation interferometer
{\tt ET} (configuration D) P1600143~\cite{Punturo:2010zz,Hild:2010id}.
The injected signals are generated with {\tt NRPM} and correspond
to the postmerger radiations of non-spinning BNSs with intrinsic parameters
\{$\M=1.188~\Mo$, $q=1$, $\Lambda_1=\Lambda_2 = 600$\}.
The signal is injected such that the merger occurs at GPS time 1126259462.0
with a sampling rate of 8192~Hz and a signal length of 4~s.
The sky location of the source corresponds to the position of 
maximum sensitivity for the detector {\tt ET}, 
$\{\alpha=2.640, \delta=0.762\}$.
The signals are injected at different luminosity distances 
in order to simulate different SNRs, spanning the range 
from 8 to 32, which corresponds to $D_L\in[20, 80]~{\rm Mpc}$.
We observe that, for these kind of signal, 
{\tt ET} is the most relevant
detector and it concurs in the determination of the SNR with a weight 
larger than 90\% for all the analyzed cases.

For these studies, we employ the PTMCMC  sampler
using 8 tempered ensembles with 128 chains and $2\times10^4$ burn-in iterations.
The injected strain is analyzed 
in the frequency domain from 1~kHz to 4~kHz
employing {\tt NRPM} waveforms
\footnote{
	The same template model is used for both injection 
	and recovery, in order to avoid noise contributions different from 
	the simulated detector noise.}. 
We set the chirp mass prior in $[0.8, 2]\,\Mo$ and the mass ratio in $[1,1.5]$.
The spins are kept fixed to zeros for every component.
The tidal parameters are extracted in the range $[0,5000]$.
We employ volumetric prior for the luminosity distance 
in the support $[5~{\rm Mpc},400~{\rm Mpc}]$
and the prior distributions for the remaining parameters are chosen 
according with Sec.~\ref{sec:prior}.

As shown in Ref.~\cite{Bernuzzi:2015rla},
the postmerger GW radiation of a long-lived BNS merger
is characterized by a main frequency peak in the Fourier domain 
($f_2$) that can be parametrized with quasiuniversal
  (EOS-insensitive) relations involving the tidal polarizability
  parameters.
The {\tt NRPM} model is constructed using these relations that allow one to constrain $\tLam$ from
  postmerger observations and at the same time map the properties 
of a postmerger signal into the inspiral parameters of the binary \cite{Breschi:2019srl}.
Furthermore, numerical relativity simulations~\cite{Bauswein:2012ya,Breschi:2019srl} have shown that
the postmerger frequency peak $f_2$ can be related to the radius $R_{\rm max}$ of the 
maximum mass configuration of a non-rotating neutron star.
For the injected sources, 
we get $f_2 = 2.94~{\rm kHz}$ and $R_{\rm max} = 10.8~{\rm km}$.

Tab~\ref{tab:inj_pm} shows the recovered mean values and
Figure~\ref{fig:inj_pm} shows the marginalized posterior distributions 
for $f_2$ and for $\tLam$ recovered during the survey described above and
Figure~\ref{fig:injrec_pm} presents the recovered postmerger signal recovered 
in the {\tt ET} detector for the case with SNR 16.
The Bayes factors shows evidence of signal from SNR 9;
however, in order to estimate $f_2$ with an accuracy of $O(0.1~{\rm kHz})$,
the method requires an SNR ${\gtrsim}12$.
The mean values estimated from the 
marginalized posteriors agree with the injected properties of the signal 
within the 90\% credible intervals; however, 
it is interesting to observe that, due to the correlations induced by the 
EOS-insensitive relations, the sampler explores non-trivial degeneracy
between the intrinsic parameters.
The same behavior has been shown in Ref.~\cite{Breschi:2019srl}.
These correlations strongly affect the estimation of $\tLam$
and PE of a postmerger signal 
is only capable of imposing an upper bound for this parameter.
For example, with SNR 32 it would be possible 
to constrain the value of $\tLam$
with an uncertainty of ${\sim}10^4$,
that is a much larger value compared with the uncertainties coming 
from the analysis of the inspiral data (see Tab.~\ref{tab:inj_bns}).
Nevertheless, the observation of a postmerger signal
would extraordinarily extend our knowledge regarding 
neutron star matter~\cite{Bauswein:2014qla, Radice:2016rys, Agathos:2019sah},
allowing us to verify our current models and to constrain the extreme-density properties
of the EOS, such as the radius of the maximum mass star $R_{\rm max}$
and the inference of softening effects at high-densities~\cite{Bauswein:2018bma,Breschi:2019srl}.
These constraints can be further improved with the inclusion of the 
inspiral information within a full inspiral-merger-postmerger analysis of the 
observed BNS signal.

\subsubsection{Confidence interval test}
\label{sec:citest}

\begin{figure}[t]
	\centering 
	\includegraphics[width=0.49\textwidth]{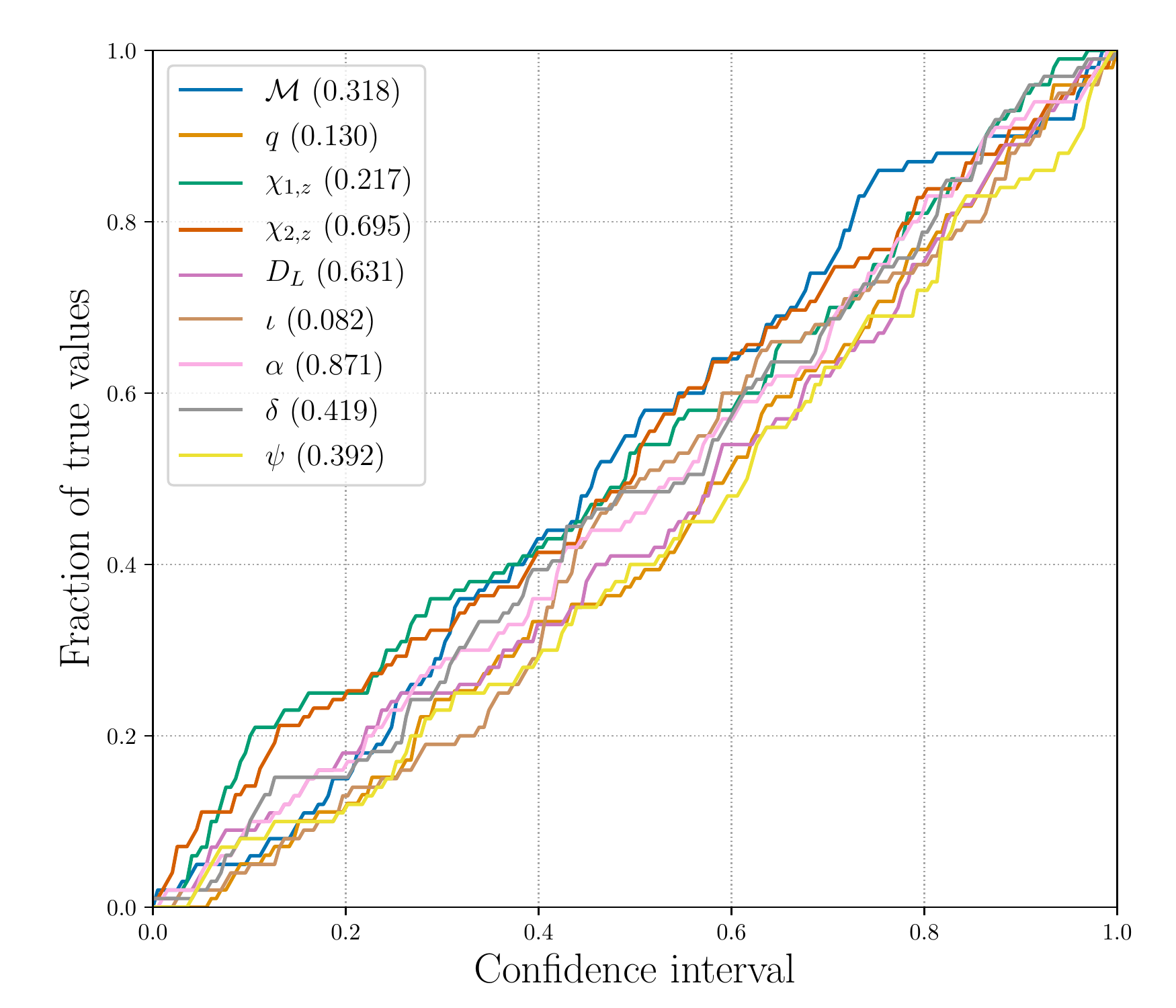}	
	\caption{Fraction of events found within a confidence interval 
					for test described in Sec.~\ref{sec:citest},
					for which a set of 100 BBH injections was used. 
					For every parameter, the label shows the $p$-value of the
					KS test. The recovered $p$-values are 
					uniformly distributed with a $p$-value of 58\%.}
	\label{fig:ci}
\end{figure}

The confidence interval (CI) test has become a standard control check to verify the reliability of a GW pipeline~\citep[e.g.][]{Veitch:2014wba,DelPozzo:2018dpu,Romero-Shaw:2020owr},
since it ensures that
the recovered probability distributions are truly 
representative of the inferred confidence levels.
For each parameter employed in the analyses,
the CI test measures the fraction of true values that follow below 
a given credible level and, if the algorithm is well-calibrated,
we expect this two quantities to be proportional. 
The test is performed using a large set of injected signals 
with parameters extracted from a population prior $\mathfrak{p}(\params_{\rm cbc})$.
Then, we conduct PE analyses in order to recover 
the posterior distributions for every injection and
the CIs can be estimated from the posterior distributions by determining the
quantiles under which the true parameters lie.
Then, the population prior
$\mathfrak{p}(\params_{\rm cbc})$ is used as input prior distribution
for the analysis of the injected signal
\footnote{We observe that, in order to perform an accurate test 
	the employed prior has to be a good representation of the population distribution. In our case this is ensured by definition since the injected 
	signals are extracted from the employed prior $\mathfrak{p}(\params_{\rm cbc})$.}.

For our CI test, we inject 100 BBH signals 
employing the prior distribution used for the parameters of 
GW150914 (see Sec.~\ref{sec:GW150914}), that includes 9 parameters.
The signals are generated using {\tt MLGW} waveform template
and injected in the two LIGO detectors, 
{\tt H1} and {\tt L1}, at design sensitivity
using segments of duration 8~s.
The analyzed frequency range goes from 20~Hz to 1~kHz.
We use the nested sampling provided by {\tt dynesty} with 1024 live points 
and tolerance equal to 0.1.
The likelihood function is marginalized over reference phase and time shift.
Figure~\ref{fig:ci} shows the recovered fractions of events found within 
an increasing confidence level.
The fraction of event is expected to be uniformly distributed
if the prior distributions is a good approximation of the 
underlying injected population distributions.

For each parameter, we compute the $p$-value of
Kolmogorov-Smirnov (KS) test, quantifying the consistency with
uniformly distributed events.
The results are shown between round brackets in Figure~\ref{fig:ci}.
From these results, we estimate the combined $p$-values
quantifying the probability that the individual $p$-values are extracted from a 
uniform distribution. We estimate an overall $p$-value of $58\%$,
according with analogous computations performed with a set of
9 random uniformly-distributed samples.
Furthermore, we compute the Jensen-Shannon (JS) divergence
 between the distribution of fraction of events with respect to 
a uniform distribution. The results lie around $2\times 10^{-3}$ for all the parameters,
in agreement with analogous estimations performed 
on a set of 100 uniformly-distributed random samples. 
These results confirm that the pipeline is well-calibrated.

\section{LIGO-Virgo transients} 
\label{sec:lvc}

In this section, we apply the {\bajes} pipeline to the 
GW transients~\cite{LIGOScientific:2018mvr} observed by the LIGO-Virgo interferometers~\cite{Aasi:2013wya,TheLIGOScientific:2014jea,TheVirgo:2014hva}.
For all the analyses, the data are extracted from the GWOSC archive~\cite{Abbott:2019ebz,Trovato:2019liz,gwosc}
with a sampling rate of 4096~Hz resorting to the {\tt gwpy} facilities.
The analyzed strains are centered around the nominal GPS time.
Subsequently, the strains are windowed and transformed in the Fourier space,
using the tools described in Sec.~\ref{sec:gw}.
PSD curves and calibration uncertainties are taken from the official 
LIGO-Virgo data release of GWTC-1~\cite{LIGOScientific:2018mvr, gwtc_psd, gwtc_cal}.
We use the nested sampling 
implemented in {\tt dynesty}, 
employing 2048 live points with a tolerance equal to 0.1.
With these settings, we collect ${\sim}5\times10^4$ samples
from each PE analysis.
The measured quantities reported in the text correspond to 
the median values and to the 90\% credible regions,
except when explicitly mentioned.

We note that 
the prior assumptions employed in {\bajes} 
slightly differs from the ones of the official LIGO-Virgo
analysis. In the latter, 
the sampling is performed imposing 
additional bounds in the mass components~\cite{Abbott:2016blz,Abbott:2016izl,TheLIGOScientific:2017qsa,Abbott:2018exr}; while, in {\bajes},
the samples are extracted from the entire square defined by the $\{\M,q\}$ bounds.
Moreover, the strains analyzed by the {\bajes} pipeline are slightly
shifted in time with respect the
official LIGO-Virgo segments due to different reference conventions.

\subsection{GW150914} 
\label{sec:GW150914}

\begin{figure}[t]
	\centering 
	\includegraphics[width=0.49\textwidth]{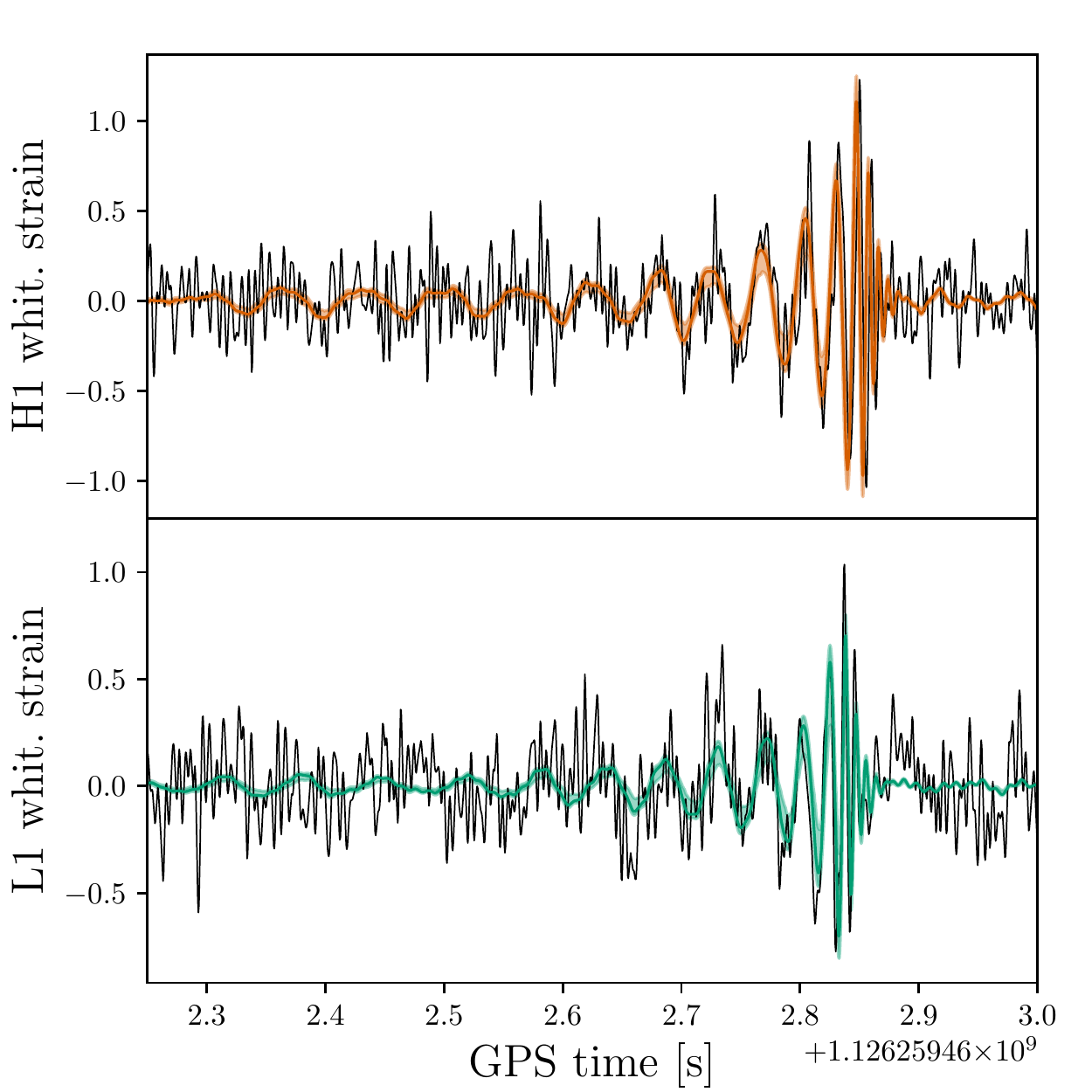}
	\caption{Waveform template reconstructed from the analysis of GW150914 wit {\tt TEOBResumS}
					compared with the LIGO-Hanford (top panel) and LIGO-Livingston (bottom panel) data.
					The black lines are the whitened strains recorded by the LIGO inteferometers,
					where we applied a band-pass filter in the frequency window $[20~{\rm Hz}, 480~{\rm Hz}]$ only for 
					visualization purposes.
					The colored lines are the median recovered template
					projected on the respective detector
					(orange for LIGO-Hanford and green for LIGO-Livingston)
					and the shadowed areas represents the 90\% credible regions.
					The estimated network SNR of the signal corresponds to 22. }
	\label{fig:150914_strain}
\end{figure}

\begin{figure*}[t]
	\centering 
	\includegraphics[width=0.49\textwidth]{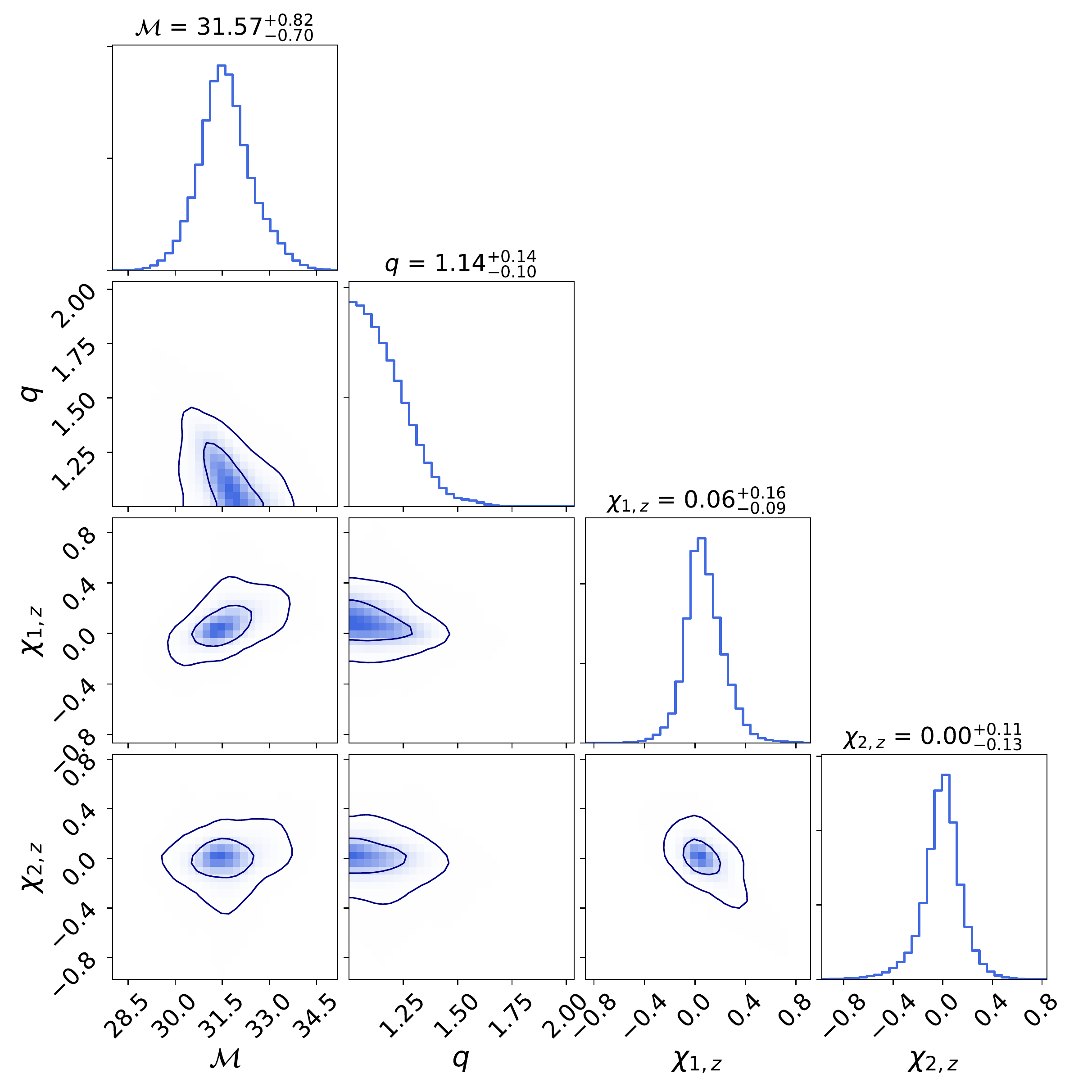}
	\includegraphics[width=0.49\textwidth]{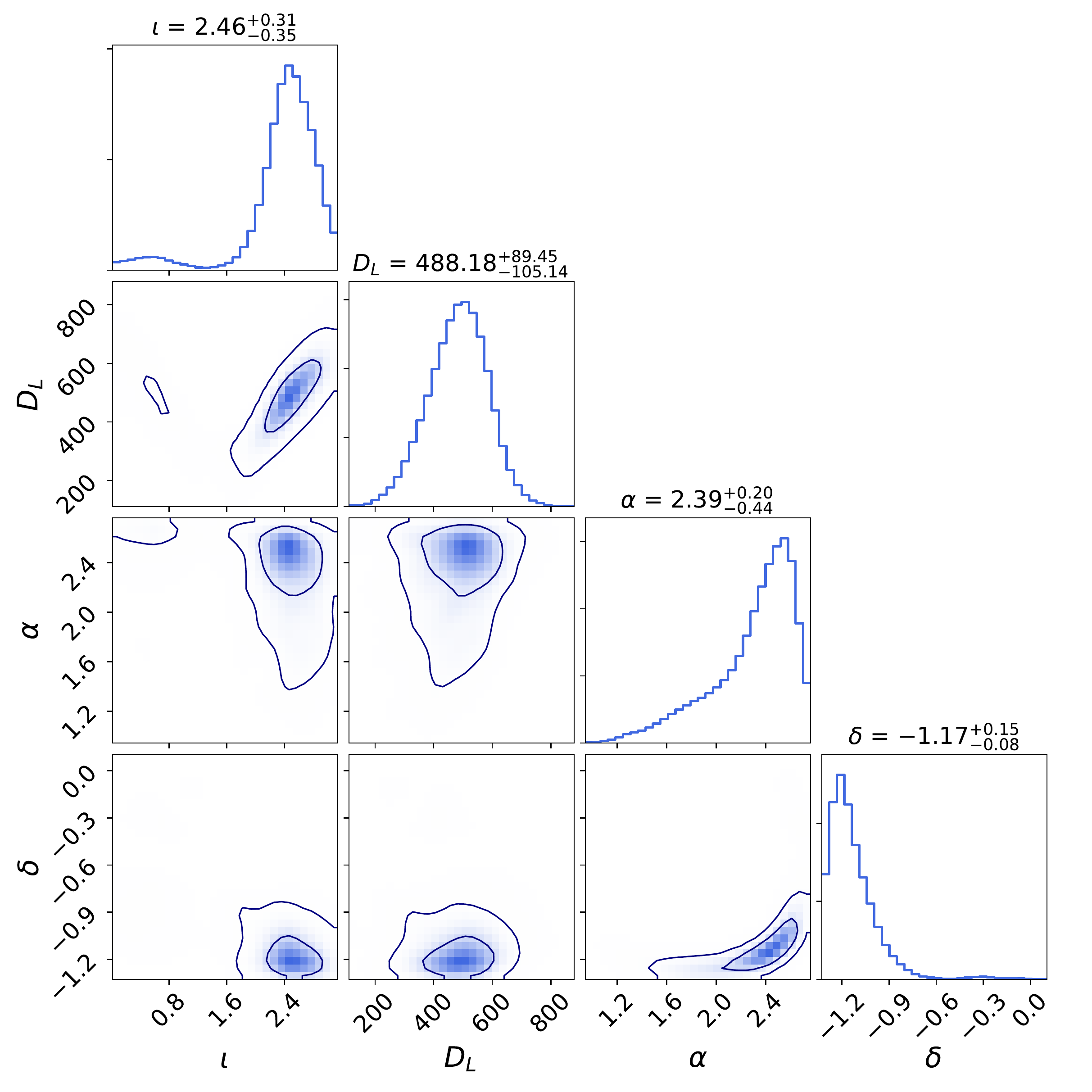}
	\caption{Posterior distributions
		for the intrinsic (left) and extrinsic (right) 
		parameters of GW150914 estimated with
		{\bajes} pipeline employing {\tt TEOBResumS} waveform model
		with aligned spin components.
		All higher-order modes up to $\ell=5$ (with $m=\pm \ell$)
		have been used to generate the waveform models.
		The chirp mass is expressed in Solar masses $\Mo$,
		the luminosity distance is expressed in megaparsec Mpc,
		while the angles $\{\iota, \alpha, \delta\}$ are in radiants.
		We report the median value and the 90\% credible regions for each parameter and the contours represent the 50\% and the 90\% credible regions.}
	\label{fig:150914_teob}
\end{figure*}

In this section, we discuss the results obtained 
from the analysis of the first GW transient observed by 
the LIGO interferometers, GW150914~\cite{Abbott:2016blz,TheLIGOScientific:2016htt,TheLIGOScientific:2016wfe, TheLIGOScientific:2016src}.
For all the discussed cases, the analyzed strains 
correspond to the GWTC-1 release~\cite{LIGOScientific:2018mvr} of LIGO-Hanford and 
LIGO-Livingston data centered around GPS time  1126259462
with a sampling rate of 4096~Hz and a duration of 8~s.
We set the lower cutoff-frequency to 20~Hz and the highest frequency to 1~kHz.
The employed prior is isotropic in spin components 
and volumetric in luminosity distance, and it
spans the ranges $\M\in[12,45]~\Mo$,
$q\in[1,8]$, $\chi_{1,2}\in [0,0.99]$ and $D_L\in [100,5000]~{\rm Mpc}$.
We include 8 spectral nodes for the calibrations of the analyzed strains.

We discuss first the PE analysis employing {\tt TEOBResumS} model with aligned spins,
including all high-order modes up to $\ell=5$ with $m=\pm \ell$, 
i.e. $(2,\pm 2)$, $(3,\pm 3)$, $(4,\pm 4)$, $(5,\pm 5)$.
Figure~\ref{fig:150914_strain} shows the recovered waveform template 
compared with the whitened strains recorded by the LIGO interferometers,
and
Figure~\ref{fig:150914_teob} shows the recovered posterior distribution.
We estimated a network SNR of 23.
The results are consistent with similar studies performed with the 
same approximant~\cite{Nagar:2018zoe,Nagar:2020pcj},
recovering the signal of a non-spinning equal-mass BBH merger 
with $\M=31.57^{+0.82}_{-0.70}~\Mo$.
The inference of the extrinsic parameters shows a source located 
at a luminosity distance of ${\sim}490~{\rm Mpc}$ and
the area covered by the 90\% isoprobability level 
of the sky location posterior distributions 
corresponds to ${\sim}670~{\rm deg}^2$.
The estimated Bayes' factor corresponds to
$\log \B^{\rm S}_{\rm N}=267.8\pm0.2$,
where the uncertainty is given by the standard deviation.

\begin{figure}[t]
	\centering 
	\includegraphics[width=0.49\textwidth]{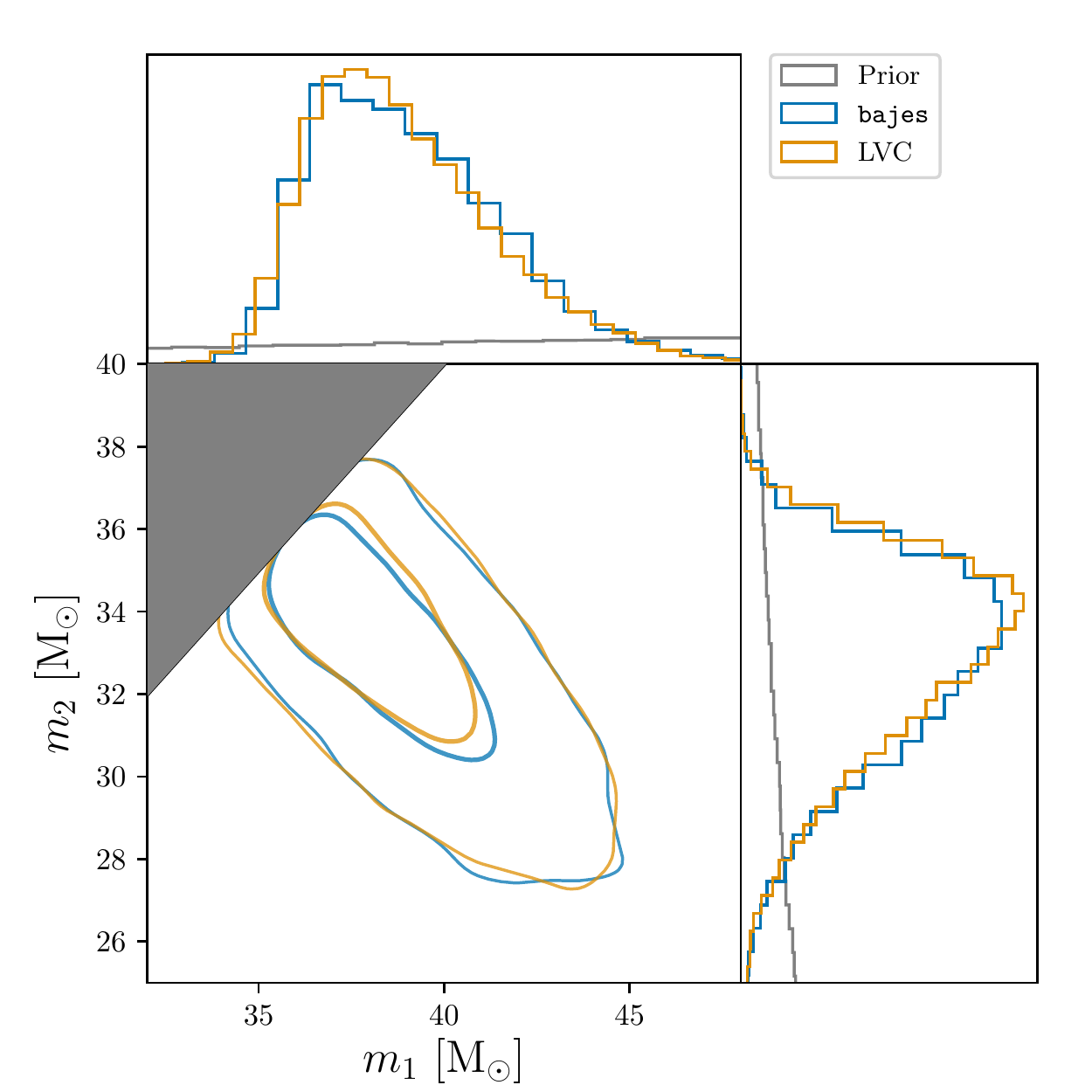}
	\caption{Posterior distributions
		of the detector-frame mass components $\{m_1,m_2\}$
		recovered in analysis of GW150914 with {\tt IMRPhenomPv2} (blue line).
		The results are compared with the official LIGO-Virgo posterior samples released with GWTC-1~\cite{LIGOScientific:2018mvr} (yellow line)
		computed using the {\tt LALInference} routines~\cite{Veitch:2014wba,lalsuite}.
		The central panel shows the 50\% and the 90\% credible regions.}
	\label{fig:150914_compare}
\end{figure}

We repeat the PE analysis with
{\tt IMRPhenomPv2} template employing precessing spin components.
In Figure~\ref{fig:150914_compare},
the marginalized posterior distribution of the recovered masses
 is compared with the official LIGO-Virgo
posterior samples release with GWTC-1~\cite{LIGOScientific:2018mvr}
performed the {\tt LALInference} routines~\cite{Veitch:2014wba,lalsuite}
using the same waveform approximant.
The two analyses are consistent with each other,
recovering a BBH signal with chirp mass $\M=31.00^{+1.52}_{-1.49}~\Mo$
and mass ratio well constrained around the equal mass case, $q=1.18^{+0.36}_{-0.17}$.
The inference of the effective spin parameter
is consistent with zero and the posterior distribution 
of the spin components does not show evidence of precession,
according to Ref.~\cite{Abbott:2016izl,TheLIGOScientific:2016wfe,TheLIGOScientific:2016htt}.
Also the extrinsic parameters show an overall good agreement with 
previous estimations performed with the same approximant,
locating the source at a
luminosity distance of $D_L=458^{+123}_{-169}$~Mpc
with a posterior sky-location area of ${\sim}640~{\rm deg}^2$ at the 90\% credible level.

The main difference between the {\tt TEOBResumS} posterior 
and {\tt IMRPhenomPv2} one is the uncertainty on the mass ratio parameter, 
for which {\tt TEOBResumS} admits a largest value of $1.28$ at the 90\% credible region,
while the {\tt IMRPhenomPv2} posterior reaches $1.94$ with the same confidence.
However, this disagreement is mainly due to the different spin assumptions employed for 
the two analyses~\cite{Abbott:2016izl,TheLIGOScientific:2016wfe}.
Moreover, the posterior distribution recovered with {\tt IMRPhenomPv2} 
shows slightly smaller $\M$ and larger $D_L$ compared with the {\tt TEOBResumS} inference,
as shown also in Ref.~\cite{Nagar:2018zoe}.

\begin{figure}[t]
	\centering 
	\includegraphics[width=0.49\textwidth]{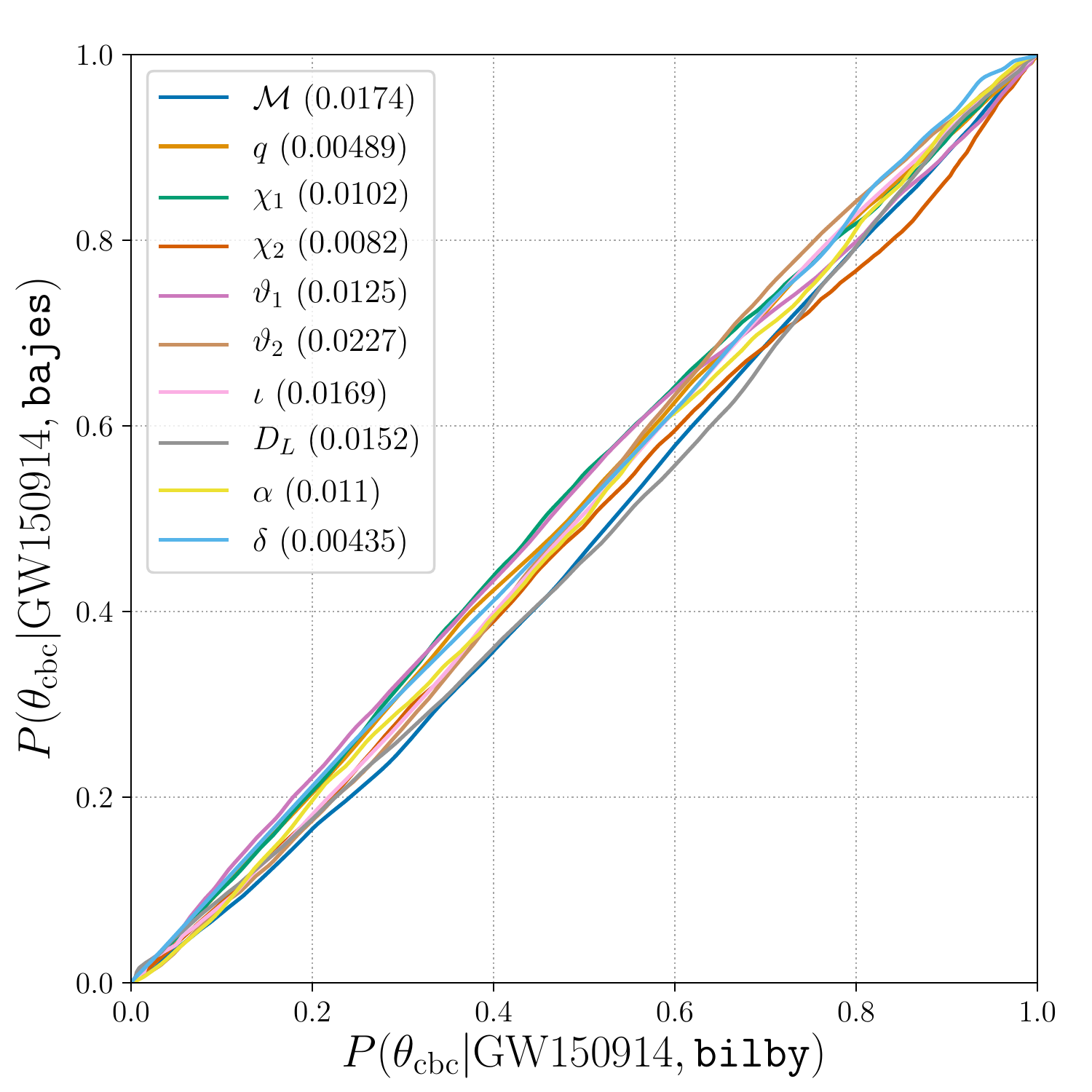}
	\caption{PP plot for the marginalized posterior distributions of GW150914 parameters. On the $x$-axis, the cumulative posterior probabilities estimated with the {\tt bilby} pipeline and on the $y$-axis the same quantities computed with the {\tt bajes} pipeline.
		Different colors refer to different parameters and the legend shows also the square root of the Jensen–Shannon divergence.}
	\label{fig:pp_gw150914}
\end{figure}

Finally, we verify the compatibility of
  the recovered posterior distribution against of existing GW pipelines.
In particular, we employ the {\tt bilby} pipeline~\cite{Ashton:2018jfp,Smith:2019ucc,Romero-Shaw:2020owr}
in order to estimate the posterior distributions 
of GW150914, using the same prior assumptions and settings
discussed above.
We observe that
GW150914 is a suitable candidate to test the statistical
significance of the results and the agreement between the pipelines:
due to the loudness of this event (corresponding to an SNR $> 20$), 
the overall impact of statistical noise fluctuations on the recovered posterior distribution is expected to be less determinant
compared with the other BBH mergers presented in GWTC-1~\cite{LIGOScientific:2018mvr}. 
Figure~\ref{fig:pp_gw150914} shows the probability-probability (PP) plot
of the marginalized posterior distributions recovered for every parameter.
A PP plot compares the cumulative distributions estimated with 
two methods, plotting one against the other. Then, if two probability
distributions are identical, the associated PP plot is represented 
by a bisector line. 
In our case, the results coming from the two pipelines 
are largely consistent between each other, 
with observed deviation fully consistent with statistical fluctuations.
This fact is confirmed by the $p$-values
computed between the marginalized distributions
of each parameter: the values are comparable to or 
larger than $0.4$ except for $\vartheta_2$ and $D_L$ parameters,
for which we respectively get $p$-values of 0.15 and 0.11.
Furthermore, the legend in Figure~\ref{fig:pp_gw150914}
shows the square root of the 
JS divergence 
of the marginalized posterior distributions for the employed parameters:
the largest recovered value corresponds to $5.2\times 10^{-4}$~bit
for the tilt angle $\vartheta_2$.

\subsection{GW170817} 
\label{sec:GW170817}

\begin{figure*}[t]
	\centering 
	\includegraphics[width=0.49\textwidth]{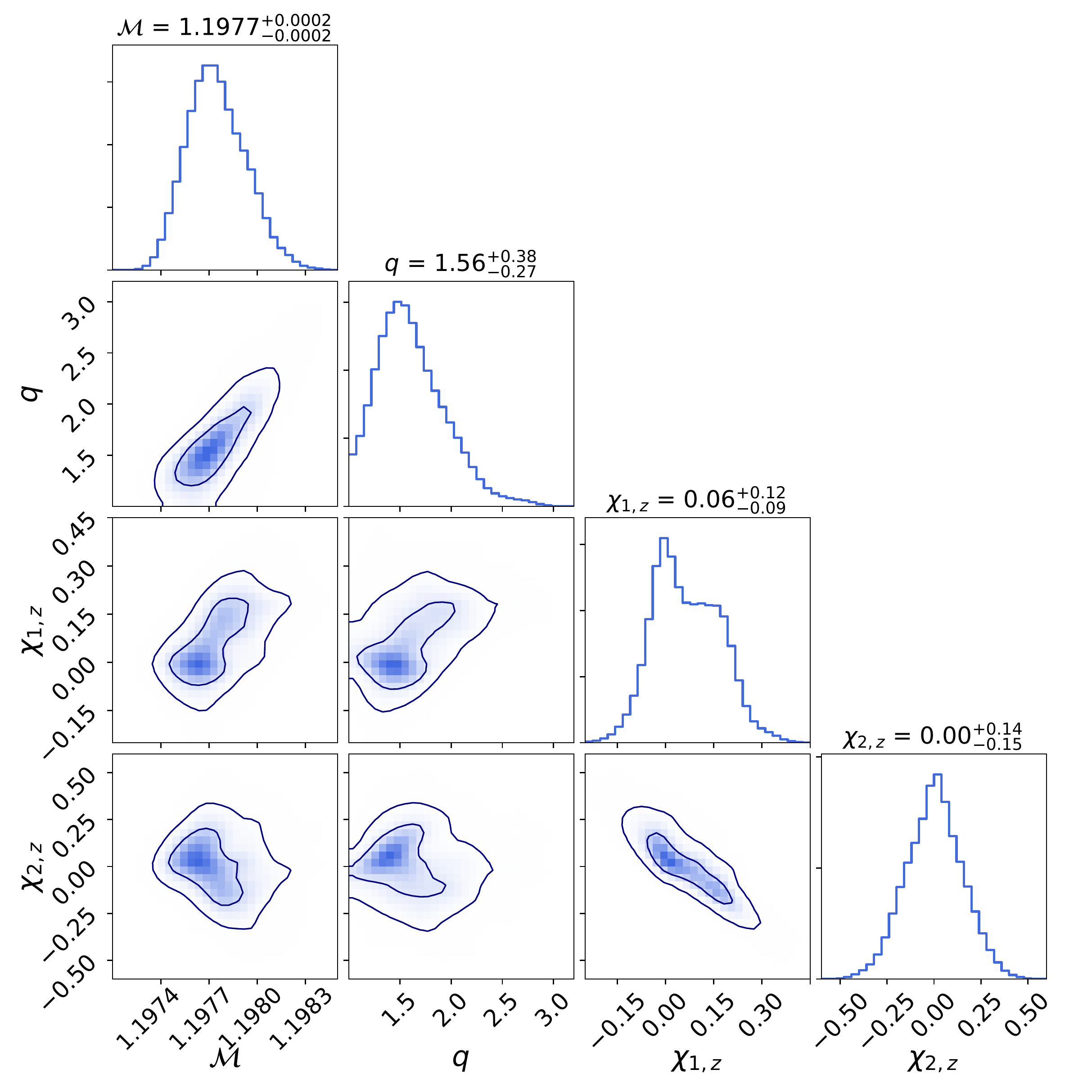}
	\includegraphics[width=0.49\textwidth]{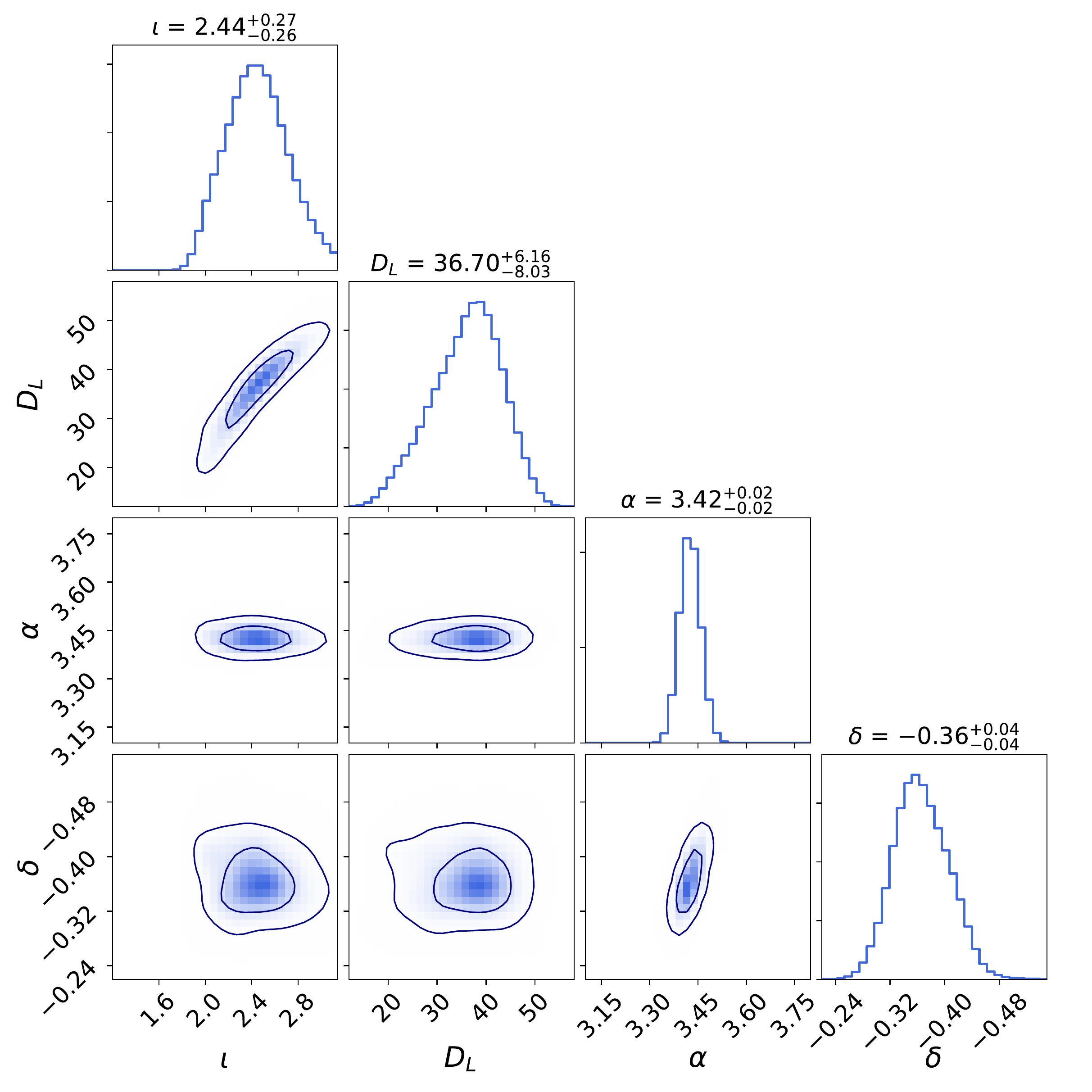}
	\caption{Posterior distributions
		for the intrinsic (left) and extrinsic (right) 
		parameters of GW170817 estimated with
		{\bajes} pipeline employing {\tt TEOBResumSPA} waveform model with aligned spin components.
		The chirp mass is expressed in Solar masses $\Mo$,
		the luminosity distance is expressed in megaparsec Mpc,
		while the angles $\{\iota, \alpha, \delta\}$ are in radiants.
		We report the median value and the 90\% credible levels for each parameter and the contours represent the 50\% and the 90\% credible regions.}
	\label{fig:170817_teob}
\end{figure*}

\begin{figure}[t]
	\centering 
	\includegraphics[width=0.49\textwidth]{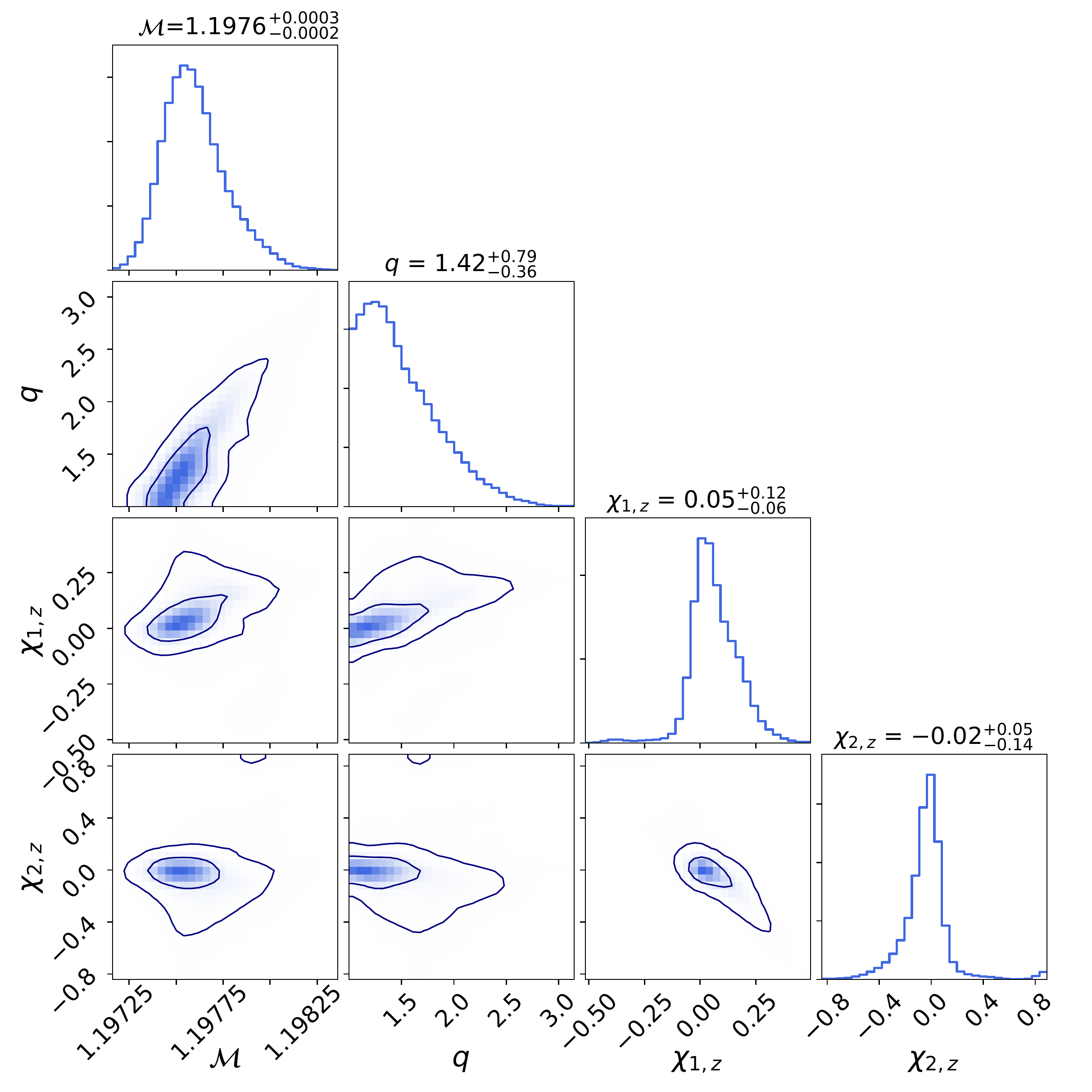}
	\caption{Posterior distributions
		for the intrinsic 
		parameters of GW170817 estimated with
		{\bajes} pipeline employing {\tt TaylorF2} (5.5PN + 7.5PN) 
		waveform model with aligned spin components.
		The chirp mass is expressed in Solar masses $\Mo$.
		We report the median value and the 90\% credible levels for each parameter and the contours represent the 50\% and the 90\% credible regions.}
	\label{fig:170817_tf2}
\end{figure}

\begin{figure}[t]
	\centering 
	\includegraphics[width=0.49\textwidth]{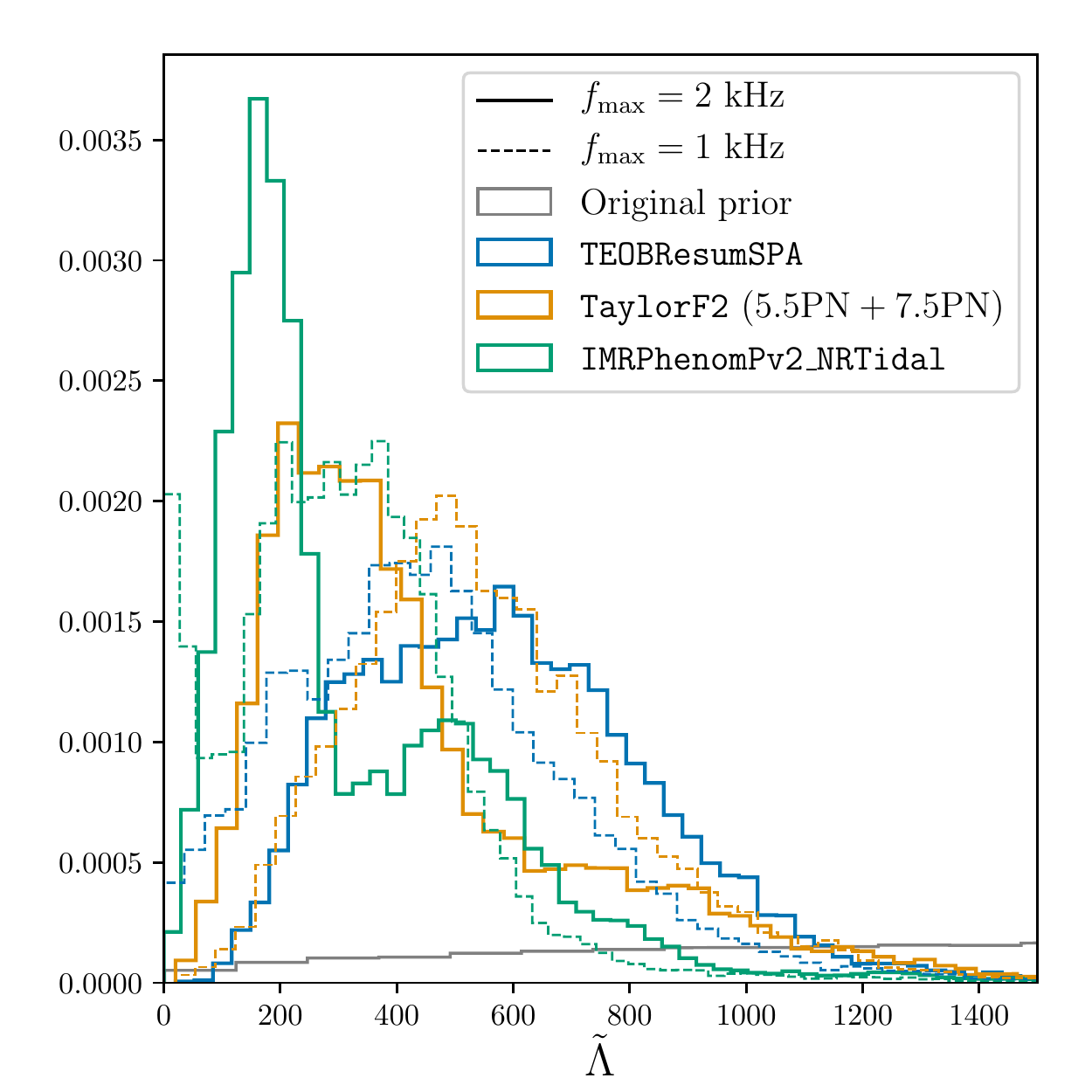}
	\caption{Posterior distributions 
		for the reduced tidal parameter $\tLam$ 
		of GW170817 
		estimated using the {\bajes} pipeline
		with different waveform approximants and different upper cutoff-frequencies.
		The posterior samples has been reweighted to uniform prior distribution and the 
		plot shows the original employed prior (gray line).
		Solid lines refer to the results with $f_{\rm max}=2~{\rm kHz}$, while 
		dashed lines are estimated with $f_{\rm max}=1~{\rm kHz}$.
	{\tt TEOBResumSPA} (blue line) and 
	{\tt TaylorF2} (yellow line) samples
	are computed employing aligned spins, 
	while precessing spin components were included
	for {\tt IMRPhenomPv2\_NRTidal} (green line).}
	\label{fig:lt_all}
\end{figure}

\begin{figure*}[t]
	\centering 
	\includegraphics[width=0.49\textwidth]{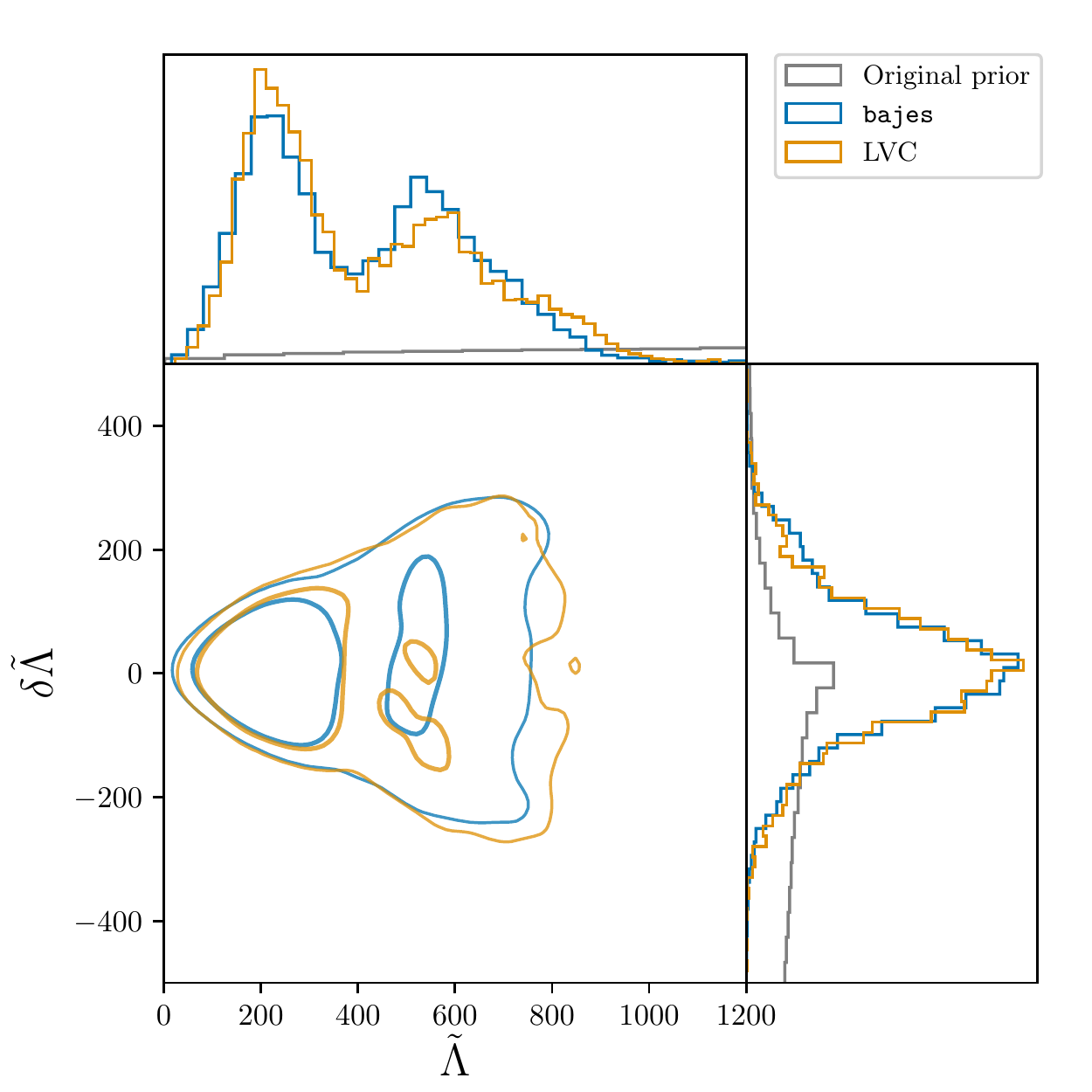}
	\includegraphics[width=0.49\textwidth]{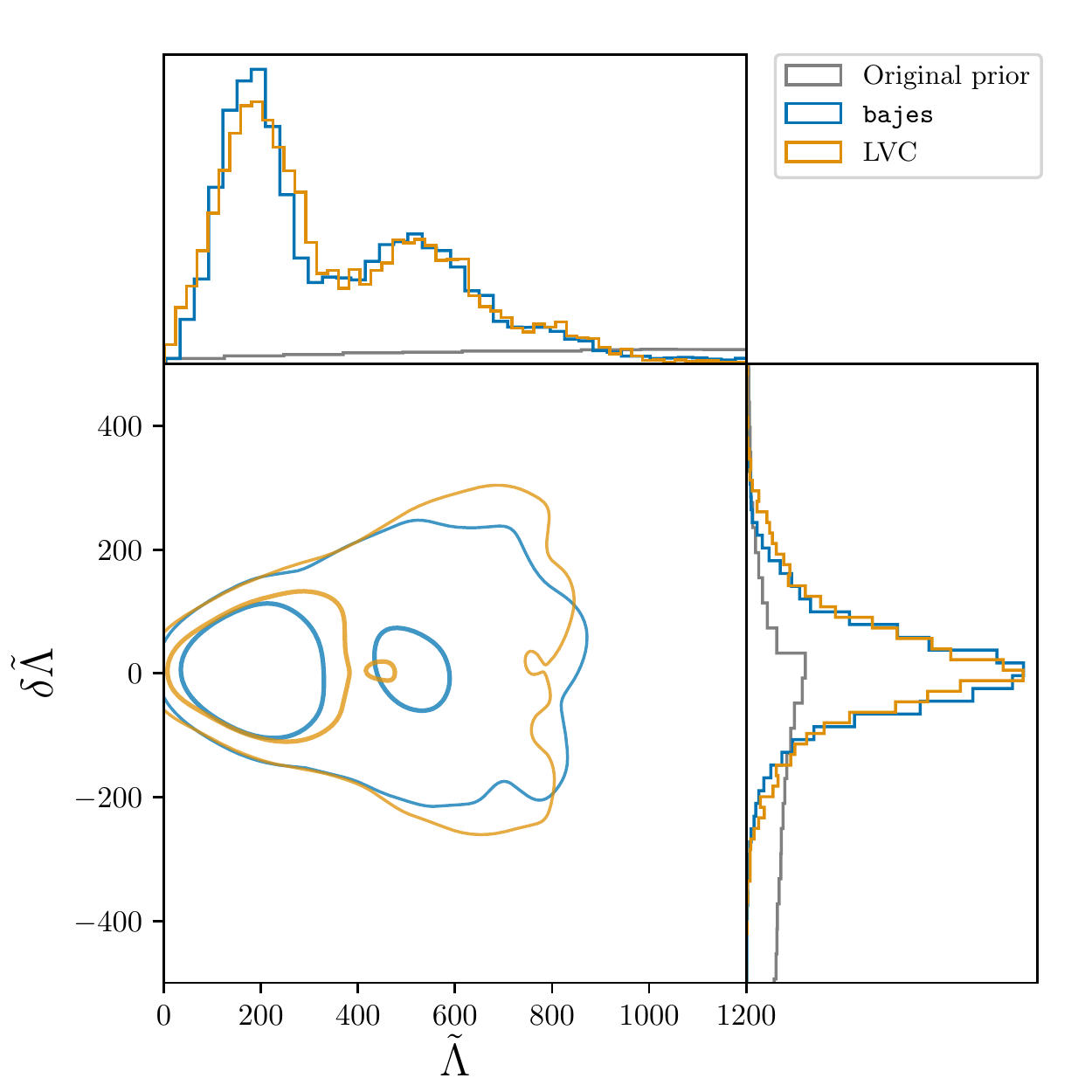}
	\caption{Comparison between the posterior distributions for the tidal parameters $\{\tLam, \dLam\}$ 
	estimated with the {\bajes} pipeline (blue lines) 
	with the official LIGO-Virgo released samples~\cite{Abbott:2018wiz}
	(yellow lines)
	computed with the {\tt LALInference} routines~\cite{Veitch:2014wba,lalsuite}.
	The analyses are performed with {\tt IMRPhenomPv2\_NRTidal}
	waveform approximant.
	The left panel shows the results with low-spin prior ($\chi_{\rm max} = 0.05$), 
	while the right panel presents 
	the high-spin prior results ($\chi_{\rm max} = 0.89$)s.
	For both panels, the central plots show the 50\% and 90\% credible regions.
	We estimate the JS divergences for the marginalized distributions,
	finding the values of $8.9\times10^{-5}$~bit and $5.3\times10^{-5}$~bit
	respectively for $\tLam$ and $\dLam$
	for the high-spin prior case.
	Regarding the low-spin prior studies, the JS divergences correspond to
	$8.3\times10^{-5}$~bit and $4.7\times10^{-5}$~bit
	respectively for $\tLam$ and $\dLam$.}
	\label{fig:lt_compare}
\end{figure*}

We analyze the LIGO-Virgo data corresponding to GW170817~\cite{TheLIGOScientific:2017qsa,Abbott:2018exr,Abbott:2018wiz,Abbott:2018hgk}, the 
first GW signal observed from a BNS merger.
The employed data correspond to the GWTC-1 release~\cite{LIGOScientific:2018mvr}
of LIGO-Virgo data centered around GPS time 1187008857
with a sampling rate of 4096~Hz and a duration of 128~s
analyzing the frequency range from 20~Hz to 2~kHz.
The employed prior is isotropic in spin components 
and volumetric in luminosity distance, and it
spans the ranges $\M\in[1.18,1.21]~\Mo$,
$q\in[1,8]$, $\chi_{1,2}\in [0,0.89]$, $\Lambda_{1,2}\in[0,5000]$ and $D_L\in [5,75]~{\rm Mpc}$.
We include 4 spectral nodes for the calibrations of the analyzed segments.

Figure~\ref{fig:170817_teob} shows the posterior distributions for the parameters
recovered employing
{\tt TEOBResumSPA} approximant (with $\ell=2$ and $|m|=\ell$)
with aligned spins. 
The recovered detector-frame chirp mass corresponds to $\M=1.1977^{+0.0002}_{-0.0002}~\Mo$ and the mass ratio 
lies around the equal mass case, $q=1.56^{+0.38}_{- 0.27}$.
The spin components do not show evidence of spin contributions,
consistently with Ref.~\cite{Abbott:2018wiz},
with an estimated effective spin of $\chi_{\rm eff} = {0.04}^{+0.06}_{-0.03}$.
The primary tidal parameter $\Lambda_1$ is constrained to be 
$\lesssim 950$ at the 90\% confidence level, while the secondary
component $\Lambda_2$ is more broadly distributed
over the prior.
The recovered tidal parameter posterior 
estimates a value of $\tLam={607}^{+477}_{-356}$,
in rough agreement with previous estimations 
obtained from EOB models~\cite{LIGOScientific:2018mvr,Gamba:2020ljo}.
The asymmetric tidal parameter $\dLam$
shows a posterior distribution centered around non-zero values, $\dLam= 92^{+200}_{-258}$; however, the hypothesis $\dLam=0$ is 
confidently included in the posterior support,
corresponding to the $27^{\rm th}$ percentile.
Moreover, the measured $\tLam$ is overall consistent with 
independent estimations coming from the analysis of
the EM counterpart AT2017gfo~\cite{Radice:2017lry,Breschi:2021tbm}.
Regarding the extrinsic parameters,
we recovered a luminosity distance of $D_L=36.7^{+6.2}_{-8.0}~{\rm Mpc}$
and a sky location at
$\{\alpha={3.42}^{+0.02}_{-0.02}~{\rm rad},
\delta={-0.36}^{+0.04}_{-0.04}~{\rm rad}\}$.
The estimation of the extrinsic parameters is generally consistent with previous 
estimations~\cite{Abbott:2018wiz,LIGOScientific:2018mvr,Gamba:2020ljo}.

Furthermore, the analysis of GW170817 is repeated with a
{\tt TaylorF2} waveform template that includes for the first time
5.5PN point-mass corrections~\cite{Messina:2019uby} and 7.5PN tidal
contributions~\cite{Vines:2011ud,Damour:2012yf}.
This analysis is performed 
using the same prior assumptions 
described above and the posterior distribution 
for the intrinsic parameters is shown in Figure~\ref{fig:170817_tf2}.
The mass parameters recover the values $\M={1.1976}^{+0.0003}_{-0.0002}~\Mo$
and $q=1.42^{+0.79}_{-0.36}$, while the effective spin 
$\chi_{\rm eff} = {0.02}^{+0.07}_{-0.04}$.
The estimated luminosity distance of 
$D_L={37.1}^{+10.3}_{-12.1}~{\rm Mpc}$. 
The inference of the tidal components roughly coincides with 
the estimations coming from the analogous analysis with PN templates~\cite{Abbott:2018wiz,LIGOScientific:2018mvr}.
The primary tidal component is constrained below
$\Lambda_1<730$ at 90\% credible region, while the 
secondary is more broadly distributed over the prior. 
The reduced tidal parameter is measured to be $\tLam = {404}^{+701}_{-246}$
and the asymmetric tidal term $\dLam$ is well constrained around zero 
with an uncertainty of ${\sim}150$ at the 90\% confidence level
The median values of the sky location angles 
coincide with the estimation performed with {\tt TEOBResumSPA}.

The GW170817 PE studies are repeated with 
{\tt IMRPhenomPv2\_NRTidal} template employing precessing spin components with high-spins prior ($\chi_{\rm max}=0.89$) and low-spins prior ($\chi_{\rm max}=0.05$),
in order to compare our results with the official LIGO-Virgo posterior
samples presented in Ref.~\cite{Abbott:2018wiz}.
The recovered posterior distributions for the mass parameters
for the low-spin case show a chirp mass of 
$1.1975^{+0.0002}_{-0.0002}~\Mo$ and the mass ratio is constrained 
below $1.46$ at the 90\% credible region.
Regarding the high-spin prior analysis, 
we recovered a chirp mass of 
$1.1976^{+0.0002}_{-0.0002}~\Mo$ 
and the mass ratio favors more asymmetric values, $q=1.49^{+0.35}_{-0.32}$.
Focusing on the tidal parameters,
Figure~\ref{fig:lt_compare} shows the comparison of the 
posterior distribution in the $\{\tLam,\dLam\}$ plane:
the marginalized distributions and the 90\% credible regions
coming from {\bajes} are largely
consistent with the official LIGO-Virgo samples,
with estimated JS divergences below $10^{-4}$.

Finally, Figure~\ref{fig:lt_all} shows the recovered 
reduced tidal parameters $\tLam$, where the posterior
distributions have been reweighted to uniform prior distribution.
The figure includes the results computed 
employing the same waveform models and using a smaller upper cutoff-frequency $f_{\rm max}=1~{\rm kHz}$.
The main differences between the analyses with different $f_{\rm max}$ 
lie in the results of the tidal sector.
Overall the recovered tidal parameters with $f_{\rm max}=2~{\rm kHz}$
appear more constrained with respect to the cases with $f_{\rm max}=1~{\rm kHz}$.
This behavior is expected considering that the tidal information
is gathered in high-frequency regimes~\cite{Damour:2012yf,Gamba:2020wgg}.
On the other hand,
the choice of $f_{\rm max}=2~{\rm kHz}$ enlarges 
multimodal and asymmetric behaviors in the posterior distribution of the 
reduced tidal parameter and systematic effects appear to be more relevant
between different template families,
as well known from previous studies~\cite{Dai:2018dca,Narikawa:2018yzt,Gamba:2020wgg}. 
The differences in the $\tLam$
parameters can be led back to the modeling choices
of the employed approximants~\citep[see][]{Abbott:2018wiz,Messina:2019uby,LIGOScientific:2018mvr}.
The results estimated with {\tt TEOBResumSPA} at $1~{\rm kHz}$
show a posterior distribution slightly shifted toward lower values with respect to 
the analysis at $2~{\rm kHz}$,
consistently with what has been observed in the BNS injection study (Sec.~\ref{sec:bnsinsp}).
The posterior distributions for EOB and PN approximants show a 
good agreement with $f_{\rm max}=1~{\rm kHz}$.

 \subsection{EOB catalog} 
\label{sec:gwtc1}

\begin{table*}[t]
	\centering    
	\caption{Prior and posterior information for the analyses of 
		the BBH events of GWTC-1 with {\tt TEOBResumS}. 
		The GPS time refers to the
		central value of the time axis.
		For all studies, we assume aligned spins with isotropic prior.
		The inferred values refer to the medians 
		of the marginalized posterior distributions
		and the uncertainties are 90\% credible regions,
		except for the log-Bayes' factors $\log\B^{\rm S}_{\rm N}$,
		for which we report the standard deviations.
	}
	\resizebox{0.99\textwidth}{!}{
		\begin{tabular}{ccc|cccc|ccccccc}        
			\hline
			\hline
			\multicolumn{3}{c|}{Data Information}&
			\multicolumn{4}{c|}{Prior bounds}&
			\multicolumn{7}{c}{Inferred values}\\
			\hline
			Event & 
			GPS time & 
			Duration&
			$\M$& 
			$q$&
			$\chi_{1,2}$&
			$D_L$&
			$\M$& 
			$q$&
			$\chi_{1,z}$&
			$\chi_{2,z}$&
			$\chi_{\rm eff}$&
			$D_L$&
			$\log\B^{\rm S}_{\rm N}$\\
			& $[{\rm s}]$ & $[{\rm s}]$ &$[\Mo]$&&&$[{\rm Mpc}]$ &$[\Mo]$& & & & &$[{\rm Mpc}]$&\\
			\hline
			GW150914 &1126259462& 8&$[12,45]$&$[1,8]$&$[0,0.99]$&$[100,5000]$&
			${31.9}^{+1.1}_{-1.5}$&
			${1.20}^{+0.29}_{-0.17}$&
			${0.07}^{+0.39}_{-0.28}$&
			${0.00}^{+0.40}_{-0.42}$&
			${0.05}^{+0.10}_{-0.13}$&
			${471}^{+167}_{-231}$&	
			$267.1^{+0.2}_{-0.2}$\\
			GW151012 &1128678900& 16&$[12,45]$&$[1,8]$&$[0,0.99]$&$[100,5000]$&
			${18.3}^{+1.8}_{-1.0}$&
			${1.86}^{+2.86}_{-0.76}$&
			${0.05}^{+0.33}_{-0.28}$&
			${0.11}^{+0.53}_{-0.48}$&
			${0.09}^{+0.22}_{-0.17}$&
			${1039}^{+627}_{-626}$&	
			$16.0^{+0.2}_{-0.2}$\\
			GW151226 &1135136350& 16&$[6.5,15]$&$[1,8]$&$[0,0.99]$&$[50,5000]$&
			${9.71}^{+0.07}_{-0.07}$&
			${2.04}^{+1.59}_{-0.93}$&
			${0.38}^{+0.23}_{-0.24}$&
			${0.16}^{+0.58}_{-0.47}$&
			${-0.05}^{+0.45}_{-0.61}$&
			${490}^{+222}_{-240}$&	
			$36.4^{+0.2}_{-0.2}$\\
			GW170104 &1167559936& 16&$[12,45]$&$[1,8]$&$[0,0.99]$&$[100,5000]$&
			${25.6}^{+1.8}_{-2.1}$&
			${1.56}^{+0.86}_{-0.46}$&
			${0.00}^{+0.29}_{-0.32}$&
			${-0.04}^{+0.40}_{-0.40}$&
			${-0.03}^{+0.21}_{-0.25}$&
			${1069}^{+423}_{-446}$&	
			$58.4^{+0.2}_{-0.2}$\\
			GW170608 &1180922494& 16&$[5,10]$&$[1,8]$&$[0,0.99]$&$[50,5000]$&
			${8.49}^{+{0.05}}_{-{0.04}}$&
			${1.48}^{+{1.22}}_{-{0.42}}$&
			${0.06}^{+{0.30}}_{-{0.28}}$&
			${0.03}^{+{0.58}}_{-{0.39}}$&
			${0.06}^{+{0.27}}_{-{0.09}}$&
			${298}^{+{146}}_{-{128}}$&
			$80.3^{+0.2}_{-0.2}$\\
			GW170729 &1187529256& 4&$[25,175]$&$[1,8]$&$[0,0.99]$&$[100,7000]$&
			${51.4}^{+9.1}_{-9.6}$&
			${1.84}^{+{0.95}}_{-{0.77}}$&
			${0.47}^{+{0.39}}_{-{0.48}}$&
			${-0.05}^{+{0.83}}_{-{0.36}}$&
			${0.30}^{+{0.28}}_{-{0.28}}$&
			${2495}^{+{1600}}_{-{1300}}$&
			$27.1^{+0.2}_{-0.2}$\\
			GW170809 &1185389807& 16&$[12,45]$&$[1,8]$&$[0,0.99]$&$[100,5000]$&
			${30.3}^{+{2.3}}_{-{2.0}}$&
			${1.45}^{+{0.72}}_{-{0.39}}$&
			${0.07}^{+{0.33}}_{-{0.27}}$&
			${0.17}^{+{0.58}}_{-{0.21}}$&
			${0.17}^{+{0.24}}_{-{0.21}}$&
			${999}^{+{473}}_{-{483}}$&
			$41.8^{+0.2}_{-0.2}$\\
			GW170814 &1186302519& 16&$[12,45]$&$[1,8]$&$[0,0.99]$&$[100,5000]$&
			${26.8}^{+{1.3}}_{-{1.0}}$&
			${1.29}^{+{0.52}}_{-{0.26}}$&
			${0.07}^{+{0.39}}_{-{0.28}}$&
			${0.02}^{+{0.49}}_{-{0.38}}$&
			${0.08}^{+{0.16}}_{-{0.12}}$&
			${540}^{+{224}}_{-{189}}$&
			$99.6^{+0.2}_{-0.2}$\\
			GW170818 &1186741861& 16&$[12,45]$&$[1,8]$&$[0,0.99]$&$[100,5000]$&
			${31.8}^{+{3.4}}_{-{2.9}}$&
			${1.48}^{+{0.96}}_{-{0.43}}$&
			${-0.08}^{+{0.27}}_{-{0.35}}$&
			${0.00}^{+{0.45}}_{-{036}}$&
			${-0.06}^{+{0.33}}_{-{0.27}}$&
			${1190}^{+{594}}_{-{438}}$&
			$29.7^{+0.2}_{-0.2}$\\
			GW170823 &1187058327& 4&$[25,175]$&$[1,8]$&$[0,0.99]$&$[100,7000]$&
			${37.4}^{+{ 5.5}}_{-{5.1}}$&
			${1.57}^{+{0.94}}_{-{0.51}}$&
			${-0.01}^{+{0.42}}_{-{0.33}}$&
			${0.06}^{+{0.56}}_{-{0.55}}$&
			${0.03}^{+{0.30}}_{-{0.28}}$&
			${1690}^{+{1081}}_{-{880}}$&
			$39.5^{+0.2}_{-0.2}$\\
			\hline
			\hline
		\end{tabular}
	}
	\label{tab:bbh_gwtc_prior}
\end{table*}

As full scale application,
we reproduce the analyses of the BBH mergers 
published in GWTC-1~\cite{LIGOScientific:2018mvr}
employing  the {\bajes} pipeline and
the time-domain EOB waveform model {\tt TEOBResumS}, including only
the dominant $(2,2)$ mode.
Table~\ref{tab:bbh_gwtc_prior} shows the priors used for each event, where
the nominal GPS time refers to the
central value of the analyzed time axis.
For all the studies, the analyzed frequency range goes from 20~Hz 
to 1~kHz and
we assume aligned spin components 
with isotropic prior distribution and volumetric prior for the luminosity 
distance. The prior distributions for the other parameters are chosen
accordingly with Sec.~\ref{sec:prior}.
For these studies, we employ 8 calibration envelope nodes 
for each detector, the phase $\phi_0$ is marginalized during the
likelihood evaluations and the time-shift parameter $t_0$ is sampled 
from a $2~{\rm s}$ window centered around the nominal GPS time.

\begin{figure*}[t]
	\centering 
	\includegraphics[width=0.49\textwidth]{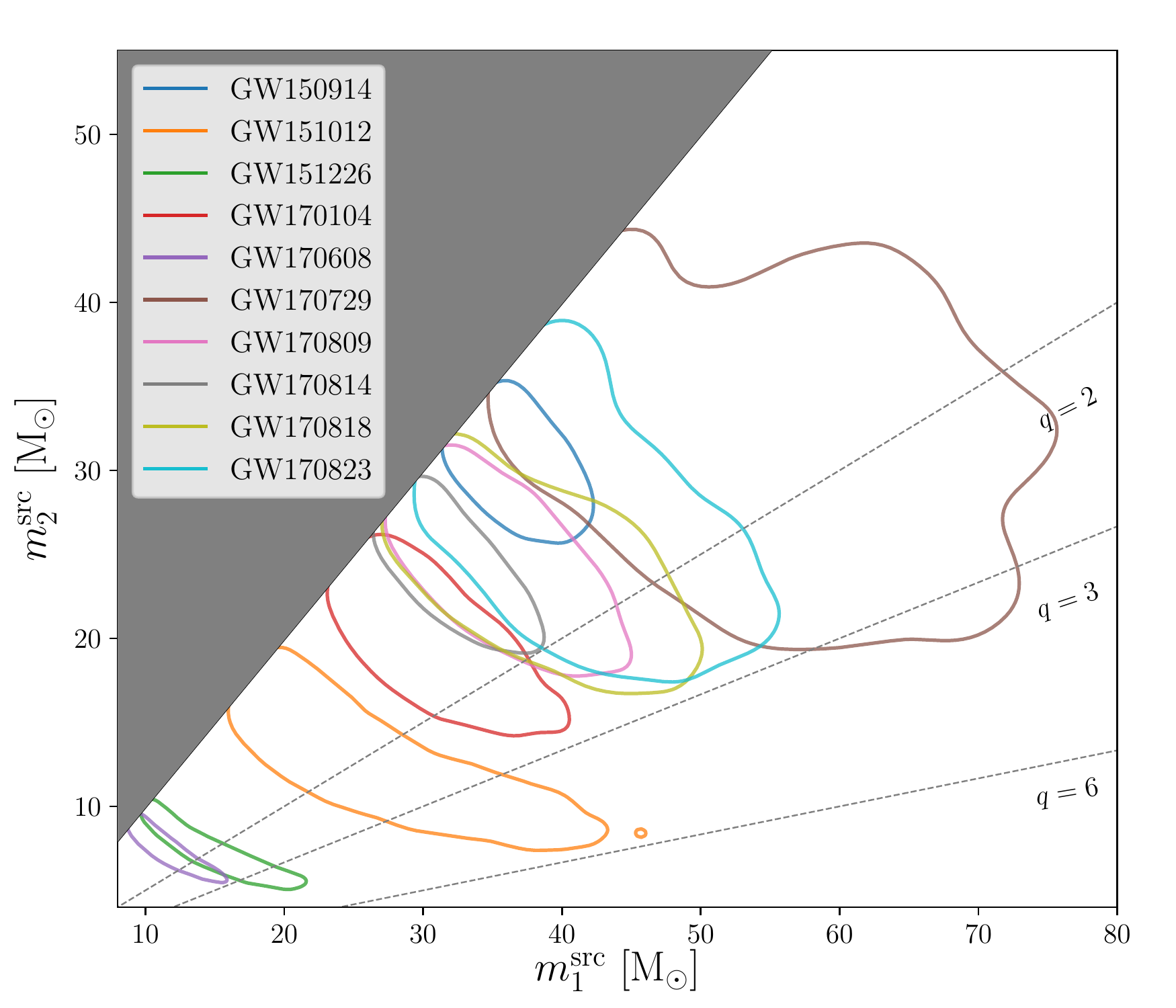}
		\includegraphics[width=0.49\textwidth]{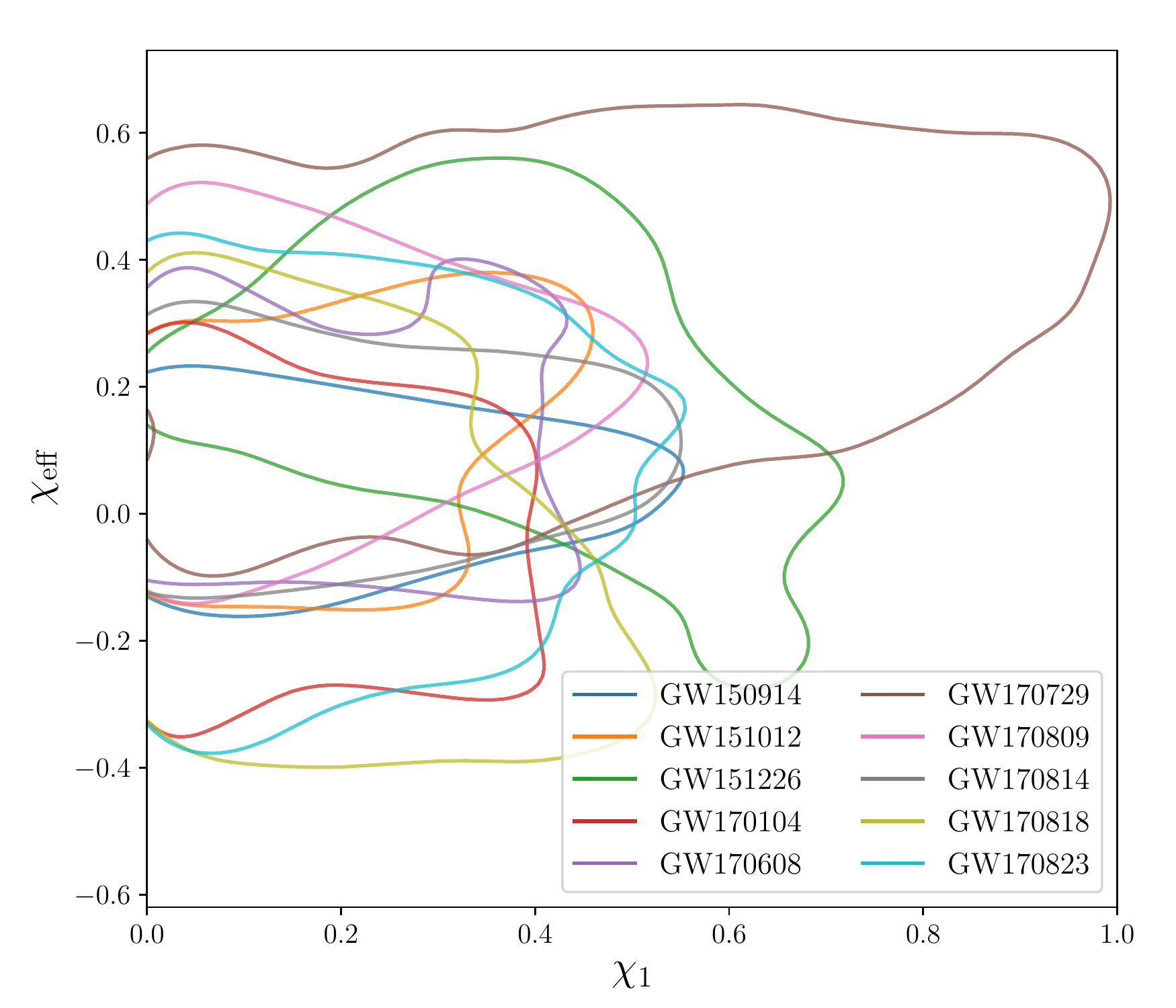}
	\caption{Marginalized posterior distribution for the source-frame mass components $\{m_1,m_2\}$ (left)
		and the spins $\{\chi_{1},\chi_{\rm eff}\}$ (right)
					of the BBH mergers presented in GWTC-1~\cite{LIGOScientific:2018mvr}.
										The PE studies have been
					performed with {\tt TEOBResumS} model.
					The masses are expressed in the binary source-frame 
					employing the cosmological model proposed in Ref.~\cite{Aghanim:2018eyx}.
					The contours refer to the 90\% credible regions.}
	\label{fig:m1m2_s1s2}
\end{figure*}

Figure~\ref{fig:m1m2_s1s2} shows the posterior distributions 
marginalized in the mass components and spin magnitudes planes
for all the analyzed BBH events.
The mass components are expressed in the source-frame 
of the binaries assuming the cosmological model 
presented in Ref.~\cite{Aghanim:2018eyx}.
The detector-frame mass components $m_i$ can be 
estimated in the source-frame of the binary as
\be
\label{eq:srcmass}
m_i^{\rm src} = \frac{m_i}{1+z}\,,\quad i=1,2\,,
\ee
where $z$ is the cosmological redshift of the source.
In general, the recovered mass parameters
show a predominance of equal mass binaries with
mass ratio well constrained below $q\lesssim 3$,
except for the low-mass binary GW151012,
that admits values of $q\simeq 5$ at 90\% credible level.
The recovered mass components are distributed between $7~\Mo$
and $70~\Mo$, with an abundance in the range $[20~\Mo,50~\Mo]$.
In terms of spin contributions, 
the most interesting events are GW151226,
whose posterior distribution excludes the non-spinning case at 90\%
confidence level, consistently with Ref.~\cite{LIGOScientific:2018mvr,Schmidt:2020yuu},
and GW170729, which recovers an effective spin of $\chi_{\rm eff}\approx{0.3}$ and admits 
values up to $\chi_1\gtrsim 0.9$.                                    
The other GW transients show mitigated spin contributions,
with $\chi_1 \lesssim 0.5$ at the 90\% credible level.
Generally, the posterior distributions for the secondary spin component 
$\chi_2$ are more broadly distributed and less informative 
than those for the primary component $\chi_1$.

\begin{figure}[t]
	\centering 
	\includegraphics[width=0.49\textwidth]{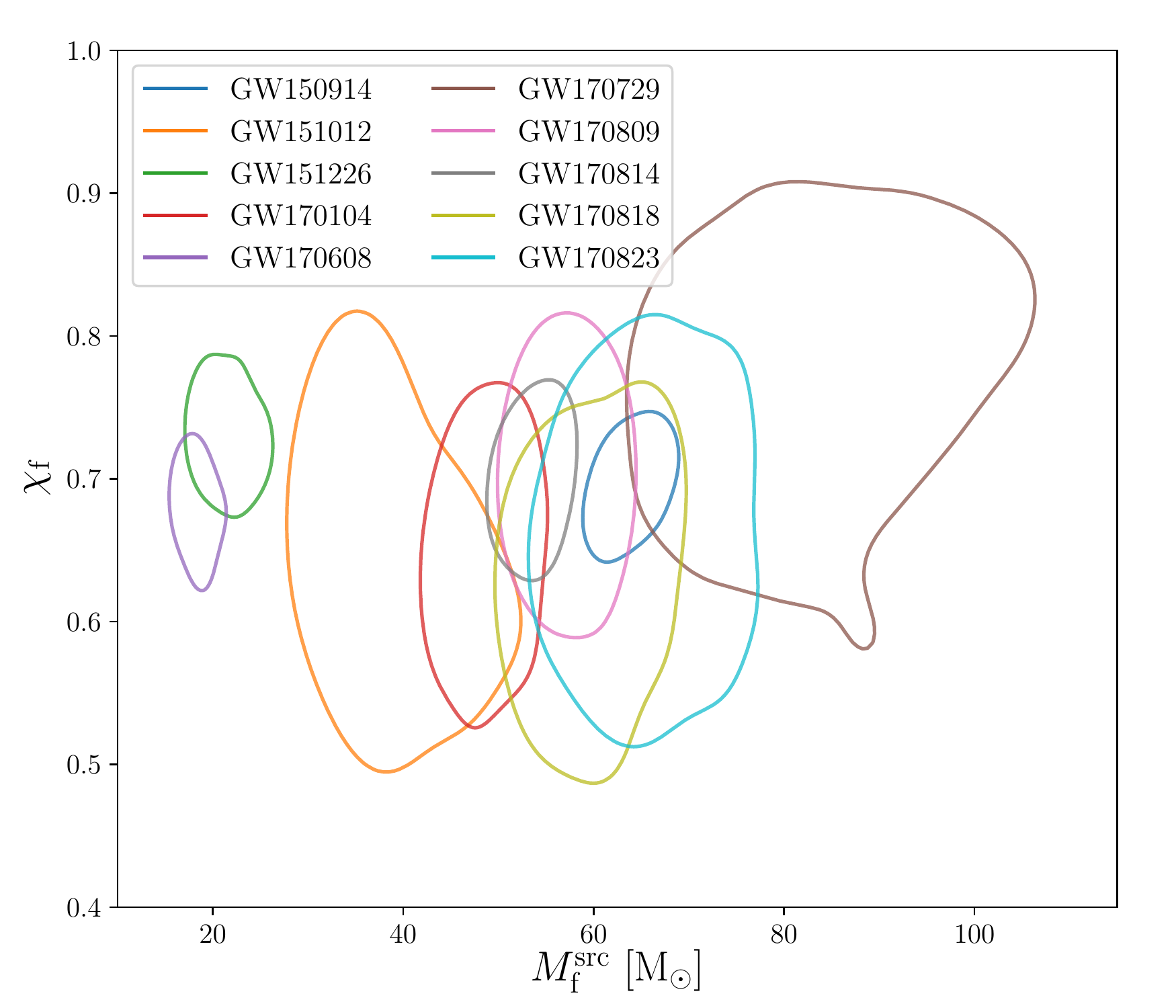}
	\caption{Marginalized posterior distribution for the final
		masses and spins $\{M_{\rm f}^{\rm src},\chi_{\rm f}\}$ 
		for the remnants of the BBH mergers presented in GWTC-1~\cite{LIGOScientific:2018mvr}.
		The analyses are performed using the {\bajes} pipeline
		and {\tt TEOBResumS} waveform approximant.
		The values have been computed from the posterior samples 
		employing NR formulae presented in Ref.~\cite{Jimenez-Forteza:2016oae}.
		The final masses are expressed in the source-frame of the binary.
		The contours refer to the 90\% credible regions.}
	\label{fig:mfsf}
\end{figure}

\begin{figure*}[t]
	\centering 
	\includegraphics[width=0.99\textwidth]{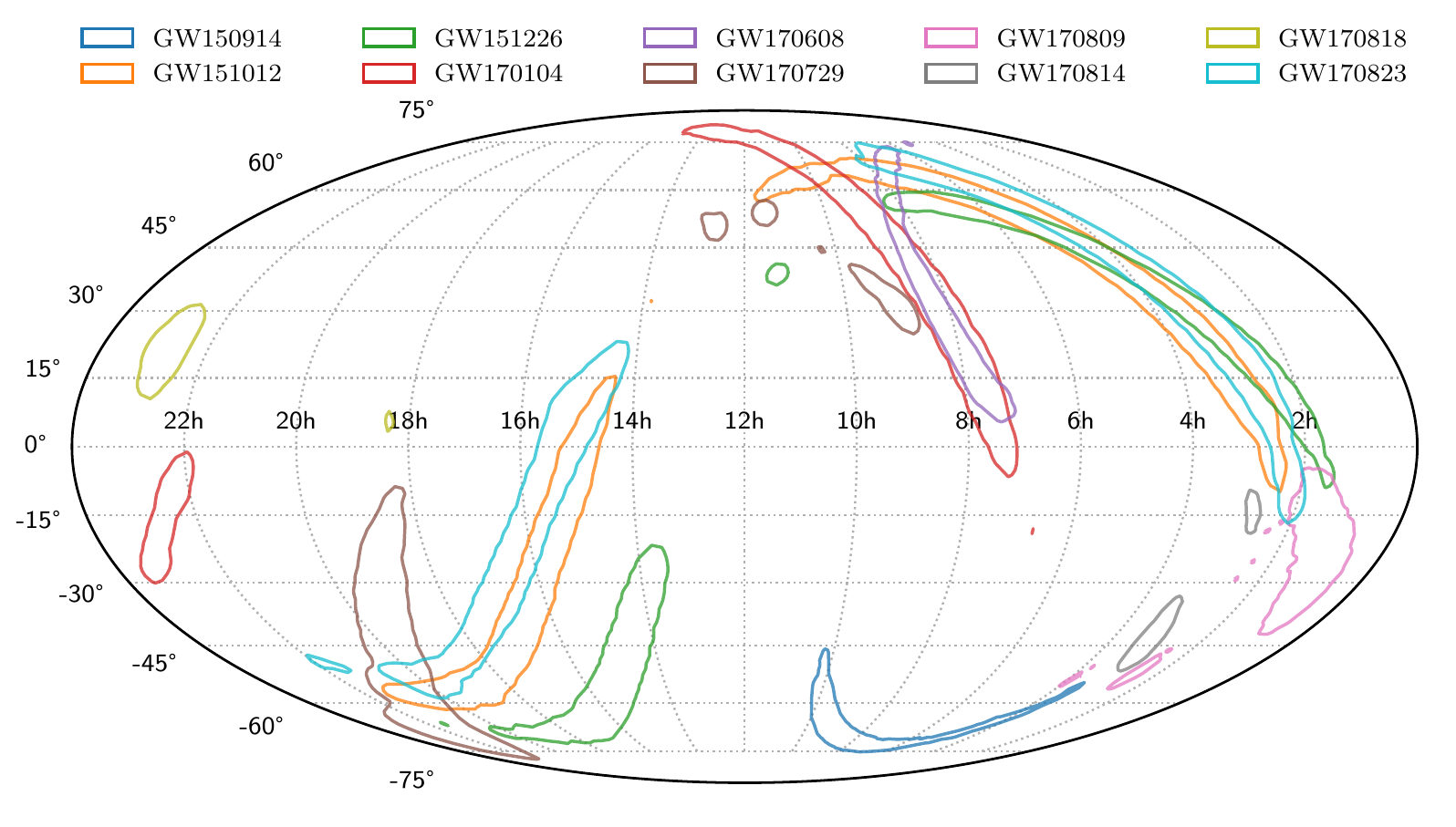}
	\caption{Marginalized posterior distribution for
		right ascension and declination angles $\{\alpha,\delta\}$
		(represented using a Mollweide projection)
		for the BBH mergers presented in GWTC-1~\cite{LIGOScientific:2018mvr}. 
		The analyses are performed using the {\bajes} pipeline
		and {\tt TEOBResumS} waveform approximant.
		The right ascension $\alpha$ is expressed in hours,
		while the declination $\delta$ is reported in degrees.
		The contours refer to the 90\% credible regions.}
	\label{fig:skyloc_catalog}
\end{figure*}

\begin{figure*}[t]
	\centering 
	\includegraphics[width=0.49\textwidth]{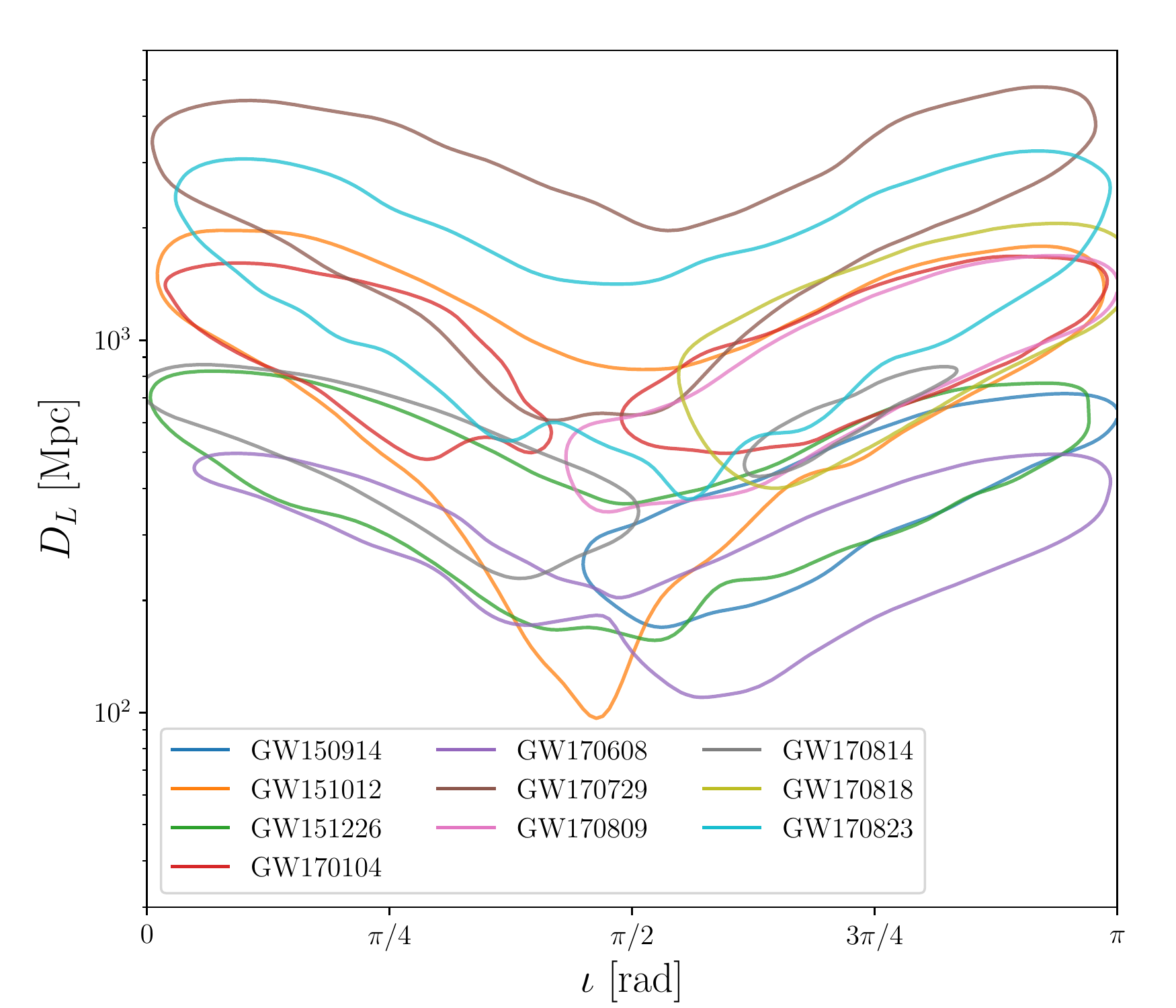}
	\includegraphics[width=0.49\textwidth]{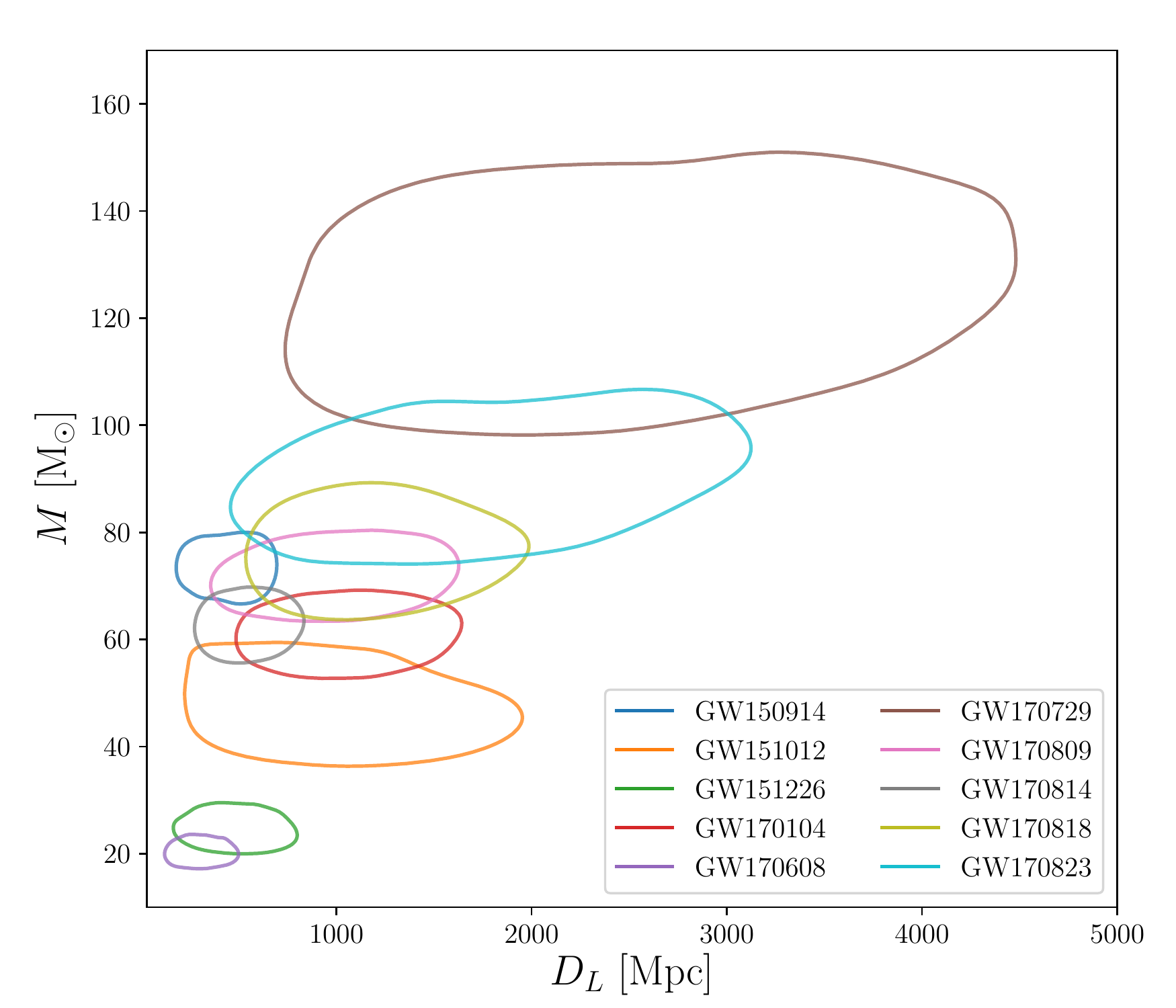}
	\caption{Marginalized posterior distribution for
		inclination angle and luminosity distance $\{\iota,D_L\}$ (left)
		and for luminosity distance and total mass 
		$\{D_L,M\}$ (right)
		for the BBH mergers presented in GWTC-1~\cite{LIGOScientific:2018mvr}. 
		The analyses are performed using the {\bajes} pipeline
		and {\tt TEOBResumS} waveform approximant.
		The contours refer to the 90\% credible regions.}
	\label{fig:dist_catalog}
\end{figure*}

Furthermore, 
Fig.~\ref{fig:mfsf} shows the posterior distributions of final masses 
$M_{\rm f}^{\rm src}$ (estimated in the source-frame)
and final spin magnitudes $\chi_{\rm f}$ 
expected for the merger remnants. 
The values are
computed resorting to NR formulae presented in Ref.~\cite{Jimenez-Forteza:2016oae},
calibrated with aligned-spin BBH simulations. 
The majority of the recovered final spins $\chi_{\rm f}$ lie 
around ${\sim}0.67$ due to the moderated spin contributions 
of the observed mergers.
Regarding the extrinsic parameters, 
Fig.~\ref{fig:skyloc_catalog} shows 
the posterior distributions of the sky location.
The sky maps for GW170814 and GW170818
show slightly larger bimodal behavior compared with the
results presented in GWTC-1~\cite{LIGOScientific:2018mvr}.
On the other hand, the 90\% credible region for GW170104 appears to be 
more constrained.
Note that the aligned-spin assumption affects
the overall SNR and, then, the recovered posterior distributions~\cite{Canton:2014ena,Canton:2014uja}.
The measurements for the sky locations of the other events
do not show considerable deviations from the 
GWTC-1 estimations~\cite{LIGOScientific:2018mvr}.
Fig.~\ref{fig:dist_catalog} shows the correlations of the
the luminosity distance $D_L$ with the inclination angle $\iota$
and with the total mass $M=m_1+m_2$ (estimated in the detector frame).
The luminosity distances are in agreement with the GWTC-1 estimations,
while the inclination angles show slightly  
wider posterior supports due to 
the degeneracy introduced with the aligned-spins assumption~\cite{PhysRevD.49.6274}.

For the GW150914 case, 
we can compare the analysis with higher-order modes (presented in Sec.~\ref{sec:GW150914}) with the results estimated using only dominant mode.
First of all, the results with $\ell \ge 5$ show narrower uncertainties,
consistently with the inclusion of a larger amount of physical information~\cite{Bustillo:2016gid,Breschi:2019wki,Mills:2020thr}.
On the other hand, the estimated Bayes' factors do not show
strong evidence in favor or against the inclusion of higher-order modes,
as expected for this kind of source;
higher-order modes are expected to be more
relevant for large mass ratios and edge-off
 binaries~\cite{Varma:2014jxa,Bustillo:2015ova}.
Overall, the median values of the recovered
parameters are consistent between each other
except for the mass ratio, that appears to be more constrained around $q=1$ 
including higher-modes.

In conclusion,
we shown that {\tt TEOBResumS} can be effectively applied to 
BBH signals, obtaining robust and consistent results~\cite{Nagar:2018zoe}.
The main limitation of the presented results comes from the aligned-spins
assumption, that introduces degeneracy with other parameters and 
induces biases in population inferences~\cite{Stevenson:2017dlk,Farr:2017uvj,Tiwari:2018qch}.
We are planning to extend the presented catalog
including precessing spin terms~\cite{Akcay:2020qrj} 
and verifying the importance of eccentric contributions~\cite{Chiaramello:2020ehz}.
In terms of computational cost, 
{\tt TEOBResumS} shows an overall good behavior,
performing the analysis of a typical BBH (with length of $8~{\rm s}$) 
in ${\sim} 14~{\rm hours}$ on 32 CPUs.

\section{Conclusion} 
\label{sec:conclusion}

In this paper we presented {\bajes},
a parallel, lightweight infrastructure 
for Bayesian inference, whose
main application is the data analysis of gravitational-wave and 
multimessenger transients.
{\bajes} is implemented in {\py} and
 comes with a versatile framework for Bayesian inference and different
 state-of-art 
 samplers.
Furthermore, it provides methods for the analysis of GW and
EM transients emitted by compact binary coalescences.
We benchmarked {\bajes} by means of
  injection-recovery experiments with BBH merger, BNS inspiral-merger
  and postmerger signals.
The injection studies and statistical tests show that the {\bajes} pipeline 
is well calibrated and it provides robust results,
within the expected statistical fluctuations.

The injections of BNS postmerger signals also offered the
  first detectability study with a 
five-detectors-network including LIGO, Virgo, KAGRA 
and third-generation ET~\cite{Punturo:2010zz,Hild:2010id}.
We find BNS postmerger signal
  will be detectable for optimally oriented  sources located at
${\lesssim}80~{\rm Mpc}$.
This result is largely merit of the ET  sensitivity 
  \cite{Punturo:2010zz,Hild:2010id}, that contribute to 90\% of the SNR.
According with recent
population studies~\cite{Abbott:2020gyp} and
using the distance threshold estimated from our survey
with third-generation network (${\sim}$80~Mpc), the detection rate of these
sources is expected to be $0.5{-}2$ events per year.  
As discussed in Sec.~\ref{sec:bnspm},
the detection of such a transient, combined with the 
knowledge of EOS-insensitive relations, can reveal essential properties 
of the nuclear matter at high densities, improving significantly
the EOS constraints~\cite{Breschi:2019srl}.

We demonstrated the reliability of {\bajes} in analyzing the observational data recorded by the LIGO-Virgo interferometers~\cite{LIGOScientific:2018mvr,Abbott:2019ebz}.
The posterior distributions for the parameters
 of GW transients computed with the {\bajes} are in agreement with
 the results from other GW pipelines~\cite{Veitch:2014wba,Lange:2018pyp}.
 The direct comparison of the {\bajes} results on GW150914 with 
 the ones obtained with the {\tt bilby}~\cite{Ashton:2018jfp,Smith:2019ucc,Romero-Shaw:2020owr} pipeline shows a maximum JS divergence 
 of $5.2{\times}10^{-4}$~bit for the tilt angle $\vartheta_2$ and the marginalized
 posterior distributions are largely consistent between each other.
 Furthermore, the analyses of GW150914 with {\tt TEOBResumS} approximant
slightly emphasize the relevance 
 of higher-order modes for improving the accuracy of the 
 binary properties estimations.
 
We performed PE studies on GW170817 using {\tt TEOBResumSPA}, {\tt IMRPhenomPv2\_NRTidal} 
	and, for the fitrst time, {\tt TaylorF2} including 5.5PN point-mass contributions
  and 7.5PN tidal terms.
  The novel analysis with the extended PN model shows a good agreement with
  previous estimations performed with the same template family~\cite{LIGOScientific:2018mvr,Abbott:2018wiz}.
  Using {\tt IMRPhenomPv2\_NRTidal} template,
  we found full consistency with previous results and 
  the official LIGO-Virgo posterior samples~\cite{LIGOScientific:2018mvr,Abbott:2018wiz}.
 Posterior distributions for the reduced tidal parameter $\tLam$
 recovered with upper cutoff-frequency $f_{\rm max}=2~{\rm kHz}$ 
 show larger systematic biases between different waveform templates
 compared with the $1~{\rm kHz}$ analyses.
 However, using a larger $f_{\rm max}$, it is possible to take into account 
 a larger amount of tidal information that leads to more constrained measurements. 
 The tidal parameter $\tLam$ estimated with {\tt TEOBResumSPA} 
 has slightly larger values compared with the measure of {\tt TaylorF2}
 and {\tt IMRPhenomPv2\_NRTidal} with $f_{\rm max}=2~{\rm kHz}$.
 The results of EOB and PN models are in 
 overall good agreement if $f_{\rm max}=1~{\rm kHz}$ is employed and also with independent analysis of AT2017gfo~\cite{Breschi:2021tbm}.
  We note that parallelization methods are key for the PE of BNS
  signals associated to long data segments ${\gtrsim} 100\,$s.
  The {\bajes} runs discussed in Sec.~\ref{sec:GW170817} were
  efficiently performed on 128 CPUs with total
  execution-time of ${\sim}$1~day. 

Future work will present the validation and the application of
{\bajes} to multimessenger analyses, including EM counterparts like
kilonovae and $\gamma$-ray burst~\citep[e.g.][]{Hayes:2019hso,Breschi:2021tbm}.
Moreover, we are implementing 
reduced-order-quadrature~\cite{Canizares:2014fya,Smith:2016qas,Morisaki:2020oqk}
and the relative-binning~\cite{Zackay:2018qdy,Dai:2018dca} in order 
to speedup the likelihood evaluations in the GW studies.
Inferences on the properties of neutron star matter will be supported
with the inclusion of a parametrized EOS sampling
method~\cite{Read:2008iy,Raithel:2016bux}.
Moreover, future {\bajes} releases will include an extended set of 
nested samplers, in particular 
algorithms based on machine learning~\citep[e.g.][]{albert2020jaxns} 
and efficiently parallelizable routines~\citep[e.g.][]{buchner2021ultranest}.\\

{\bajes} is publicly available at
\begin{center}
\href{https://github.com/matteobreschi/bajes}{\tt https://github.com/matteobreschi/bajes}
\end{center}
and contributions from the community are welcome.
The posterior samples presented in Sec.~\ref{sec:gwtc1}
and the configuration files to reproduce
the runs are available on {\tt Zenodo} \cite{zenodo_cite_doi}.

\begin{acknowledgments}
   The authors would like to thank Walter Del Pozzo for useful discussions.
  M.B. and S.B. acknowledge support by the EU H2020 under ERC Starting
  Grant, no.~BinGraSp-714626.  
  R.G. and M.B. acknowledge support from the Deutsche Forschungsgemeinschaft
  (DFG) under Grant No. 406116891 within the Research Training Group
  RTG 2522/1.
  The computational experiments were performed on the ARA cluster at
  the Friedrich Schiller University Jena supported in part by DFG grants INST
  275/334-1 FUGG and INST 275/363-1 FUGG and and ERC Starting Grant,
  no.~BinGraSp-714626.
  This research has made use of data obtained from the Gravitational
  Wave Open Science Center~\cite{gwosc}, a
  service of LIGO Laboratory, the LIGO Scientific Collaboration and
  the Virgo Collaboration. LIGO is funded by the U.S. National Science
  Foundation. Virgo is funded by the French Centre National de
  Recherche Scientifique (CNRS), the Italian Istituto Nazionale della
  Fisica Nucleare (INFN) and the Dutch Nikhef, with contributions by
  Polish and Hungarian institutes. 
\end{acknowledgments}

\appendix

\section{MCMC and PTMCMC} 
\label{app:mcmc}

A generic Markov-chain Monte Carlo (MCMC) 
algorithm explores the parameter space moving forward region
with increasing value of the probability
and returns a set of independent samples  
representative of the target probability density.
The MCMC samplers implemented in {\bajes} 
is based on {\tt emcee}~\cite{Foreman_Mackey_2013}:
this routine represents a simple and complete
implementation of a Metropolis-Hastings sampling 
that takes advantage of parallel chains.

The MCMC algorithm can be summarized as follows.
An arbitrary number of chains, say $n_{\rm chain}$,
 are initialized with as many random prior sample.
For each chain, the last stored sample is evolved and 
a new sample $\params^*$ is proposed according to 
predefined proposal methods (see App.~\ref{app:proposal}).
The new sample $\params^*$
is accepted with probability 
\be
\label{eq:mcmc}
{\rm min}\left[\,1,\,\frac{p(\params^*|\data,H)}{p(\params_i|\data,H)}\,\frac{q(\params_i|\params^*)}{q(\params^*|\params_i)}\right]\,,
\ee
where $\params_i$ is the last sample of the chain
and $q(\params_i|\params_j)$ is the proposal density function 
computed between $\params_i$ and 
$\params_j$. 
This procedure is iterated for every chain of the ensemble
and samples are collected during the exploration.
Note that, according with prescription Eq.~\eqref{eq:mcmc},
the probability of the proposed sample is not required to
 be strictly greater than that of the current sample.
The initial exploration is called {\it burn-in}, in which
the chains randomly explores the surrounding prior volume.
The algorithm spends these iterations in order to localize the maximum-probability region.
After a certain amount of iterations, depending on the complexity of the parameter space, 
the chains can converge and the actual collection 
of posterior samples starts.
Subsequently,
when the algorithm reaches the stopping condition,
the burn-in portion is removed, 
the samples from different chains are joined together, 
the autocorrelation length (ACL) is computed in order to estimate the effective 
number of independent posterior samples,
and the final set of samples is extracted from the joined set according with the value of ACL.
The stopping criterion implemented in {\bajes} for the MCMC algorithms 
is defined by the requested number of output samples $n_{\rm out}$,
\be
\label{eq:mcmcstop}
\frac{\left( i- n_{\rm burn}\right)\, n_{\rm chain}}{{\rm ACL}} \ge n_{\rm out}\,,
\ee 
where $i$ is the current iteration, 
$n_{\rm chain}$ is the total number of chains, 
$n_{\rm burn}$ is the number of iterations required for burn-in
and ACL is computed on the set of post-burn-in samples
\footnote{We are planning to modify the current MCMC stopping 
condition implementing the Gelman-Rubin diagnostic test~\cite{gelman1992,Brooks:1998}.}.

The MCMC algorithm disposes of a light-weighted settings
and it is a fast and versatile algorithm.
However, when the parameter space becomes large or the distributions have multimodalities,
this method could have many issues; 
such as insufficient and inaccurate exploration of the parameter space,
some chains could get trapped in a local minima or the 
ensemble might not be able to reach convergence.
These issues can be mitigated resorting to a large number of parallel chains
or to specific proposal methods (see App.~\ref{app:proposal}). 
Moreover, given a set of posterior samples, it is possible to estimate the evidence 
using the approximation
\be
\label{eq:mcmcapprox}
p(\data|H) \approx \frac{1}{n_{\rm samples}}\sum_{i=1}^{n_{\rm samples}} p(\params_i|\data,H)\,,
\ee
where the index $i=1,\dots,n_{\rm samples}$ runs over the posterior samples.
However, in general, Eq.~\eqref{eq:mcmcapprox} is unable to perform accurate estimations of the evidence,
since the MCMC algorithm is not designed to minutely inspect all the parameter space.

On the other hand, the parallel tempering MCMC (PTMCMC)~\cite{B509983H,PhysRevLett.57.2607,Neal1996}
performs improved exploration of the parameter space
and it provides a more accurate 
estimations of the evidence integral compared to
standard MCMC techniques.
The PTMCMC sampler implemented in {\bajes} 
is inspired by {\tt ptemcee}~\cite{Vousden_2015}.
The PTMCMC
introduces an inverse temperature coefficient $\beta=1/T\in [0,1]$ 
in the computation of posterior distribution, such that 
\be
\label{eq:temp-post}
p_\beta(\params|\data,H) \propto \big[ p(\data|\params,H) \big]^\beta p(\params|H)\,.
\ee
The set of all chains is grouped in equally-populated sub-ensembles
and a different value of $\beta$ is associated to each tempered ensembles.
The default $\beta$ ladder is geometrically-spaced in the range $[0,1]$. 
The algorithm proceeds as the usual MCMC for every chain
using the tempered posterior distribution Eq.~\eqref{eq:temp-post}.
For $T=1$, the tempered posterior is identical to the original one
and low-temperature chains 
will move toward regions with large likelihood values
focusing on the estimation of the volume of the bulge.
However, the contribution of the likelihood function 
is mitigated by increasing values of $T$, 
up to the limit $T\to\infty$, where the posterior is identically equal to the prior.
Then, high-temperature chains will be able 
to freely explore the majority
the prior support,
inspecting the tails of the targeted posterior distribution 
and providing a good coverage of the entire prior volume.
Furthermore, the algorithm proposes swaps  
between consecutive pairs of chains, received with acceptance
\be
\label{eq:ptmcmc}
{\rm min}\left[\,1,\,\left(\frac{p(\data|\params_i,H)}{p(\data|\params_j,H)}\right)^{\beta_j-\beta_i}\right]\,,
\ee
where $\params_i$ and $\beta_i$ are respectively the last sample and the inverse temperature of the $i$-th chain,
and analogously for $j$.
If the swap is accepted the position of the two samples are exchanged in the different selected chains.
This procedure allows the information of the high-temperature chains
to propagate to the low-temperature ones and viceversa,
improving the correlation between the samples.
Another key feature of parallel tempering is that it satisfies the detailed balance condition~\cite{Sambridge:2013},
required for convergence of the MCMC chains.

Finally, the stopping criterion Eq.~\eqref{eq:mcmcstop} is estimated for the untempered chains;
when it is satisfied, the sampler stops and reproduces the posterior samples using 
only the chains of the $T=1$ sub-ensemble.
Furthermore, using the auxiliary coefficient $\beta$ 
and thermodynamic integration~\cite{Goggans:2004,Lartillot:2006},
it is possible to write the evidence as
\be
\label{eq:ptZ}
\log p(\data|H) = \int_0^1E_\beta\big[ \log p(\data|\params,H)\big] \,\d\beta\,.
\ee
where the expectation value is estimated using the tempered posterior, i.e.
\be
\label{eq:tempexp}
E_\beta\left[ f(\params)\right] = \int_{\paramspace} f(\params )\, \big[ p(\data|\params,H) \big]^{\beta} p(\params|H) \, \d\params\,.
\ee
Eq.~\eqref{eq:ptZ} can be estimated through numerical integration.
The terms $E_\beta\left[ \log p(\data|\params,H)\right]$ 
are estimated over the initial $\beta$ ladder
applying Eq.~\eqref{eq:mcmcapprox} to the tempered posterior samples
and the integral is approximated using the trapezoidal rule.
The PTMCMC represents an improved version of a standard MCMC technique, 
that aims to provide much accurate estimations of the evidence.
However, the accuracy of the estimation Eq.~\eqref{eq:ptZ} strongly depends on the number of employed temperatures: in complex situations, 
the total number of chains needed to accurately estimate the evidence
could overcome the number of available processes, affecting the efficiency of the sampler~\cite{Gupta:2018}.

\section{Nested sampling} 
\label{app:nest}

The nested sampling~\cite{Skilling:2006,Sivia2006} 
is a finely-designed 
Bayesian technique designed to accurately estimate the evidence integral
and, nevertheless, it provides a set of posterior samples as final product of the algorithm.
The strength of this technique is the capability to succeed even in cases of high-dimensional parameter space or multimodal distributions.
Nested sampling computes the evidence identifying nested isoprobability contours
and estimating the amount of prior volume enclosed by each level.
The main advantage of this technique is the reduction of 
the multidimensional integral Eq.~\eqref{eq:evidence} to a
one-dimensional problem~\cite{Chopin_2010,Betancourt_2011}, introducing the variable
\be
\label{eq:nsX}
X(\lambda) = \int_{p(\data|\params,H)>\lambda}p(\params|H) \,\d\params\,.
\ee
The quantity $X(\lambda)$ is usually labeled as {\it prior mass} 
and it is the cumulative prior volume covering all likelihood values greater than $\lambda$. 
The prior mass takes values in the range $[0,1]$, where
$X=1$ corresponds to the entire prior volume.
Then, we can rewrite the likelihood as function of the prior mass,
i.e. $p(\data|X(\lambda),H) = \lambda$, from which follows,
\be
\label{eq:nsZ}
p(\data|H)= \int_0^1 p(\data|X,H) \,\d X\,.
\ee
Eq.~\eqref{eq:nsZ} has a further advantage: 
by definition, the likelihood $p(\data|X,H)$
is a monotonic decreasing function of $X$.
Then, for $X\to 0$, the likelihood tends to its maximum value.
Accomplishing the transformation $\params \to X$ 
involves dividing the unit prior mass range into small bins and sorting them by likelihood.

A standard nested sampling routine requires 
an input number of live points $n_{\rm live}$
and a real positive number $\zeta$ 
representing the final tolerance of the computation.
The live points are samples of the parameter space 
that are evolved during the routine: 
starting from a set of $n_{\rm live}$ initial samples 
(usually extracted from the prior distribution),
the live point with lowest likelihood value, say $\params_i$, is discarded and 
replaced with a new point $\params^*$ extracted from the prior distribution
that satisfies the relation $ p(\data|\params^*,H) > p(\data|\params_i,H) $.
The new point $\params^*$ is usually proposed using internal MCMC routines (see App.~\ref{app:proposal}).
The procedure is repeated taking the lowest-likelihood 
live point at every iteration, such that the algorithm starts 
inspecting the entire prior volume ($X_0=1$),
and it moves toward lower value of the prior mass,
\be
\label{eq:nsXseries}
0 < X_i <\dots < X_2<X_1<X_0=1 \,,
\ee
up to the the most likely value(s), 
where the majority of the likelihood volume is located.
At the $n$-th iteration, the evidence is approximated from Eq.~\eqref{eq:nsZ} 
using trapezoidal rule,
\be
\label{eq:nsZapprox}
p_n(\data|H)\approx \frac{1}{2} \sum_{i=1}^{n} \left( X_{i-1} - X_{i+1}\right)\, p(\data|\params_i,H)\,,
\ee
where $X_i$ is estimated with the expectation value~\cite{Betancourt_2011},
\be
\label{eq:nsXexp}
E[X_i] = e^{-i/n_{\rm live}}\,.
\ee
From Eq.~\eqref{eq:nsXexp}, we can deduce that 
the average volume occupied by a live point corresponds to
the $n_{\rm live}$-th part of the current prior mass.
Then, increasing $n_{\rm live}$,
the sampling will perform a finer grained inspections of the 
prior volume. 
In the nested sampling context,
the $i$-th extracted sample is taken as representative element of the respective 
 isoprobability level of likelihood and, since the algorithm accepts strictly increasing likelihood values, 
 it ensures that each level is nested in the previous one.
Then, the overall evidence is computed summing all the likelihood contributions from each nested
level weighted on the expected difference in prior mass, according with Eq.~\eqref{eq:nsXexp}.
The specific stopping condition depends on the requested sampler.
In general, 
if the algorithm converged to the global maximum-likelihood value,
at the $n$-th iteration the evidence is expected to vary at most of 
\be
\label{eq:nsdelta}
\Delta_n = \max_{i\le n}\big[ p(\data|\params_i,H)\big]\cdot X_n \,,
\ee
where $\max_{i\le n}[ p(\data|\params_i,H)]$ 
is the maximum likelihood discovered up to the $n$-th iteration.
Then, the general stopping criterion 
requires that the estimated evidence is not expected to 
change more than a factor $e^\zeta$, i.e.
\be
\label{eq:nsstop}
\zeta \ge \log \left(1+ \frac{\Delta_n}{p_n(\data|H)}\right) \,.
\ee
When the stopping condition is satisfied, the sampler stops and 
it includes the contributions of the remaining live points to the overall evidence.
Then, the posterior distribution can be reconstructed by the chain of collected samples, 
weighting each point according with its probability distribution,
\be
\label{eq:nspost}
p(\params_k|\data,H)\approx \frac{\left( X_{k-1} - X_{k+1}\right)}{2} \, \frac{p(\data|\params_k,H)}{p(\data|H)}\,,
\ee
where the index $k$ runs over the extracted samples~\cite{Hol2006_resampling}.

The nested sampling routine offers a much better architecture for 
evidence estimation than MCMC techniques.
In general, the estimated log-evidence carries a statistical uncertainty
inversely proportional to $n_{\rm live}$
due to the marginalization over the prior mass; while,
numerical errors are dominated by the use of point estimates and by
the length of the MCMC sub-chains $n_{\rm MCMC}$ used to propose new samples,
as shown in Ref.~\cite{Veitch:2009hd}. 
This inefficiency can be suppressed estimating the ACL 
of the MCMC sub-chains and proposing a new sample independent of the previous one. 
Note that also the estimation of the posterior samples Eq.~\eqref{eq:nspost} is affected by
statistical and numerical uncertainty~\cite{Higson_2018}.

 The {\tt cpnest}~\cite{cpnest} 
 software represents an exemplary implementation of a standard nested sampling:
 the code is designed to be nicely interfaced with user-defined models
 and its sampling methods can be easily customized.
 On the other hand, {\tt dynesty}~\cite{Speagle_2020}
 takes advantage of flexible bounding
 methods~\cite{2014arXiv1407.5459B,2017arXiv170704476B},
 that aim to define isoprobability contours 
 in order to exclude least likely regions of the parameter space
 improving the robustness of the algorithm.
  Moreover, {\tt dynesty} provides an implementation of 
 dynamic nested sampling~\cite{dynamic_ns}:
 this technique allocates an adaptive number of live points at each iteration iteration $i$,
 i.e. $n_{\rm live}\equiv n_{\rm live}(i)$. 
 Since the change in prior volume at a given iteration depends on the number of live points,
 as shown in Eq.~\eqref{eq:nsXexp}, 
the possibility of varying $n_{\rm live}$ gives the algorithm the freedom to 
control the effective resolution of the sampling as a function of prior mass,
adapting it to the shape of the posterior in real time and 
improving the evaluation of the posterior density.
Since the architecture of the dynamic nested sampling differs from the standard,
it requires modified methods in order to compute the evidence and estimate the posterior.
By default, the importance of each sample  for the evidence computation
is proportional to the amount of the posterior density 
enclosed in the prior mass probed by that point.

 \section{Proposal methods} 
\label{app:proposal}

The exploration of the parameter space is defined 
by proposal methods which aim to move a sample 
toward a more likely position in the parameter space independent from the previous.
The efficiency of the proposals determines the rate of acceptance 
of new samples and it affects the final ACL and subsequently the efficiency of the whole sampler.
It follows that these tools are fundamental for the chains progress 
and a generic, broad and varied combination of proposal methods is needed to
accurately inspect the entire parameter space.
For this reason, {\bajes} implements an assorted combination of proposal methods.

Before discussing the specific proposals implemented in {\bajes},
we observe that a generic proposal method requires the introduction of a proposal distribution $q$, 
in order to satisfy the detailed balance.
A proposal distribution $q(\params^*|\params_i)$
 quantifies the probability of proposing $\params^*$ given $\params_i$.
A symmetric proposal is such that the proposed point is fully independent from the initial sample,
i.e. $q(\params^*|\params_i) \propto 1$ for every $(\params^*,\params_i)$.
The interested reader might 
look at Ref.~\cite{Hastings1970,roberts_rosenthal_2007}
for details on Markovian process.

\begin{itemize}
\item {\it Prior proposal}: A new point is extracted from the prior distribution
generating a sample uniformly distributed over an hyper-cube 
and projecting it in the current parameter space according with 
the prescribed prior.
This method could show low acceptance 
on long-time scales,
especially for complex posterior distribution. However,
it can improve the exploration of the parameter space
and it does not require expensive computations.	
\item {\it Stretch proposal}: This method is introduced in Ref.~\cite{random_walk} and it make use of an auxiliary 
random sample $\params_a$ 
extracted from the history
of the chains. A new sample $\params^*$
is proposed
from the initial position $\params_i$
according with
\be
\label{eq:str_prop}
\params^* = \params_a + \xi\,(\params_i-\params_a)\,.
\ee
where $\xi$ is a scale factor. 
As pointed out in Ref.~\cite{stretch}, this proposal is symmetric
if the probability density of scaling factor $g(\xi)$ satisfy 
the condition $g(1/\xi)=\xi\,g(\xi)$.
The {\bajes} implementation adopts the settings presented in Ref.~\cite{Foreman_Mackey_2013}.
In order to satisfy the detailed balance, 
this proposal method requires an acceptance factor,
\be
\label{eq:stretch_f}
\frac{q(\params_i|\params^*)}{q(\params^*|\params_i)} =\xi ^{n_{\rm dim}-1}\,,
\ee							
where $n_{\rm dim}$ is the number of dimensions
of the parameter space.
Eq.~\eqref{eq:stretch_f} 
is computed conditioning of the target distribution on the
trajectory described by Eq.~\eqref{eq:str_prop}.
The method shows a good adaptation to 
arbitrary shapes of the distributions~\cite{random_walk} but 
it might become inefficient for multimodal cases.
\item {\it Random walk proposal}: This method is introduced in Ref.~\cite{random_walk}.
The new sample $\params^*$ is extracted from 
a multivariate normal distribution,
centered in the initial sample $\params_i$
and with covariance defined using a subset of $N$ auxiliary points randomly extracted from the history of the chains.
By default, {\bajes} walk proposal uses $N=25$.
This symmetric method is efficient with 
unimodal distributions and it can
improve the correlation between the samples. However, it becomes 
inefficient in case of complex posterior distribution
since it is not capable to arbitrarily adapt its shape.
\item {\it Replacement proposal}: This method is introduced in Ref.~\cite{random_walk}.
A subset of $N$ auxiliary points 
$\bar \paramspace\equiv\{\bar\params_k\}$ is randomly chosen from the 
history of the chains and it is used to identify a probability distribution $p(\params|\bar \paramspace)$ from which the new samples will be extracted.
The idea is to estimate $p(\params|\bar \paramspace)$ such that is it capable to 
approximate the target distribution, increasing the acceptance.
Moreover, the estimation can be refined and adapted during the 
exploration of the parameter space.
In order to estimate $p(\params|\bar \paramspace)$, the replacement proposal
implemented in {\bajes} employs a Gaussian kernel density estimation
with $N=25$.
However, this method does not access to unexplored regions of 
the parameter space, with the possibility of leading to highly correlated chains. 
Furthermore, this method is not symmetric and the proposal distribution
is described by $p(\params|\bar \paramspace)$.
\item {\it Differential evolution proposal}: This method is introduced in Ref.~\cite{Nelson_2013} and it aims to solve problems due to multimodal distributions
using a differential move based on the information on the explored samples.
Two auxiliary random samples $\{\params_{a,b}\}$ 
are extracted from the 
from the history of the chains
and a new sample $\params^*$
is proposed from $\params_i$ as 
\be
\label{eq:de_prop}
\params^* = \params_i + \gamma \left(\params_a - \params_b\right)\,,
\ee
where $\gamma$ is a scale factor
whose value is randomly extracted when a new sample is proposed.
The differential evolution proposal of {\bajes} assign $\gamma=1$ 
with 50\% of probability in order to
to improve the mixing between different modes. 
The remaining 50\% ot the time, the scale factor is extracted from a 
normal distribution such that $\gamma\sim {\rm N}(0,2.38 /\sqrt{2n_{\rm dim}})$, 
where $n_{\rm dim}$ is the dimension of the parameter space.
This choice has been proved to increase the acceptance of the algorithm~\cite{DEprop1,DEprop2}.
In general, differential evolution is capable to capture 
linear correlations and improve mixing between 
different modes, however it can perform poorly in
more complicated scenarios.
\item {\it Eigenvector proposal}: This method computes the covariance
from history of the chains and estimates the relative 
eigenvectors. Then, the new point is proposed moving the
initial sample along a random eigenvector of the covariance
with a scale prescribed by the respective eigenvalue.
As shown in Ref.~\cite{Veitch:2014wba}, this method can
improve the efficiency of the sampler and decrease the correlation
of the chains.
\item {\it Ensemble slice proposal}: This method has been introduced in
Ref.~\cite{Karamanis:2020zss}
and it represents an ensemble-based 
improvement of the standard slice 
proposal~\cite{neal2003}.
Let us call $\params_i$ and $p_i$ respectively the initial sample and its probability.
The method extracts a value $y\sim{\rm U}(0,p_i)$ and
estimates a direction in the parameter space $\boldsymbol{\eta}$,
resorting to the information
of the ensemble samples and
using differential and Gaussian estimations.
Then, the initial sample is moved along the slide defined by the direction $\boldsymbol{\eta}$
and a new point $\params^*$ is proposed when the associated probability value $p^*$ is
greater than $y$.
With respect to standard slice sampling~\cite{neal2003},
this method takes advantage of adaptive scale factors refined 
during the evolution of the sampler that increase the efficiency of 
the proposal and drastically reduce the correlation between the 
collected samples.
On the other hand, the ensemble slice proposal requires multiple
likelihood evaluations; then,
when the likelihood is computationally expensive,
it might affect the computational cost 
of the whole algorithm.
\item {\it GW-targeted proposal}: 
As discussed in Ref.~\cite{Veitch:2009hd,Veitch:2014wba},
generic posterior distributions of GW signal parameters 
show many multimodalities and 
large correlations between different parameters.
Moreover, its shape is usually elongated and non-regular.
Then, in order to properly and efficiently explore the parameter 
space, it is useful to inform the proposal method with 
known structures expected in the posterior distributions
and it is possible to take advantage of the
determined analytical dependencies on the extrinsic parameters.
The GW-targeted proposal implemented in {\bajes}
explores most of the GW specific methods introduced
in Ref.~\cite{Veitch:2009hd,Veitch:2014wba};
such as sky reflection,
Gibbs sampling for the luminosity distance,
and specific methods to explore the 
phase-polarization and the					
distance-inclination correlations.								
\end{itemize}

\section{Simple example}
\label{app:example}

\begin{figure}[t]
	\centering 
	\includegraphics[width=0.49\textwidth]{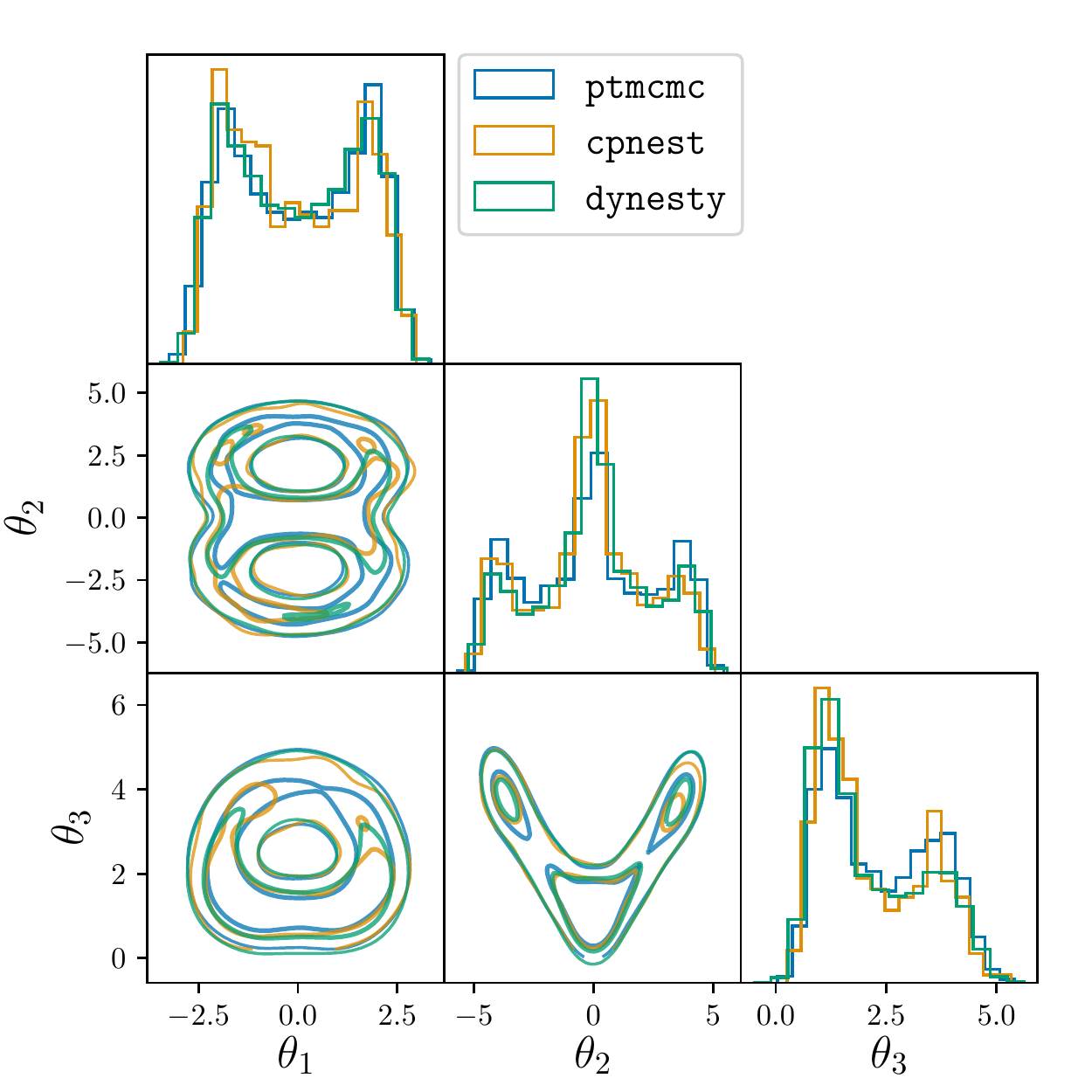}
	\caption{Posterior distribution for the parameters $\{\theta_1,\theta_2,\theta_3\}$
		discussed in App.~\ref{app:example}. The blue lines correspond to the 
		results obtained with PTMCMC sampler, while the yellow and green lines refer 
		respectively to {\tt cpnest} and {\tt dynesty} results. 
		The marginalized contours are the 50\% and 90\%
		credible regions.}
	\label{fig:exe}
\end{figure}

In this section we show a simple example of Bayesian inference 
performed with the samplers implemented in {\bajes}.
We use a 3-dimensional parameter space $\params = \{\theta_1,\theta_2,\theta_3\}$
bounded to $\theta_i\in[-8,+8]$ for $i=1,2,3$ with an uniformly distributed prior distribution.
For the sake of simplicity, we employ a fully-analytical likelihood function;
however, we include multimodalities and non-regular shapes in order to 
test the behaviour of the samplers. Introducing the auxiliary variables
\be
\label{eq:exe:r}
r_\pm = \sqrt{\theta_1^2 + (\theta_2\mp2)^2}\,,\quad
\zeta = \theta_3-\sqrt{1+\theta_2^2}\,,\\
\ee
we define the likelihood as
\be
\label{eq:exe:like}
p(d|\params,H) = \left[ e^{-5(r_+ - 2)^2} + e^{-5(r_- - 2)^2}\right] e^{-5\zeta^2}\,.
\ee
To give an idea, the isoprobability contours described by this function
are roughly similar to the union of two toroidal surfaces 
where the second is identical to the first except for a rotation of $\pi/2$.
The likelihood function in Eq.~\eqref{eq:exe:like}
can be numerically integrated using the 
quadrature rule, resulting in the evidence $\log p(d|H)\approx -5.5583$
with an error of the order of $O(10^{-5})$.

In order to infer the described model, 
{\bajes} provides an
user-friendly and simple-to-use interface for generic Bayesian inference.
In order to define the prior distributions, 
it is sufficient to write a prior configuration file
specifying the name of the parameters and the bounds.
For the case discussed above, we can write
the following {\tt prior.ini} file.
		\begin{algorithm}[H]
	\caption{{\tt prior.ini}}
	\begin{algorithmic}[1]
		\State {\tt [x1]}
		\State {\tt min=-8}
		\State {\tt max=+8}
		\State
		\State {\tt [x2]}
		\State {\tt min=-8}
		\State {\tt max=+8}
		\State
		\State {\tt [x3]}
		\State {\tt min=-8}
		\State {\tt max=+8}
	\end{algorithmic}
\end{algorithm}

Then, the likelihood function can be written 
in an auxiliary {\py} file
defining a {\tt log\_like} method.
This method will be imported by the {\bajes} routine and used 
to determine the log-likelihood function for each sample.
We observe that the only argument of the 
customized {\tt log\_like} method has to be a dictionary  
whose keywords are specified by the prior file.
This procedure easily allows the user the 
make use of the {\bajes} inference
introducing arbitrary external data or packages.
Following our example, we write the following pseudo-code.
		\begin{algorithm}[H]
	\caption{{\tt like.py}}
\begin{algorithmic}[1]
	\State {\tt from math import sqrt, exp, log}
	\State 
	\State {\tt def log\_like(p):} 
%	\State  $\qquad${\tt x1 = p['x1']}
%	\State  $\qquad${\tt x2 = p['x2']}
%	\State  $\qquad${\tt x3 = p['x3']}
	\State  $\qquad${\tt rp = sqrt(p['x1']**2 + (p['x2'] - 2)**2)}
	\State  $\qquad${\tt rm = sqrt(p['x1']**2 + (p['x2'] + 2)**2)}
	\State  $\qquad${\tt zt = p['x3'] - sqrt(1 + p['x2']**2)}
	\State  $\qquad${\tt G1 = exp(-5*(rp - 2)**2) }
	\State  $\qquad${\tt G2 = exp(-5*(rm - 2)**2)}
	\State  $\qquad${\tt return log(G1 + G2) - 5*zt**2}
\end{algorithmic}
\end{algorithm}

Once the model is defined, the PE job can be submitted
with the command
\begin{center}
	{\tt python -m bajes -p prior.ini -l like.py}\,
	\footnote{The full list
				of input arguments 
				can be visualized with 
				the command {\tt python -m bajes --help}.}.
\end{center}
For our exercise we employ three samplers:
the PTMCMC, the nested sampling with {\tt cpnest}
and the dynamic nested sampling with {\tt dynesty}.
The PTMCMC algorithm estimated a log-evidence equal to $-6.4\pm5.0$,
where the reported uncertainty is the standard deviation.
The estimation agrees with the numerical result; however, 
its uncertainty is of the same order of the measurement.
This reflects the inability of MCMC methods to meticulously integrate 
the features of the targeted parameter space.
On the other hand, 
{\tt cpnest} estimated a log-evidence equal to $-5.50\pm0.09$ and
the dynamic nested sampler of {\tt dynesty} 
returned the value of $-5.58\pm0.14$. 
These results highlight the strength of the nested sampling with respect
to MCMC techniques in the evidence evaluation.
Figure~\ref{fig:exe} shows the marginalized posterior distributions
extracted from the posterior samples. 

\bibliography{local,references}

\end{document}